\documentclass[10pt]{article}
\usepackage[margin=0.615in]{geometry}
\usepackage{cite}
\usepackage{hyperref}
\usepackage{microtype}
\DisableLigatures[f]{encoding = *, family = * }
\usepackage[utf8]{inputenc}
\usepackage{graphicx}% Include figure files
\usepackage{adjustbox}% adjust sizes
\usepackage{dcolumn}% Align table columns on decimal point
\usepackage{bm}% bold math
\usepackage{ulem}
\usepackage{xcolor}
\usepackage{amsmath}
\usepackage{amssymb}
\usepackage{mathtools}
\usepackage{tabularx}
\usepackage{wrapfig}
\usepackage{nameref}
\usepackage{siunitx}
\usepackage{ulem}
\usepackage{lscape}
\usepackage{tocloft}

\newcommand{\Rho}{\mathrm{P}}
\newcommand{\NN}[1]{ \langle #1 \rangle}
\newcommand{\avgrhoN}[1]{ \left\langle #1 \right\rangle^{(N)}_{\vec{\rho}}}
\newcommand{\avgrho}[1]{ \left\langle #1 \right\rangle_{\rho}}
\newcommand{\avg}[1]{ \left\langle #1 \right\rangle}
\newcommand{\avgcp}[1]{ \left\langle #1 \right\rangle_{cp}}

\newcommand{\avgphi}[1]{ \langle #1 \rangle_{\phi}}
\newcommand{\avgvarphi}[1]{ \left\langle #1 \right\rangle_{\varphi}}
\newcommand{\vecexp}[2]{\vec{#1}^{(#2)}}
\newcommand{\rhoN}{\vec{\rho}^{(N)}}
\newcommand{\rhoNminus}{\vec{\rho}^{\ (N{-}1)}}
\newcommand{\rhocpvec}{\vec{\rho}^{\ (cp)}}
\newcommand{\rhocp}{\rho^{(cp)}}
\newcommand{\avgrhoNcp}[1]{ \left\langle #1 \right\rangle_{\rho^{(cp)}}}
\newcommand{\fNminus}{f_{N{-}1}}

\newcommand{\supbracket}[2]{#1^{(#2)}}
\newcommand{\minom}[1]{\min_{\supbracket{\omega}{#1}}}
\newcommand{\minnu}[1]{\min_{\supbracket{\upsilon}{#1}}}
\newcommand{\cpp}{\supbracket{C}{{+}{+}}}
\newcommand{\cpm}{\supbracket{C}{{+}{-}}}
\newcommand{\cmp}{\supbracket{C}{{-}{+}}}
\newcommand{\cmm}{\supbracket{C}{{-}{-}}}

\begin{document}

\begin{center}
{\Large
\textbf{Thermodynamic stability and critical points in multicomponent mixtures with structured interactions}
}
\bigskip
\\
Isabella R.\ Graf$^{1\star}$,
Benjamin B. Machta$^{1,2}$
\\
\medskip
\noindent\upshape$^{1}$Department of Physics, Yale University, New Haven, Connecticut 06511, USA
\\
\noindent\upshape$^{2}$Quantitative Biology Institute, Yale University, New Haven, Connecticut 06511, USA
\\
\noindent\upshape$^{\star}$ isabella.graf@yale.edu

\end{center}

\section*{Abstract}
Theoretical work has shed light on the phase behavior of idealized mixtures of many components with random interactions.
But typical mixtures interact through particular physical features, leading to a structured, non-random interaction matrix of lower rank.
Here we develop a theoretical framework for such mixtures and derive mean-field conditions for thermodynamic stability and critical behavior.
Irrespective of the number of components and features, this framework allows for a generally lower-dimensional representation in the space of features and proposes a principled way to coarse-grain multicomponent mixtures as binary mixtures.
Moreover, it suggests a way to systematically characterize different series of critical points and their codimensions in mean-field.
Since every pairwise interaction matrix can be expressed in terms of features, our work is applicable to a broad class of mean-field models.

\vspace{20pt}

\section*{Main text}

Determining the phase behavior of mixtures is an important goal of Statistical Physics.
But while the thermodynamics of mixtures with few components is well understood theoretically~\cite{Mao2019}, most functional mixtures are made up from a large number of distinct components, and the principles underlying the phase and critical behavior of such multicomponent mixtures are less clear.  There have been substantial steps towards understanding these systems, but only in limiting cases.
Sear and Cuesta~\cite{Sear2003} and subsequent followups~\cite{Jacobs2013, Jacobs2017} determined conditions for phase separation in idealized mixtures with random, independent pairwise interactions.
Taking a very different limit, Sollich and others~\cite{Sollich2001, Sollich2002, DeCastro2017} have made progress for polydisperse mixtures interacting through a continuous distribution of attributes.

Different from these theoretical studies, many physical examples are made up of defined components whose interaction structure is governed by the physical details that underpin them.
In phase-separation prone lipid membranes, though there are thousands of chemical species, interactions are thought to be primarily driven by just a few features -- interactions between headgroups, the degree of acyl-chain saturation, and the mismatch between hydrophobic height ~\cite{Bigay2012, Quinn2012, BernardinodelaSerna2016, Cebecauer2018, Sezgin2017}.
In protein condensates, interactions are likely mediated by a combination of specific motifs like repetitive binding domains, and less specific electrostatic and hydrophobic interactions~\cite{Banani2017, Ditlev2018, Berry2018}.
This observation suggests that in both cases the resulting effective pairwise interaction matrix is non-random in a particular way: its rank, given by the number of independent features, can be considerably smaller than its dimension, given by the number of components.
In other examples it might be less clear what features mediate interactions, but an effectively low-rank interaction matrix is likely common to most mixtures made up of many components.
One class of examples are fluids like petroleum for which an approximation in terms of lower-dimensional interaction parameters has been successfully applied~\cite{Hendriks1992}.

To systematically investigate the role of such a low-dimensional interaction structure for phase behavior, in this work we develop a theoretical framework to study the phase behavior of mixtures with many components but structured interactions.
We show that the stability of phases and critical behavior can be understood in a ``feature space", which is typically much lower-dimensional than the space of component densities.

\begin{figure*}[t]
\begin{center}
\includegraphics[width=0.8\textwidth]{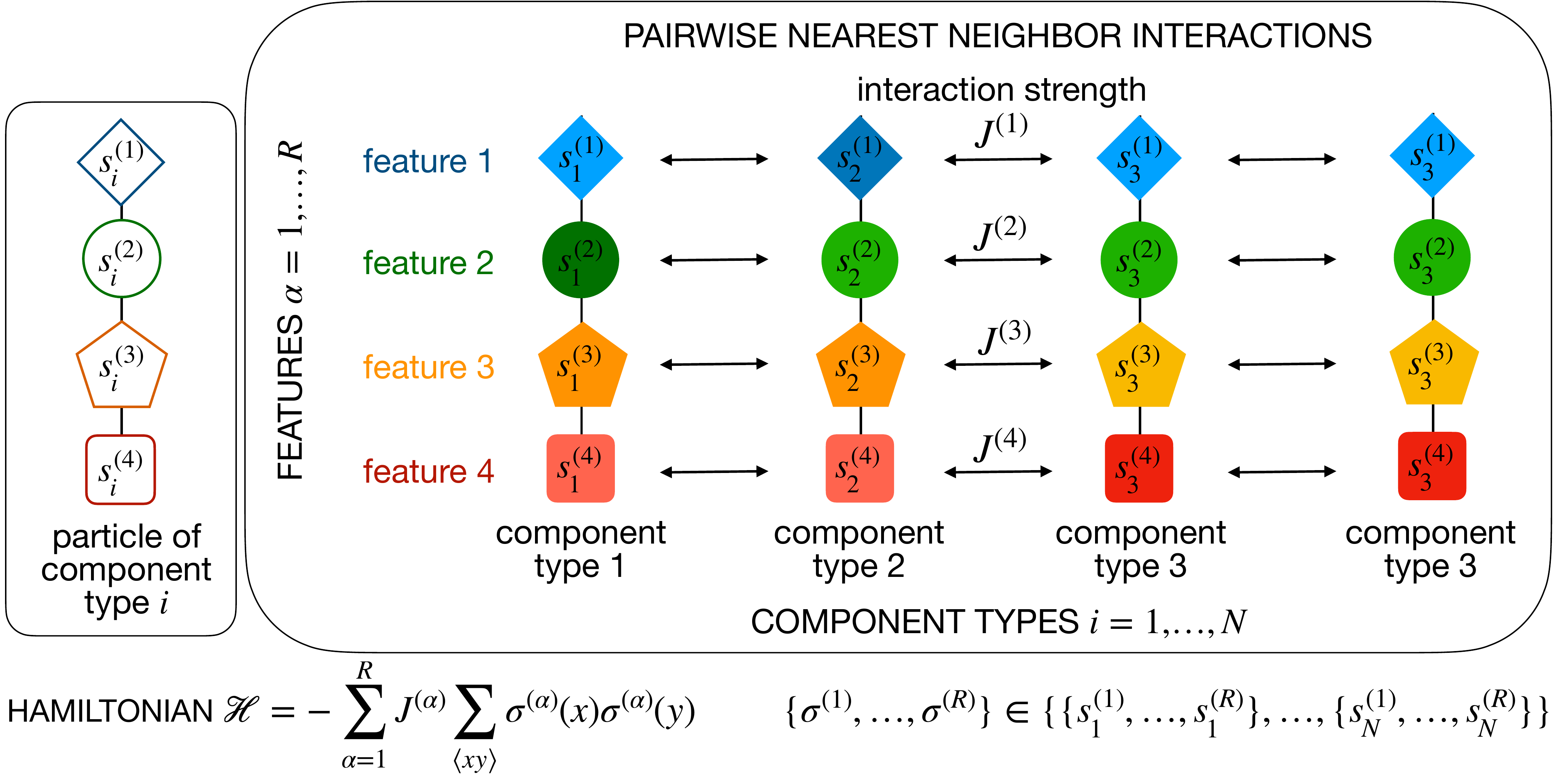}%
\end{center}
\caption{\label{fig:model} Model: The multicomponent mixture comprises $N$ different component types (``components").
Component $i=1,{...},N$ is characterized by $R$ features $\supbracket{s}{\alpha}_i, \alpha=1,{...},R$, each of which conveys an additive pairwise interaction with interaction strength $\supbracket{J}{\alpha}$ between neighboring components.
}
\end{figure*}

\noindent \textbf{Mean-field model.}  We specifically consider a family of multicomponent models with a pairwise interaction matrix of variable rank (see Fig.~\ref{fig:model}).
The mixture is made up of $N$ different component types.
Component type $i$ is characterized by a ``feature vector" $\vec{s}_i$ composed of $R$ real features $\supbracket{s}{\alpha}_i, \ \alpha{=}1,\ldots, R$.
Each feature conveys an additive, Ising-like interaction with interaction strength $\supbracket{J}{\alpha} {\neq} 0$~\footnote{Note that in cases where the different components do not share all features, one can set $\supbracket{s}{\alpha}_i=0$ for components $i$ that do not possess a particular feature $\alpha$.}.
The corresponding lattice Hamiltonian reads $\mathcal{H} {=} {-}\sum_{\alpha=1}^R \supbracket{J}{\alpha} \sum_{\NN{xy}} \supbracket{\sigma}{\alpha} (x) \supbracket{\sigma}{\alpha} (y)$ where $\sum_{\NN{xy}}$ is the sum over all neighboring sites $x, y$ on the lattice and  the spins take the values $\supbracket{\sigma}{\alpha} (x) {=} \supbracket{s}{\alpha}_i {\in} \mathbb{R}$ if site $x$ is occupied by component type $i$ (see~\cite{Wilding2008} for a related model with a single feature).
In a mean-field approximation, our system is described by a Flory-Huggins-like free energy density per $k_B T$~\cite{Huggins1941, Flory1941} 
$
f_N {=} 
\sum_{i=1}^N \rho_i \log \rho_i {-} \sum_{i,j=1}^N \rho_i \chi_{ij} \rho_j 
$
where the densities $\rho_i$ are subject to the incompressibility constraint $\sum_i \rho_i {=} 1$.
The interaction matrix is given by
\begin{align}
\chi_{ij}  := \sum_{\alpha=1}^R \frac{z \supbracket{J}{\alpha}}{2 k_B T} \supbracket{s}{\alpha}_i \supbracket{s}{\alpha}_j =: \sum_{\alpha=1}^R \supbracket{C}{\alpha} \supbracket{s}{\alpha}_i \supbracket{s}{\alpha}_j, \label{eq:interaction_matrix_rank}
\end{align}
for a lattice coordination number $z$. 
For $R$ features, the interaction matrix is of rank $r{\leq} R$.
While this decomposition into features may be motivated by the physics of interactions, any real and symmetric interaction matrix $\chi_{ij}$ can be decomposed in this way, with $R{\leq} N$, eigenvectors $\supbracket{s}{\alpha}_i$ and eigenvalues $\supbracket{C}{\alpha} |\supbracket{\bm{s}}{\alpha}|^2$; see~\cite{Hendriks1992} for a related (eigen)decomposition in the context of petroleum.
A (precise) way to think about the features is thus as eigenvectors of the interaction matrix.
Furthermore, as long as interactions are pairwise and meaningfully described by mean-field, our choice of representing components in terms of additive Ising-like features is entirely general.
For now, we assume all eigenvalues to be positive, but discuss the general case in the SM and briefly below.

\noindent \textbf{Thermodynamic stability and critical points.} 
The main challenge in working with mixtures with $N{\gg} 1$ components is that they are embedded in a very high-dimensional space of densities.
We now develop an analytic framework, wherein the mixtures are instead represented in the corresponding, potentially much lower-dimensional feature space.
To this end, we use matrix inversion techniques and will then successively derive conditions for local thermodynamic stability and the occurrence of different series of critical points. 

In general, the thermodynamic behavior of the multicomponent mixture is determined by the free-energy landscape in the $N$-dimensional space of densities $\vec{\rho}^{(N)}$.
Due to the incompressibility constraint, the densities are not independent, $\rho_N{=}1{-}\sum_{i=1}^{N-1} \rho_i$, leaving the free energy density $f$ a function of $N{-}1$ densities and temperature.
 The mixture is (locally) thermodynamically stable if the Hessian matrix ($i,j{=}1{,}{\ldots}{,}N{-}1$)
\begin{align}
H_{ij} &:= \frac{\partial^2 f}{\partial \rho_i \partial \rho_j} = \underbrace{\delta_{ij} \frac{1}{\rho_i} + \frac{1}{\rho_N}}_{=:K_{ij}} - \underbrace{\sum_{\alpha{=}1}^R r^{(\alpha)}_i r^{(\alpha)}_j}_{:=(UU^T)_{ij}}, \label{eq:Hessian} 
\end{align}
is positive definite.
Here 
$r^{(\alpha)}_i:=\sqrt{2C^{(\alpha)}} \left(s^{(\alpha)}_i - s^{(\alpha)}_N\right)$
is the rescaled and shifted feature vector.
While in the presence of a solvent, component $N$ is most straightforwardly associated with this solvent, all physical results are ultimately independent of the choice of reference point.
At high temperatures (large $T$ $\rightarrow$ small $\supbracket{C}{\alpha}$ $\rightarrow$ small $\supbracket{r}{\alpha}$), the system is dominated by entropy  ($H {\approx} K$) and all eigenvalues of the Hessian matrix are positive; the system is thermodynamically stable to perturbations which are local in composition.
At lower temperatures one or more of the eigenvalues can become negative, implying that the system can spontaneously lower its free energy by phase separating along the corresponding eigenvector.
The boundary of local thermodynamic stability is called the spinodal.
It corresponds to the submanifold of compositions and temperature where the smallest eigenvalue of the Hessian matrix is zero.
Since a matrix is invertible if and only if all of its eigenvalues are non-zero,  the spinodal is also the submanifold where the matrix becomes singular for the first time, starting from all positive eigenvalues.
Here we take advantage of the fact that $H$ is the sum of a positive definite matrix $K$ with inverse $K^{-1}_{ij} = \delta_{ij} \rho_i - \rho_i \rho_j$ and a lower rank contribution $UU^T$ arising from interactions (Eq.~\ref{eq:Hessian}).
Analogously to Ref.~\cite{Hendriks1988}, we use the Woodbury matrix identity~\cite{woodbury1950} on $K$ and $U U^T$ to invert $H$ and find that the Hessian matrix is invertible if and only if
$\mathbf{1} - U^T K^{-1} U =: \mathbf{1} - \text{Cov}$ is invertible; see SM.
Here, $\text{Cov}$ is the covariance matrix between the (rescaled) features:
\begin{align}
    \text{Cov}_{\alpha \beta} = \avgrhoN{r^{(\alpha)} r^{(\beta)}} - \avgrhoN{r^{(\alpha)}} \avgrhoN{r^{(\beta)}},
\end{align}
where the averages are taken with respect to the probability measure given by the mixture composition $\vec{\rho}^{(N)}$: $\avgrhoN{X} = \sum_{i=1}^N \rho_i X_i$.
The rank of the covariance matrix $R_{\mathrm{Cov}}$ corresponds to the maximal number of linearly independent feature vectors, $R_{\mathrm{Cov}}\leq R$; see SM.

\noindent \textit{Thermodynamic stability}
These results imply that the mixture becomes unstable when the largest eigenvalue of the covariance matrix $\supbracket{\lambda}{1} {=}1$.
Importantly, this condition is independent of the number of component types, including as limits the two-component mean-field Ising model and the infinite-component limit as discussed in the context of polydisperse systems~\cite{Sollich2001, DeCastro2018}.
\begin{figure*}[ht!]
\begin{center}
\includegraphics[width=0.8 \textwidth]{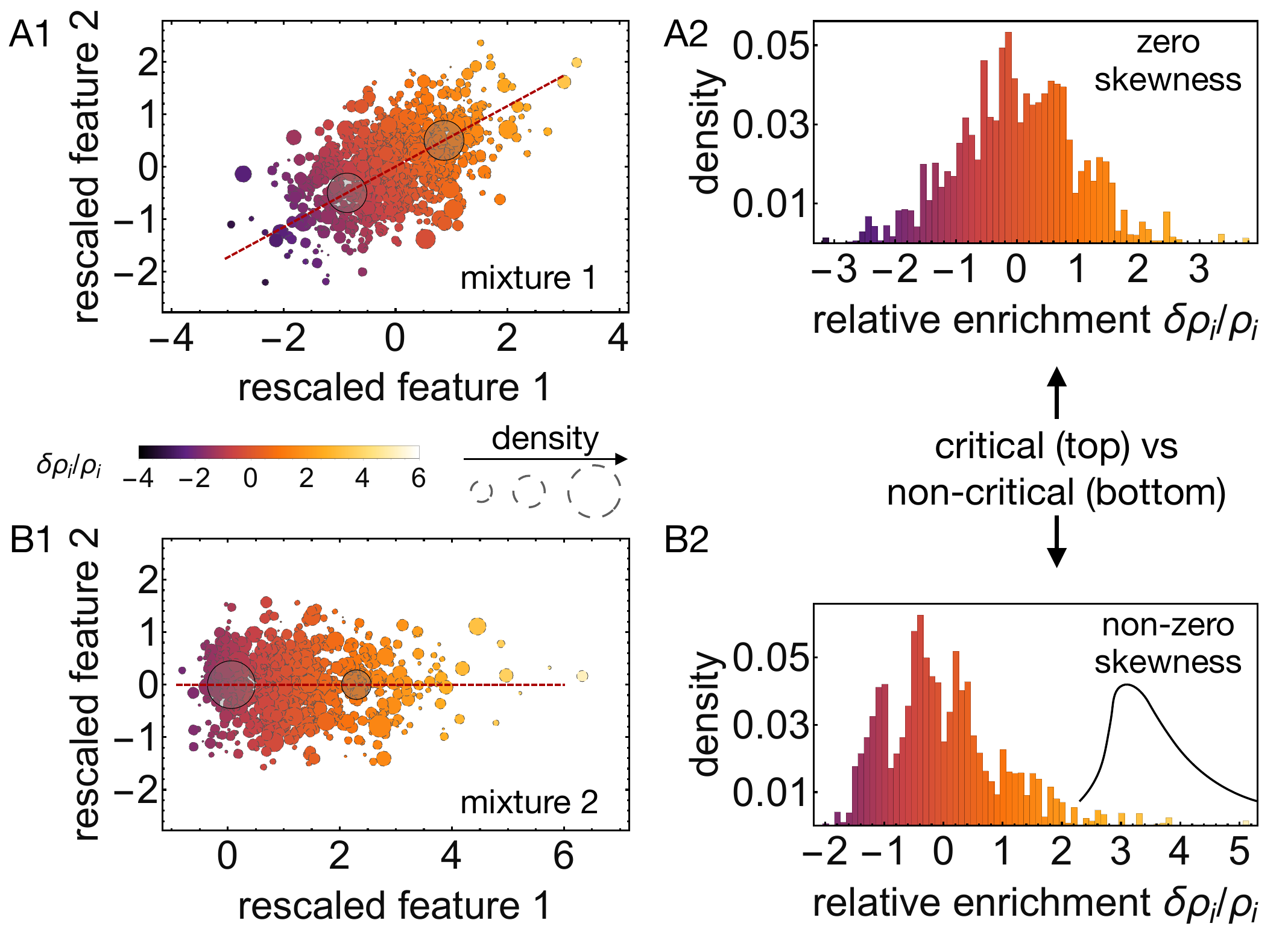}% 
\end{center}
\caption{\label{fig:spinodal_variance_equivalency_random} Left: Illustration of two multicomponent mixtures with $N{=}1000$ different component types (small, colorful disks; area of disk$_i$ ${\sim}$ density $\rho_i$) in $R{=}2$-dimensional feature space, together with the first principal component (1st PC; red dashed line) of their respective covariance matrix $\mathrm{Cov}$.
Both mixtures exhibit a variance of 1 along the 1st PC and are thus located on the spinodal.
The relative enrichment $\delta \rho_i/\rho_i$ of component $i{=}1, {\ldots}, 1000$ along the initial direction of phase separation right at the spinodal (color code; arbitrary units) is determined by the projection $\supbracket{E}{1}_i$ of the feature vector (relative to the mean) onto the 1st PC of $\mathrm{Cov}$,  see Eq.~\ref{eq:definition_Rgamma}.
Coarse-graining the multicomponent mixture as an effective binary mixture that preserves the location of the system with respect to the spinodal and critical manifold (by conserving the second and third cumulant along the 1st PC of $\mathrm{Cov}$) leads to a composition as shown by the large, translucent disks; see SM for details.
For a critical mixture (A), the composition of the binary mixture is symmetric, for a non-critical one (B), the densities of the two components are different. Note that all results are independent of global rotations or translations of the feature vectors or reflections $\supbracket{s}{\alpha}_i {\rightarrow} {-}\supbracket{s}{\alpha}_i \forall i$; see SM.
Right: For the multicomponent mixture to be critical, the skewness of the distribution of relative enrichments has to be zero (mixture 1 in A1 vs.\ mixture 2 in B1), see Eq.~\ref{eq:condition_critical}.}
\end{figure*}

\noindent \textit{Direction of instability}
In order to find the initial direction of phase separation at the spinodal we next determine the eigenvector corresponding to eigenvalue 0 (1) of the Hessian (covariance) matrix:
If $\mathbf{1} {-}\text{Cov}$ is invertible, the inverse of the Hessian matrix is given by
$H^{{-}1} = K^{{-}1} {+} K^{{-}1} U (\mathbf{1}{-}\text{Cov})^{-1} U^T K^{{-}1}$.
Using the eigendecomposition of the covariance matrix in terms of its (descending) eigenvalues $\supbracket{\lambda}{\gamma}, \gamma=1,{...}, R$ and corresponding orthonormal eigenvectors $\supbracket{V}{\gamma}$ (whose dependency on $\supbracket{\vec{\rho}}{N}$ we drop for conciseness)
$\text{Cov}_{\alpha \beta} = \sum_{\gamma=1}^R \supbracket{\lambda}{\gamma} \supbracket{V}{\gamma}_{\alpha}  \supbracket{V}{\gamma}_{\beta}$,
the inverse of the Hessian is
$H^{{-}1}_{ij} = \delta_{ij} \rho_i -  \rho_i \rho_j + \sum_{\gamma=1}^R \frac{1}{1-\supbracket{\lambda}{\gamma}} e^{(\gamma)}_i e^{(\gamma)}_j$,  with
\begin{align}
    \label{eq:definition_Rgamma}
    e^{(\gamma)}_i := \rho_i \supbracket{E}{\gamma}_i := \rho_i \sum_{\alpha=1}^R \supbracket{V}{\gamma}_{\alpha} \left( r^{(\alpha)}_i {-} \avgrhoN{r^{(\alpha)}} \right).
\end{align}
Close to the spinodal ($\supbracket{\lambda}{1} {\approx} 1$), the dominant term is $\frac{1}{1{-}\supbracket{\lambda}{1}} e^{(1)} \left(e^{(1)} \right)^T$.
Correspondingly, on the spinodal $e^{(1)}$ is the eigenvector of the Hessian with eigenvalue 0~\footnote{
Note that unless stated otherwise we assume that the covariance matrix has a non-degenerate maximal eigenvalue, $\supbracket{\lambda}{2} < 1$.}  and coincides with the direction of instability $H e^{(1)} = 0$ (see also Ref.~\cite{Sollich2001} for polydisperse systems).
Equation~\ref{eq:definition_Rgamma} implies that the relative enrichment $\delta \rho_i/\rho_i \sim \supbracket{e}{1}_i/\rho_i=\supbracket{E}{1}_i$ of component $i$ along the initial direction of phase separation at the spinodal (``partition coefficient")  is given by the deviations of the features from their mean, projected onto the first principal component (PC) of the feature distribution; see Fig.~\ref{fig:spinodal_variance_equivalency_random}. 

\noindent \textit{Ordinary critical points}
The spinodal marks the edge of local thermodynamic stability.  Except at special points, the spinodal lies within the binodal, the region of global thermodynamic stability.  
Points where the spinodal and binodal make contact are critical points (cp). 
At a usual critical point $\rhocpvec$, two phases become indistinguishable, corresponding to two minima and one maximum of the tilted Landau free energy $f{\rightarrow} f{-}\sum_i \rho_i \left. \partial_i f \right|_{cp}$ merging into one minimum. 
This merging occurs when the change in free energy along the direction of instability $\delta f{=}f(\rhocp {+} \epsilon \supbracket{e}{1}) {-} f(\rhocp)$ is zero up to order $\leq 3$ in $\epsilon$.
The first order term of the tilted free energy is zero by definition and $\left.(\partial_i \partial_j f) e^{(1)}_i \right|_{cp}{=}0$ as $\rhocpvec$ lies on the spinodal, yielding the following additional condition for the critical point:
\begin{align}
   \left.(\partial_i \partial_j \partial_k f)e^{(1)}_i e^{(1)}_j e^{(1)}_k\right|_{cp} =  0 
\rightarrow  \avgcp{ (\supbracket{E}{1})^3}  {=}  0  \label{eq:condition_critical},
\end{align}
where the average is with respect to the density at the critical point $\rhocpvec$; see SM (compare also~\cite{Sollich2001}).
Thus, the third cumulant (skewness) of the partition coefficient needs to be zero at a critical point; see Fig.~\ref{fig:spinodal_variance_equivalency_random}.
This condition on the third cumulant extends and substantiates the notion that binary systems and systems composed of ideal random copolymers are critical if their mixture composition is symmetric~\cite{Nesarikar1993, Binder1994}.

Notably, the conditions themselves ($\supbracket{\lambda}{1} {=} 1$ on spinodal, skewness ${=}0$ at a critical point) are valid irrespective of the mixture composition or feature distribution.
To illustrate this generality, for Fig.~\ref{fig:spinodal_variance_equivalency_random} we randomly generated the components' features ($R{=}2$) either (A) via a multivariate Gaussian with zero mean or (B) as two independent features following a Poisson distribution (plus Gaussian noise) with non-zero mean and a Gaussian, respectively.
In both cases, the mixture composition is drawn from a uniform distribution over the $N{-}1$-simplex; see SM.
This procedure results in interaction matrices of rank 2, while ``usual"
random matrices have full rank~\cite{Feng2007}.

\noindent \textit{Higher-order critical points}
At an $n$-th order critical point $\rhocpvec$, $n$ phases become indistinguishable.
For a single order parameter (density) $\rho$, this condition corresponds to the merging of $n$ minima and $n{-}1$ maxima into a single minimum of the tilted Landau free energy.
The free energy expansion around the critical point is then of the order $2n$: $\delta f {\sim}  \mathcal{O} \left( \delta \rho^{2n} \right)$.
In a high-dimensional density space, the phases that become indistinguishable when crossing the $n$-th order critical point $\rhocpvec$ do not necessarily lie on a straight line.
Instead, the phases merge along a more general smooth curve $\rho_i (\epsilon) {=} \rhocp_i {+} \delta \rho_i (\epsilon)$ in density space, parameterized by $\epsilon$; see also~\cite{Sollich2001}: 
$
\delta \rho_i (\epsilon) {=} \sum_{m=1}^{\infty} \frac{\epsilon^m}{m!} \supbracket{\Upsilon}{m}_i,
$
for some vectors $\vecexp{\Upsilon}{m}, m{\in} \mathbf{N}$, with $\sum_k \vecexp{\Upsilon}{m}_k {=} 0$ to conserve the incompressibility constraint.
The (tilted) free energy change $\delta f (\epsilon) {=} f(\vec{\rho}_c {+}\delta \vec{\rho} (\epsilon) ) {-} f(\vec{\rho}_c)$, whose first order term vanishes, should be of order $2n$ in $\epsilon$:
$\delta f (\epsilon) {=} \sum_{k{=}2}^\infty \frac{1}{k!} \sum_{i_1,{\ldots}, i_k} \left. \frac{\partial^k f}{\partial \rho_{i_1} {\ldots} \partial \rho_{i_k}}\right|_{cp} \delta \rho_{i_1} {\ldots} \delta \rho_{i_k} {=}  \mathcal{O}(\epsilon^{2n})
$.
At the same time, the system has to be stable against fluctuations in orthogonal directions.
Thus, we determine the curve $\delta \vec{\rho}$ around $\rhocpvec$ in a way that it minimizes the free energy change $\delta f$ up to the respective order; see SM.
Systematically minimizing and setting the coefficients in front of $\epsilon^m$ to zero, we find the following conditions for an $n$-th order critical point in terms of the partial exponential Bell polynomials $B_{m,l}$; see SM:
\begin{align}
    &\frac{1}{2} \sum_{\alpha=1}^R \sum_{k=1}^{m-1} \binom{m}{k} \avgcp{\supbracket{r}{\alpha}\supbracket{\Omega}{k}} \avgcp{\supbracket{r}{\alpha}\supbracket{\Omega}{m-k}} =  \\
    {=}\sum_{l=2}^m  &({-}1)^l (l{-}2)! \avgcp{B_{m,l} (\supbracket{\Omega}{1},  \supbracket{\Omega}{2}, \ldots)}, 2{\leq}m{\leq}2n{-}1 \nonumber,
\end{align}
which only depend on the vectors $\supbracket{\Upsilon}{m} {=:}\rhocp \supbracket{\Omega}{m}$, $m{=}1,{...},n{-}1$ determined recursively: $\supbracket{\Omega}{1} =\supbracket{E}{1}$ and
\begin{align}
    \supbracket{\Omega}{m} {=} \sum_{l=2}^{m}   [ \tilde{B}_{m,l} {-} \avgcp{\tilde{B}_{m,l}} {+} \sum_{\alpha=2}^R \frac{\supbracket{E}{\alpha}}{1{-}\supbracket{\lambda}{\alpha}}  \avgcp{\supbracket{E}{\alpha} \tilde{B}_{m,l}}]\nonumber{,} 
\end{align}
where $\tilde{B}_{m,l}:= (-1)^l (l{-}1)! B_{m,l} (\supbracket{\Omega}{1}, {...}, \supbracket{\Omega}{m-l+1})$.
In the case of a single feature, $R{=}1$, this recursion is solved by $\supbracket{\Omega}{m} {=} \partial^m_\epsilon \left. (e^{\epsilon \sqrt{2C} s}/\avgcp{e^{\epsilon \sqrt{2C} s}}) \right|_{\epsilon=0}$, and the conditions for an $n$-th order critical point reduce to $\supbracket{\kappa}{s}_2 {=}k_B T/(zJ)$ (spinodal) together with $
\supbracket{\kappa}{s}_m {=}  0 \ {\forall} m{=}3,{\ldots},2n{-}1$.
Here $\supbracket{\kappa}{s}_m$ is the $m$-th cumulant of the spin $s$ with respect to $\rhocpvec$.
Thus, the more cumulants (order $m\geq3$) of the spin distribution are zero, the more phases become indistinguishable and the higher the order of the critical point can be.
Merging of phases necessarily happens along the single direction of instability and there is only one series of higher-order critical points -- the one just discussed.
This is not true for $R{>}1$, which we discuss next.

\noindent \textit{Multiple directions of instability}
A series of critical points distinct from the previously discussed higher-order critical points occurs when the largest eigenvalue 1 of $\mathrm{Cov}$ is $D$-fold degenerate; see SM.
To ensure stability along any direction in the corresponding $D$-dimensional subspace of eigenvectors, the third cumulant of all vectors in the subspace then needs to equal 0.
For example, a system has a critical point with two unstable directions if it has a two-fold degenerate maximal eigenvalue, $\supbracket{\lambda}{1}{=}\supbracket{\lambda}{2}{=}1$, and if the four distinct third cumulants $\supbracket{\kappa}{\alpha\beta\gamma}{:=} \avgcp{\supbracket{E}{\alpha} \supbracket{E}{\beta} \supbracket{E}{\gamma}}$ with $\alpha\beta\gamma{=}111,112,122,222$ are zero.
In general, a $D{+}1$-th order critical point with $D$ degenerate unstable directions has codimension $\binom{D+1}{2}+\binom{D+2}{3}$: it requires tuning $\binom{D+1}{2}$ parameters for $\mathrm{Cov}$ to have its $D$ largest eigenvalues equal to 1~\cite{Keller2008} and $\binom{D+2}{3}$ for the third cumulants; see SM.

The emergent symmetry of these critical points is reminiscent of the order-parameter symmetry in $q$-state Potts models, with $q{=}D{+}1$.
In two dimensions there are known to be critical transitions in the $q$-state Potts model for $q {\leq} 4$~\cite{Baxter1973}, but in mean field these transitions are first order~\cite{Straley1973, Mittag1974} except for the case of the Ising model, $q{=}2$~\cite{Wu1982, Wipf2013}.
Our results show that it may be possible to have critical transitions in mean field models that have the symmetry of $q$-state Potts models, but only in models with sufficient flexibility.
For $q{=}3$, an example of an $N{=}6$ component model with $R{=}2$ features is given explicitly in the SM.

\noindent \textbf{\label{sec:model_extensions}Model extensions}
So far we have focused on a mean-field free energy whose interaction matrix has positive eigenvalues and whose components are of the same size.
We now briefly discuss how the previously discussed conditions generalize to the case of negative interaction strengths, and comment on components with different sizes as in the original Flory-Huggins theory~\cite{Huggins1941, Flory1941} in the SM.
If the feature interaction strengths $J^{(\gamma)}$ are all positive, the pairwise interactions satisfy 
$2\chi_{ij} {-} (\chi_{ii} {+} \chi_{jj}) {<} 0 \ \forall i,j$,
and interactions between alike components are always energetically preferred compared to dislike components.
To resolve this limitation, we consider the general case with $R^+$ ``positive", attractive features and $R{-}R^+$ ``negative", repulsive ones:
$
J^{(\alpha)} {>} 0 \ \forall \alpha {=} 1, {\ldots}, R^+ $ and $
J^{(\alpha)} {<} 0 \ \forall \alpha {=} R^+ {+}1, {\ldots}, R
$.
Performing a conceptually similar but more intricate analysis as before, the spinodal criterion is
$\supbracket{\lambda}{1}_{\bar{C}} = 1$.
Here $\supbracket{\lambda}{1}_{\bar{C}}$ is the largest eigenvalue of the real, symmetric matrix $\bar{C} {=}  \cpp {-} \cpm (\mathbf{1}{+}\cmm)^{{-}1} \cmp$, which is determined by the covariances among the subsets of positive (${+}$) and negative (${-}$) features; see SM.
$\bar{C}$ has dimensions $R^+{\times} R^+$ and can be interpreted as representing a multicomponent system with $R^+$ positive features and effective, reduced interactions.
The extent to which the negative features influence the phase behavior depends on the relative correlations between all features.
If for each dominant positive feature, there is a highly correlated negative feature of similar strength, their effects will roughly cancel and the mixture will not phase separate.
Conversely, if the dominant positive features driving phase separation correlate weakly with the negative features, thermodynamic stability is barely modified by the presence of the latter.
At the spinodal, the direction of instability is
$
    \supbracket{\bar{e}}{1}_i = \rho_i 
    \sum_{\alpha=1}^{R^+} \supbracket{\phi}{1}_{\alpha} \left[ \supbracket{\pi}{\alpha}_i - \sum_{\beta,\gamma=1}^{R{-}R^+} \cpm_{\alpha \beta} (\mathbf{1} +\cmm)^{-1}_{\beta \gamma} \supbracket{\nu}{\gamma}_i \right]$
in  terms of the first eigenvector $\supbracket{\phi}{1}$ of $\bar{C}$ and the deviations of positive ($\pi$) and negative ($\nu$) features from the mean; see SM.
We observe that this direction of instability again corresponds to (a combination of) feature deviations from the mean, projected onto the first principal component $\supbracket{\phi}{1}$ (now of the ``effective covariance matrix" $\bar{C}$).
Roughly speaking, the relative sign of the contributions of the negative and positive features depends on whether they are correlated or anti-correlated (negative or positive sign).
Finally, performing the same analysis as for the original model, we find an analogous condition for the ordinary critical point:
$
    \avgcp{ \left(\bar{E}^{(1)}\right)^3}  {=}  0
$,
where $\supbracket{\bar{e}}{1}_i {=:} \rho_i  \supbracket{\bar{E}}{1}_i$.

\noindent \textbf{\label{sec:discussion}Discussion}
In this work, we consider a general mean-field model for multicomponent mixtures with an arbitrary pairwise interaction matrix $\chi_{ij}$ of variable rank which we decompose in terms of different ``features" mediating additive interactions between the components.
The analytic conditions we derive for the spinodal and (higher-order) critical points only depend on the distribution of components in feature space.
Specifically, the spinodal and submanifold of ordinary critical points are determined exclusively by the variance and third cumulant of the component distribution projected along the first principal component of the feature covariance matrix; 
Fig.~\ref{fig:spinodal_variance_equivalency_random}.

This representation in feature space is reminiscent of the dimensional reduction obtained for polydisperse systems whose excess free energy only depends on a few generalized moments of the attributes~\cite{Sollich2001, Sollich2002, DeCastro2017}. 
While the derivation of the ``moment free energies" relies on either a division of density space into a subspace of moments and its ``transverse" space or on combinatorial arguments~\cite{Sollich2001, Sollich2002}, 
here we instead exploit that the condition for the Hessian matrix to become singular only depends on an $R$-dimensional matrix originating from the interaction structure.
A related simplification of the spinodal condition  in terms of a lower-dimensional matrix has been achieved for Flory-Huggins models with an excess free energy depending only on a finite number of moments of the molecular weight distribution~\cite{Gordon1977, Irvine1981, Beerbaum1986}.

The representation in feature space also suggests a principled method for finding coarse-grained binary mixtures with similar properties.
By choosing the composition and interaction strength of the binary mixture so as to preserve the second and third cumulant along the first principal component, the coarse-grained binary mixture maintains the location of the multicomponent system with respect to the spinodal and critical manifold; see Fig.~\ref{fig:spinodal_variance_equivalency_random} and SM.

In addition, our analysis allows for a systematic identification of the codimension of different series of critical points in multicomponent systems; see also~\cite{Griffiths1975, DeOliveira2013}.
For instance, we find that, in the absence of symmetries, a tricritical point has codimension four in mean-field.
Furthermore, higher-order critical points with symmetry reminiscent of the $q$-state Potts model require tuning of $\binom{q}{2}{+}\binom{q{+}1}{3}$ parameters.
For the $q{=}3$-states Potts model, this counting suggests a codimension of seven for the critical point, which is larger than the one accessible with just $N{=}3$ components but feasible for a mean-field model with $N{=}6$ components and $R{=}2$; we explicitly construct such a $3$-states-Potts-like model containing a critical point; see SM.

Our results offer an appealing avenue towards  understanding intracellular liquid-liquid phase separation~\cite{Berry2018} and the critical phase behavior observed in cell-derived plasma membranes~\cite{Veatch2008}.
These mixtures are composed of thousands of proteins (and lipids), and depending on the conditions, small domains form spontaneously.
The number of coexisting domains appears to be orders of magnitude smaller than the number of components and is thus well below the limit set by Gibb’s phase rule~\cite{Harmon2017}.
In cell-derived plasma membranes, while true phase separation occurs when cooling them below the critical temperature~\cite{Baumgart2007}, nanoscopic domains observed at physiological temperatures~\cite{Li2020} have been suggested to be critical fluctuations close to a thermodynamic critical point in the 2D Ising universality class~\cite{Veatch2008}.
Strikingly, specific lipids and proteins robustly partition into specific phases -- seemingly under fairly broad conditions~\cite{Levental2020}.
Our work offers an interpretation of this experimental observation: 
phase behavior is determined by just a few important features.
Looking for such a low-dimensional feature space representation might help to make sense of the growing amount of experimental data generated by proximity-labeling techniques~\cite{Bracha2019}, and should provide important insights into the physical characteristics underlying intracellular phase separation.
In these biological systems, effects of finite dimension (two or three) and  sequence-dependent interaction patterns~\cite{Lin2018,Statt2020} will likely quantitatively, but not qualitatively change the mean field picture we present here.
Finally, our analytic theory only makes predictions about local thermodynamic properties but cannot now make statements about the global phase behavior, which would require knowledge of the full free energy landscape~\cite{Jacobs2013, Jacobs2017, Mao2019, Shrinivas2021, Jacobs2021}.
Whether global phase behavior can be understood in feature space is an interesting question for future research.

\vspace{10pt}

\section*{Acknowledgments}
We thank Michael Abbott, Samuel Bryant, William Jacobs, Henry Mattingly, Mason Rouches, Thomas Shaw and Peter Sollich for helpful comments on the manuscript. 
This work was supported by NIH R35GM138341 (IRG, BBM), a Simons Investigator award (IRG, BBM), a biosciences postdoctoral fellowship from Yale University (IRG) and by the Deutsche Forschungsgemeinschaft (DFG, German Research Foundation) – Projektnummer 494077061 (IRG).

\providecommand{\noopsort}[1]{}\providecommand{\singleletter}[1]{#1}%

\clearpage

\begin{center}
\textbf{\large Supplemental Material to\\ Thermodynamic stability and critical points in multicomponent mixtures with structured interactions} \\
\vspace{10pt}
Isabella R. Graf \\%
\textit{Department of Physics, Yale University} \\
\vspace{10pt}
Benjamin B. Machta \\
\textit{Department of Physics, Yale University
}\\
\textit{
 Systems Biology Institute, West Campus, Yale University
}%
\end{center}
\setcounter{equation}{0}
\setcounter{figure}{0}
\setcounter{table}{0}
\setcounter{section}{0}
\makeatletter
\renewcommand{\theequation}{S\arabic{equation}}
\renewcommand{\thefigure}{S\arabic{figure}}
\renewcommand{\thesection}{S\arabic{section}}

In this Supplemental Material we present in detail the derivations of the conditions for the spinodal and critical points as outlined in the main text and explain how we generated the distribution of features in the multicomponent mixtures depicted in Fig. 2 of the main text. 
We first discuss the mean-field approximation of the lattice Hamiltonian in terms of a free energy of Flory-Huggins type.
In this approximation, we then derive the condition for the spinodal and the intial direction of phase separation, the direction of instability.
To find the conditions for (higher-order) critical points, we further systematically expand the mean-field free energy around the critical density.
After deriving these conditions, we look at two extensions of our original model, systems with components of different sizes and mixtures with partially negative and partially positive interaction strengths, and derive the corresponding conditions for the spinodal and ordinary critical points.
To show that our result do not depend on choosing the features linearly independently, we then comment on the case of linearly dependent features.
In a similar spirit, we further show that the system is invariant under global rotations and translations of the rescaled feature vectors.
While our focus is on the case where the largest eigenvalue of the covariance matrix is non-degenerate, we then dedicate one section to the case with several directions of instability.
Based on this discussion, we further propose an effective mean-field model for the two-dimensional $q{=}3$-states Potts model which exhibits a critical point, in contrast to the regular mean-field theory of the Potts model.
We then briefly discuss our idea how to coarse-grain multicomponent mixtures in terms of an effective binary mixture.
Finally, we provide details about how we generated the multicomponent mixtures illustrated in Fig. 2 of the main text.

\vspace{30pt}

\noindent
Note that a glossary containing the most important symbols and definitions can be found at the end of the SM.

\clearpage

\renewcommand*{\contentsname}{Table of Contents for SM}
\addtocontents{toc}{\setlength{\cftsecnumwidth}{3em}}
\addtocontents{toc}{\setlength{\cftsubsecnumwidth}{3em}}
\setcounter{tocdepth}{2}
\tableofcontents

\clearpage

\section{Mean-field approximation of the lattice Hamiltonian}
\label{SMsec:MF_description}

The Hamiltonian from the main text is
\begin{align}
    \mathcal{H} = - \sum_{\gamma=1}^R \supbracket{J}{\gamma} \sum_{\NN{xy}} \supbracket{\sigma}{\gamma} (x) \supbracket{\sigma}{\gamma} (y),
\end{align}
where $\gamma=1,\ldots,R$ denotes the different features, $x$ and $y$ label the sites on the lattice and $\NN{xy}$ denotes the set of all nearest neighbors on the lattice.
If component type $i \in \{1,\ldots,N\}$ occupies site $x$, $\supbracket{\sigma}{\gamma} (x) = \supbracket{s}{\gamma}_i$.
Since we assume that each lattice site is occupied by exactly one component (particle), the overall number of lattice sites equals the number of particles.

In a mean-field approximation, we neglect correlations between the different sites on the lattice and the different sites are not distinguished anymore.
The probability that component type $i$ occupies any site $x$ is $\rho_i$. 
Denoting by $z$ the coordination number of the lattice, there are $z/2$ pairs of nearest neighbors per particle. 
For each pair of nearest neighbors $\NN{xy}$ with component type $i$ on $x$ and $j$ on $y$, the energetic contribution is $\sum_{\gamma=1}^R \supbracket{J}{\gamma} \supbracket{s}{\gamma}_i \supbracket{s}{\gamma}_j$.
The probability for this configuration is $\rho_i \rho_j$.
Thus, the MF contribution per particle is $(z/2) \sum_{i,j=1}^N \rho_i \rho_j \sum_{\gamma=1}^R \supbracket{J}{\gamma} \supbracket{s}{\gamma}_i \supbracket{s}{\gamma}_j$.
Overall, the energetic contribution per particle is
\begin{align}
    U_{\text{MF}}/M = -\frac{z}{2} \sum_{\gamma=1}^R \supbracket{J}{\gamma}  \left(\sum_{i=1}^N \rho_i \supbracket{s}{\gamma}_i \right)^2,
\end{align}
where $M$ denotes the total number of particles/lattice sites.
Combining the energetic contribution with the entropic contribution, $k_B T \sum_{i=1}^N \rho_i \log \rho_i$, per particle, we find the following mean-field free energy per particle and $k_B T$
\begin{align}
    \tilde{f}_N:= \frac{F_N}{M k_B T} =  \underbrace{\sum_{i=1}^N \rho_i \log \rho_i}_{\text{entropic term}}  \underbrace{-\sum_{\gamma=1}^R \left(\sum_{i=1}^N \rho_i \sqrt{\supbracket{C}{\gamma}} \supbracket{s}{\gamma}_i \right)^2}_{\text{pairwise interactions}},
\end{align}
with
\begin{align}
    \supbracket{C}{\gamma}:=\frac{z \supbracket{J}{\gamma}}{2 k_B T}.
\end{align}
The subscript $N$ indicates that $\tilde{f}$ is a function of all $N$ densities $\rhoN$.

\subsection{Flory-Huggins interaction matrix}
\label{SMsubsec:FH_matrix}

Comparing the energetic contribution from the pairwise interactions with a Flory-Huggins term, $-\sum_{i,j=1}^N \rho_i \chi_{ij} \rho_j$, we identify the interaction matrix as
\begin{align}
    \chi_{ij}= \sum_{\gamma=1}^R \supbracket{C}{\gamma} \supbracket{s}{\gamma}_i \supbracket{s}{\gamma}_j.
\end{align}

\subsection{Effective reduction to an  $N{-}1$-component system}
\label{SMsubsec:Nminus_component_system}

Due to the incompressibility constraint, not all the densities can be changed independently. 
Here, we directly impose this constraint by replacing 
\begin{align}
   \rho_N \rightarrow 1-\sum_{i=1}^{N-1} \rho_i =: 1- \Rho 
\end{align}
in the free energy.
Instead of $\tilde{f}_N$, we consider the ``restricted" free energy
\begin{align}
    \fNminus &= \tilde{f}_N (\rho_1, \ldots, \rho_{N-1},1-\Rho) = \nonumber \\
    &=\sum_{i=1}^{N-1} \rho_i \log \rho_i + (1-\Rho) \log(1-\Rho) -\sum_{\gamma=1}^R \supbracket{C}{\gamma} \left( \sum_{i=1}^{N-1} \rho_i \supbracket{s}{\gamma}_i + (1-\Rho) \supbracket{s}{\gamma}_N \right)^2,
    \label{SMeq:free_energy_Nminus}
\end{align}
which only depends on the first $N{-}1$ densities $\rhoNminus$.
Note that as expected and as we will see explicitly below, the ``physical results" do not depend on this choice of integrating out component $N$ (instead of any other or a combination of them).

\section{Derivatives of the free energy $\fNminus$ and the Hessian matrix}
\label{SMsec:free_energy_derivatives}

Taking the derivative of the restricted free energy $\fNminus$, Eq.~\ref{SMeq:free_energy_Nminus}, with respect to the density $\rho_i, i=1, \ldots, N-1$ yields
\begin{align}
    \partial_i \fNminus:= \frac{\partial \fNminus}{\partial \rho_i} &= 1+ \log \rho_i - 1- \log(1-\Rho) - 2 \sum_{\gamma=1}^R \supbracket{C}{\gamma} \left( \sum_{j=1}^{N-1} \rho_j \supbracket{s}{\gamma}_j + (1-\Rho) \supbracket{s}{\gamma}_N \right) \left(  \supbracket{s}{\gamma}_i - \supbracket{s}{\gamma}_N \right) = \nonumber \\
    &= \log \rho_i - \log(1-\Rho) - 2 \sum_{\gamma=1}^R \supbracket{C}{\gamma}\left( \supbracket{s}{\gamma}_N +  \sum_{j=1}^{N-1} \rho_j (\supbracket{s}{\gamma}_j - \supbracket{s}{\gamma}_N) \right) \left(  \supbracket{s}{\gamma}_i - \supbracket{s}{\gamma}_N \right). \label{SMeq:first_deriv_free_energy}
\end{align}

For the Hessian matrix $H_{ij}, i,j=1,\ldots, N-1$, we find
\begin{align}
    H_{ij} := \partial_i \partial_j \fNminus &:= \frac{\partial^2}{\partial \rho_i \partial \rho_j} \fNminus = \delta_{ij} \frac{1}{\rho_i} + 1_{ij} \frac{1}{1-\Rho} - 2\sum_{\gamma=1}^R \supbracket{C}{\gamma} \left(  \supbracket{s}{\gamma}_i - \supbracket{s}{\gamma}_N \right) \left(  \supbracket{s}{\gamma}_j - \supbracket{s}{\gamma}_N \right) =: \nonumber \\
    &=: K_{ij} - \sum_{\gamma=1}^R \supbracket{r}{\gamma}_i \supbracket{r}{\gamma}_j =: \supbracket{A}{2}_{ij},
    \label{SMeq:Hessian}
\end{align}
where we define $1_{ij} := 1 \ \forall i,j$ in order to keep track of sums over repeated indices.
\begin{align}
    K_{ij} := \delta_{ij} \frac{1}{\rho_i} + 1_{ij} \frac{1}{1-\Rho}
\end{align}
corresponds to the Hessian of an incompressible $N$-component system without interactions.
Furthermore, 
\begin{align}
    \supbracket{r}{\gamma}_i = \sqrt{2 \supbracket{C}{\gamma}} (  \supbracket{s}{\gamma}_i - \supbracket{s}{\gamma}_N )
\end{align}
denotes the rescaled and shifted features. 

Similarly, the higher-order derivatives $n\geq 3$ are
\begin{align}
    \supbracket{A}{n}_{i_1 i_2 {\ldots} i_n} := \partial^n_{i_1 i_2 {\ldots} i_n} \fNminus := \frac{\partial^n}{\partial \rho_{i_1} \ldots \partial \rho_{i_n}} \fNminus =  (n-2)! \left[ \delta_{i_1 i_2 {\ldots} i_n} (-1)^n \frac{1}{\rho_{i_1}^{n-1}} + 1_{i_1 i_2 {\ldots} i_n} \frac{1}{(1-\Rho)^{n-1}} \right],
\end{align}
where $\delta_{i_1 i_2 {\ldots} i_n} = \prod_{k=1}^{n-1} \delta_{i_k i_{k+1}}$ and $1_{i_1 i_2 {\ldots} i_n} = 1 \ \forall i_1,{\ldots},i_n$.

\section{Spinodal condition and direction of instability}
\label{SMsec:spinodal}

For a non-interacting multicomponent mixture, which is entirely dominated by entropic effects and accordingly does not phase separate, the Hessian matrix equals $K$.
All eigenvalues of the Hessian are thus positive:
For each vector $w \in \mathbb{R}^{N{-}1}$, there is a vector $v \in \mathbb{R}^{N{-}1}$ with $w_i = \rho_i v_i$ (assuming all $\rho_i >0$).
Thus,
\begin{align}
    w_i K_{ij} w_j = \rho_i v_i K_{ij} \rho_j v_j = \sum_{i=1}^{N-1} \rho_i v_i^2 + \frac{1}{1-\Rho} \left( \sum_{i=1}^{N-1} \rho_i v_i \right)^2 = \underbrace{\left( 1-\Rho\right)}_{> 0} \underbrace{\left( \sum_{i=1}^{N-1} \frac{\rho_i}{1-\Rho} v_i^2 + \left(\sum_{i=1}^{N-1} \frac{\rho_i}{1-\Rho} v_i \right)^2 \right)}_{=\NN{v^2}_{\tilde{\rho}} + (\NN{v}_{\tilde{\rho}})^2 = \text{Var}_{\tilde{\rho}} (v) + 2 (\NN{v}_{\tilde{\rho}})^2 \geq 0} \geq 0,
\end{align}
where in the last step, we used that the variance of $v$ with respect to the probability measure  $\tilde{\rho}_i = \rho_i/(1-\Rho), i=1,{\ldots},N{-}1$ for the $N{-}1$-component system is always larger or equal to 0.
Note that $w_i K_{ij} w_j =0$ only if $\text{Var}_{\tilde{\rho}} (v) = \NN{v}_{\tilde{\rho}}= 0$ and thus $v_i = w_i=0 \ \forall i$; the matrix $K$ is positive definite.

Introducing (positive) interactions among the component types, the system becomes more and more prone to phase separate and the eigenvalues of the Hessian decrease.
The system becomes marginally thermodynamically stable if the smallest eigenvalue of the Hessian crosses 0.
Since a square matrix is invertible if and only if all eigenvalues are non-zero, marginal stability is equivalent to the Hessian matrix becoming singular for the first time.

\subsection{Inverse of Hessian matrix}
\label{SMeq:inverse_Hessian}

From Eq~\ref{SMeq:Hessian}, we observe that the Hessian corresponds to a rank $\leq R$ correction of the matrix $K$.
The existence of the inverse of such a rank correction to $K$ and its explicit form has been determined by Woodbury~\cite{woodbury1950}:
A matrix $K - UW$ with $K: (N{-}1)\times (N{-}1)$, $U: (N{-}1)\times R$, $W: R\times (N{-}1)$ is invertible if and only if $K$ and $\mathbf{1}-W K^{-1} U$ are both invertible (here $\mathbf{1}: R \times R$).
In this case, the inverse is given by
\begin{align}
    (K-UW)^{-1} = K^{-1} + K^{-1} U \left(\mathbf{1}-W K^{-1} U \right)^{-1} W K^{-1}.
\end{align}
Identifying $U_{i\alpha}= \supbracket{r}{\alpha}_i$ and $W=U^T$, we find that the Hessian matrix, Eq~\ref{SMeq:Hessian}, is invertible if and only if $K$ and $\mathbf{1}-U^T K^{-1} U$ are invertible.
As we will see next, $K$ is always invertible and  $\mathbf{1}-U^T K^{-1} U$ is rewritten as $\mathbf{1}-\text{Cov}$, with the covariance matrix as defined in the main text: 
$\text{Cov}_{\alpha \beta} = \avgrhoN{r^{(\alpha)} r^{(\beta)}} - \avgrhoN{r^{(\alpha)}} \avgrhoN{r^{(\beta)}}$.

\paragraph{Inverse of $K$}
$K$ itself is written as a rank-$1$ correction to a diagonal matrix $M$:
\begin{align}
    K_{ij} = \delta_{ij} \frac{1}{\rho_i} + \frac{1}{1-\Rho} 1_{ij} =: M_{ij} + Q_i Q_j \hspace{10pt} \text{where} \hspace{10pt} Q_i = \frac{1}{\sqrt{1-\Rho}} \ \forall i
\end{align}
Since $M$ is always invertible with inverse $(M^{-1})_{ij} = \delta_{ij} \rho_i$, the matrix $K$ is invertible if and only if
\begin{align}
    1 + Q M^{-1} Q  = 1 + Q_i \delta_{ij} \rho_i Q_j = 1+ \sum_{i=1}^{N-1} \rho_i \frac{1}{1-\Rho} = \frac{1}{1-\Rho} \neq 0.
\end{align}
Since $\Rho \leq 1$, this condition is always satisfied and the inverse of $K$ is
\begin{align}
    (K^{-1})_{ij} = (M^{-1})_{ij} - (M^{-1})_{ik} Q_k \left(1-\Rho\right) Q_l (M^{-1})_{lj} = \delta_{ij} \rho_i - \delta_{ik} \rho_i Q_k \left(1-\Rho\right) Q_l \delta_{lj} \rho_j = \delta_{ij} \rho_i - \rho_i \rho_j.
    \label{SMeq:inverse_K}
\end{align}
It is straightforward to test that indeed $K K^{{-}1} = K^{{-}1}K = \mathbf{1}$.

\paragraph{Inverse of $\mathbf{1}-U^T K^{-1} U$}

Using the definition of $U_{i\alpha} = r_i^{(\alpha)}$ with $r_N^{(\alpha)} =0$ and the previous result $(K^{-1})_{ij} = \delta_{ij} \rho_i - \rho_i \rho_j$, we find 
\begin{align*}
(\mathbf{1}-U^T K^{-1} U)_{\alpha \beta} &= \delta_{\alpha\beta} - \sum_{i,j=1}^{N-1} r_i^{(\alpha)} \left( \delta_{ij} \rho_i - \rho_i \rho_j\right) r_j^{(\beta)} = \delta_{\alpha\beta} - \sum_{i=1}^N \rho_i r_i^{(\alpha)} r_i^{(\beta)} + \sum_{i=1}^N \rho_i r_i^{(\alpha)} \sum_{j=1}^N \rho_j r_j^{(\beta)} =\\
&=\delta_{\alpha\beta} -\left(\avgrhoN{ \supbracket{r}{\alpha} \supbracket{r}{\beta}} -  \avgrhoN{ \supbracket{r}{\alpha} } \avgrhoN{  \supbracket{r}{\beta}}\right) = (\mathbf{1}-\text{Cov})_{\alpha \beta}.
\end{align*}
Here and in the following, averages $\avgrhoN{\ldots}$ are with respect to the densities of the $N$-component system (for simplicity, we will often omit the $N$ and the vector notation):
\begin{align}
    \avgrhoN{X}:= \avgrho{X}:= \sum_{i=1}^N \rho_i X_i.
\end{align}
Taken together, we find that the Hessian matrix is invertible if and only if $\mathbf{1}-\text{Cov}$ is invertible.
Note that the Hessian matrix is an $(N-1)\times (N-1)$ matrix whereas $\mathbf{1}-\text{Cov}$ has dimensions $R\times R$. 

\vspace{12pt}

For positive interaction strengths, $\text{Cov}$ is a true covariance matrix, which is real, symmetric and positive semi-definite.
The eigenvalues of $\mathbf{1}-\text{Cov}$ are thus all real and $\leq 1$.
For high temperature or weak/no interactions, the rescaled features $\supbracket{r}{\alpha}$ are small and thus $\mathbf{1}-\text{Cov}$ has only positive eigenvalues.
Decreasing the temperature or increasing the attractive interactions, the mixture becomes marginally stable when $\mathbf{1}-\text{Cov}$ becomes singular for the first time, i.e.\ when the largest eigenvalue of $\text{Cov}$ is 1.
Writing the covariance matrix in terms of its eigenvectors $\supbracket{\vec{V}}{\gamma}$ and eigenvalues $\supbracket{\lambda}{\gamma}$, $\gamma=1,\ldots, R$: 
\begin{align}
  \text{Cov} = \sum_{\gamma=1}^R \supbracket{\lambda}{\gamma} \supbracket{\vec{V}}{\gamma} \left(\supbracket{\vec{V}}{\gamma}\right)^T,  
\end{align}
with $\supbracket{\lambda}{\alpha} \geq \supbracket{\lambda}{\beta}\ \forall \alpha \leq \beta$, this condition for marginal stability (spinodal) is
\begin{align}
    \supbracket{\lambda}{1} = 1  \hspace{10pt} \text{(condition \ for \ the \ spinodal)}.
\end{align}

If $\mathbf{1}-\text{Cov}$ is invertible, $\supbracket{\lambda}{\gamma}\neq 1 \ \forall \gamma$, the inverse of the Hessian is given by
\begin{align}
    H^{-1} = (K-U U^T)^{-1} = K^{-1} + K^{-1} U (\mathbf{1}-\text{Cov})^{-1} U^T K^{-1},
\end{align}
or written in terms of the eigenvectors and eigenvalues of the covariance matrix,
\begin{align}
    (H^{-1})_{ij} &=  (K^{-1})_{ij} + \sum_{k,m=1}^{N-1} \sum_{\alpha,\beta=1}^R (K^{-1})_{ik} U_{k\alpha} \left( \sum_{\gamma=1}^R \frac{1}{1-\supbracket{\lambda}{\gamma}} \supbracket{V}{\gamma}_{\alpha} \supbracket{V}{\gamma}_{\beta} \right) (U^T)_{\beta m} (K^{-1})_{mj}= \\
    &= \delta_{ij} \rho_i - \rho_i \rho_j +\sum_{k,m=1}^{N-1} \sum_{\alpha,\beta=1}^R \sum_{\gamma=1}^R (\delta_{ik} \rho_i - \rho_i \rho_k) \supbracket{r}{\alpha}_k \left(  \frac{1}{1-\supbracket{\lambda}{\gamma}} \supbracket{V}{\gamma}_{\alpha} \supbracket{V}{\gamma}_{\beta} \right) \supbracket{r}{\beta}_m (\delta_{mj} \rho_j - \rho_m \rho_j)= \\
    &= \delta_{ij}\rho_i - \rho_i \rho_j + \sum_{\gamma=1}^R \frac{1}{1-\supbracket{\lambda}{\gamma}} \supbracket{e}{\gamma}_i \supbracket{e}{\gamma}_j \label{SMeq:invHessian_EVas_Cov},
\end{align}
where 
\begin{align}
    \supbracket{e}{\gamma}_i := \rho_i \supbracket{E}{\gamma}_i := \rho_i \sum_{\alpha=1}^R \supbracket{V}{\gamma}_\alpha \left( \supbracket{r}{\alpha}_i - \avgrho{\supbracket{r}{\alpha}}\right)
    \label{SMeq:definition_e_general_gamma}
\end{align}
and we used that $\sum_{l=1}^{N-1} \rho_l \supbracket{r}{\alpha}_l = \sum_{l=1}^{N} \rho_l \supbracket{r}{\alpha}_l = \avgrho{\supbracket{r}{\alpha}} $.

Approaching the spinodal, the inverse of the Hessian is dominated by the term $\sim 1/(1-\supbracket{\lambda}{1})$~\footnote{or the terms $\sim 1/(1-\supbracket{\lambda}{i}), i=1,\ldots,D$ if the zero eigenvalue of the Hessian is $D$-fold degenerate}, suggesting that on the spinodal $\supbracket{e}{1}$ is the eigenvector of the Hessian to eigenvalue 0.
Using that $\supbracket{r}{\gamma}_N = 0 \ \forall \gamma$ and $\sum_{\beta=1}^R \supbracket{V}{\alpha}_{\beta} \text{Cov}_{\beta\gamma} = \supbracket{\lambda}{\alpha} \supbracket{V}{\alpha}_{\gamma}$ by definition, and
\begin{align*}
    \avgrho{\supbracket{E}{\alpha}} = 0 \ \hspace{10pt} \forall \alpha,
\end{align*}
we indeed find that
\begin{align}
    H_{ij} \supbracket{e}{\alpha}_j &= \left( \delta_{ij} \frac{1}{\rho_i} + 1_{ij} \frac{1}{1-\Rho} - \sum_{\gamma=1}^R \supbracket{r}{\gamma}_i \supbracket{r}{\gamma}_j\right) \rho_j \supbracket{E}{\alpha}_j = \supbracket{E}{\alpha}_i + \frac{1}{1-\Rho} \sum_{j=1}^{N-1} \rho_j \supbracket{E}{\alpha}_j - \sum_{\gamma=1}^R \supbracket{r}{\gamma}_i \avgrho{\supbracket{E}{\alpha} \supbracket{r}{\gamma}} \nonumber\\
    &= \supbracket{E}{\alpha}_i - \supbracket{E}{\alpha}_N - \sum_{\gamma=1}^R \supbracket{r}{\gamma}_i \sum_{\beta=1}^R \supbracket{V}{\alpha}_{\beta} \left( \avgrho{\supbracket{r}{\beta} \supbracket{r}{\gamma}} - \avgrho{\supbracket{r}{\beta}} \avgrho{\supbracket{r}{\gamma}} \right) = \nonumber\\
    &= \sum_{\beta=1}^R \supbracket{V}{\alpha}_{\beta} \supbracket{r}{\beta}_i - \sum_{\gamma=1}^R \supbracket{r}{\gamma}_i \sum_{\beta=1}^R \supbracket{V}{\alpha}_{\beta} \text{Cov}_{\beta\gamma} = (1-\supbracket{\lambda}{\alpha}) \sum_{\gamma=1}^R \supbracket{V}{\alpha}_{\gamma} \supbracket{r}{\gamma}_i. \label{SMeq:eigenvectorsHessian}
\end{align}
In particular, $H \supbracket{e}{\alpha} = 0$ if $\supbracket{\lambda}{\alpha}=1$.
On the spinodal, where $\supbracket{\lambda}{1}=1$, $\supbracket{e}{1}$ thus corresponds to the eigenvector of the Hessian to eigenvalue 0:

\begin{align}
    \supbracket{e}{1}_i = \rho_i \sum_{\alpha=1}^R \supbracket{V}{1}_\alpha \left( \supbracket{r}{\alpha}_i - \avgrho{\supbracket{r}{\alpha}}\right) \hspace{5pt} = \hspace{5pt} \text{direction \ of \ instability.}
\end{align}

\subsection{Degenerate maximal eigenvalue of $\text{Cov}$: Dimensionality of the plane of instability}

If $\text{Cov}$ has a $D$-fold degenerate maximal eigenvalue of 1, the corresponding eigenvectors can be chosen orthonormal since $\text{Cov}$ is real and symmetric.
Therefore, they span a $D$-dimensional submanifold in feature space.
One may wonder if also the corresponding plane in component space, spanned by the different directions of instability $\supbracket{e}{1}, \ldots, \supbracket{e}{D}$, is $D$-dimensional, or if these directions of instability might be linearly dependent.
Here we briefly show that the (generally non-orthogonal) vectors $\supbracket{e}{1}, \ldots, \supbracket{e}{D}$ are indeed linearly independent:

Suppose $\sum_{\alpha=1}^D c_{\alpha} \supbracket{e}{\alpha}_i = 0 \ \forall i$, then $\sum_{i=1}^N \supbracket{E}{\beta}_i \sum_{\alpha=1}^D c_{\alpha} \supbracket{e}{\alpha}_i \stackrel{!}{=} 0 \ \forall \beta$.
Furthermore, $\sum_{i=1}^N \supbracket{E}{\beta}_i \supbracket{e}{\alpha}_i= \sum_{\gamma,\delta=1}^R \supbracket{V}{\alpha}_{\gamma} \supbracket{V}{\beta}_{\delta}$ $\sum_{i=1}^N \rho_i (\supbracket{r}{\gamma}_i - \avgrho{\supbracket{r}{\gamma}}) (\supbracket{r}{\delta}_i - \avgrho{\supbracket{r}{\delta}}) = \sum_{\gamma,\delta=1}^R \supbracket{V}{\alpha}_{\gamma} \supbracket{V}{\beta}_{\delta} \text{Cov}_{\gamma \delta} = \sum_{\gamma=1}^R \supbracket{V}{\alpha}_{\gamma} \supbracket{\lambda}{\beta} \supbracket{V}{\beta}_{\gamma} = \supbracket{\lambda}{\beta} \delta_{\alpha \beta} $, where we used the orthonormality of the eigenvectors $\supbracket{V}{\alpha}$ of the covariance matrix.
As a result, $\forall \beta = 1, \ldots, D$, we find $\sum_{\alpha=1}^R c_{\alpha} \supbracket{\lambda}{\beta} \delta_{\alpha \beta} = c_{\beta} \supbracket{\lambda}{\beta} = c_{\beta} \stackrel{!}{=} 0$.
Taken together, it follows from $\sum_{\alpha=1}^D c_{\alpha} \supbracket{e}{\alpha}_i = 0 \ \forall i$ that $c_{\alpha} = 0 \ \forall \alpha=1,\ldots, D$ and, correspondingly, the $\supbracket{e}{\alpha}$ are linearly independent, thus spanning a $D$-dimensional plane of instability in component space.

This argument is indeed true for all sets of vectors $\{\supbracket{e}{\gamma}\}$, for which all corresponding eigenvalues $\supbracket{\lambda}{\gamma} \neq 0$.
As we will see later, $\supbracket{\lambda}{\gamma} = 0$ only for linearly dependent features, and we conclude that the plane in component space spanned by $\{\supbracket{e}{\gamma}\}_{\gamma=1, \ldots, R}$ has dimension equal to the number of linearly independent features.

\section{Systematic expansion of the free energy along a path in density space}
\label{SMsec:expansion_free_energy}

The spinodal corresponds to the submanifold in phase space where a homogeneous phase becomes unstable with respect to local fluctuations and the system spontaneously phase separates into two (or more) phases. 
Critical points on the spinodal occur if these phases become indistinguishable.
Here, we define an $n$-th order critical point as a point where $n$ different phases become indistinguishable and merge into one phase.
In a one-dimensional system, an $n$-th order critical point at densities $\rhocp$ manifests as a minimum of order $2n-1$ in the \textbf{tilted} Landau free energy $\fNminus \rightarrow \fNminus - \sum_{i=1}^{N-1} \partial_i \fNminus |_{\rhocp} \rho_i$. 
There $n$ minima and the $n-1$ maxima in between merge into one minimum. 
We generalize this idea to the multicomponent system by requiring that for an $n$-th order critical point, the free energy change $\Delta \fNminus$ (with respect to $\rhocp$) along the path of minimal change exhibits a minimum of order $\geq 2n-1$~\cite{Sollich2001}:
\begin{align}
	\Delta \fNminus (\epsilon):= \fNminus (\rho (\epsilon)) - \fNminus (\rhocp) \geq 0 \hspace{10pt} \text{to \ order} \hspace{10pt} \mathcal{O}(\epsilon^{2n}).
\end{align}
for all paths
\begin{align}
     \rho_i(\epsilon) := \rhocp_i + \delta \rho_i (\epsilon):= \rhocp_i \left(1+\sum_{m=1}^\infty \frac{\epsilon^m}{m!} \supbracket{\omega}{m}_i\right)
     \label{SMeq:general_path_density}
\end{align}
in density space, with equality
\begin{align}
	 \fNminus (\rho^{\ \text{opt}} (\epsilon)) - \fNminus (\rhocp) = \mathcal{O}(\epsilon^{2n})
	 \label{SMeq:critical_point_opt_path}
\end{align}
along an ``optimal" path in density space
\begin{align}
    \rho_i^{\text{opt}}(\epsilon) := \rhocp_i + \delta \rho_i^{\text{opt}} (\epsilon):= \supbracket{\rho}{cp}_i \left(1+\sum_{m=1}^\infty \frac{\epsilon^m}{m!} \supbracket{\Omega}{m}_i\right)
    \label{SMeq:min_path_density},
\end{align}
for which $\supbracket{\omega}{m} = \supbracket{\Omega}{m} \ \forall m$.
Here, the density vectors (for which we omit the vector notation) are to be understood in the $N-1$ dimensional space.
The incompressibility constraint, $\sum_{i=1}^N \rho_i =1$, is taken care of implicitly by extending the density vectors to $N$ dimensions via $\rho_N (\epsilon) = 1-\sum_{i=1}^{N-1} \rho_i (\epsilon)$.
Similarly, the changes in density need to satisfy $\sum_{i=1}^{N} \delta \rho_i (\epsilon) = 0 \ \forall \epsilon$ and, in particular, 
\begin{align}
    \sum_{i=1}^N \rhocp_i \supbracket{\omega}{m}_i = \avgrhoNcp{\omega}=0,
    \label{SMeq:average_omegas}
\end{align}
if extended to $N$ dimensions.
Note that in Eq.~\ref{SMeq:general_path_density} we have defined the ``derivatives" $\supbracket{\omega}{m}$ in a way that factors out the densities $\rhocp_i$.
Furthermore, $\supbracket{\omega}{m}$ generally depends on the density $\rhocp$.
For simplicity, we omit this dependency in the following.

Since the first derivative of the tilted free energy is zero, expanding $\Delta \fNminus (\epsilon)$ in terms of the (higher-order) derivatives of $\fNminus$ yields
\begin{align}
    \Delta \fNminus (\epsilon) = \fNminus (\rho (\epsilon)) - \fNminus (\rhocp) = \sum_{l=2}^\infty \frac{1}{l!} \left. \frac{\partial^l \fNminus}{\partial \rho_{i_1} \ldots \partial \rho_{i_l}} \right|_{\rhocp} \delta \rho_{i_1} (\epsilon) \ldots \delta \rho_{i_l} (\epsilon) =  \sum_{l=2}^\infty \frac{1}{l!} \supbracket{A}{l}_{i_1 \ldots i_l} \delta \rho_{i_1} (\epsilon) \ldots \delta \rho_{i_l} (\epsilon),
\end{align}
where we implicitly sum over repeated indices $i_1, \ldots, i_l$ and implicitly assume that $\supbracket{A}{l}$ is evaluated at the critical density $\rhocp$ here and in the following.
Grouping these terms in orders of $\epsilon$ gives
\begin{align}
    \Delta \fNminus (\epsilon) = \sum_{n=2}^\infty \epsilon^n \sum_{l=2}^n \frac{1}{l!} \supbracket{A}{l}_{i_1 \ldots i_l} \rhocp_{i_1} \ldots \rhocp_{i_l} \sum_{\substack{\{ m_1, \ldots, m_l\} \in \mathbb{N}^l \\ \sum_{k=1}^l m_k = n}} \prod_{j=1}^l \frac{\supbracket{\omega}{m_j}_{i_j}}{m_j!}.
\end{align}
Since $\supbracket{A}{l}_{i_1 \ldots i_l}$ and $ \rhocp_{i_1} \ldots \rhocp_{i_l}$ are invariant with respect to permutations of the indices $\{i_j\}_{j=1,\ldots,l}$, we do not need to explicitly keep track of the indices in $\supbracket{\omega}{m_j}$ and simplify the previous expression as
\begin{align}
    \Delta \fNminus (\epsilon) = \sum_{n=2}^\infty \epsilon^n \sum_{l=2}^n \frac{1}{l!} \supbracket{A}{l} \sum_{\substack{\{ m_1, \ldots, m_l\} \in \mathbb{N}^l \\ \sum_{k=1}^l m_k = n}} \prod_{j=1}^l \frac{\supbracket{\upsilon}{m_j}}{m_j!},
    \label{SMeq:Delta_f_contraction}
\end{align}
where we defined the full contraction 
\begin{align}
    \supbracket{A}{l} \prod_{j=1}^l \supbracket{\alpha}{j} := \supbracket{A}{l}_{i_1 \ldots i_l} \prod_{j=1}^l \supbracket{\alpha}{j}_{i_j}
\end{align}
and furthermore $\supbracket{\upsilon}{m} := \rhocp \supbracket{\omega}{m}$, or in index notation
\begin{align}
    \supbracket{\upsilon}{m}_i = \rhocp_i \supbracket{\omega}{m}_i \hspace{10pt} \forall m.
\end{align}
The last sum in Eq.~\ref{SMeq:Delta_f_contraction} contains products with $l$ factors of $
\supbracket{\upsilon}{m_j}$, whose superscripts $m_j$ add up to $n$, i.e.\ terms of the form $\left(\supbracket{\upsilon}{1}\right)^{p_1} \ldots \left(\supbracket{\upsilon}{n}\right)^{p_n}$ with $p_1, \ldots, p_n \in \mathbb{N}_0$, subject to the constraints $\sum_{j=1}^n p_j = l$ and $\sum_{j=1}^n j p_j = n$.
The number of combinatorial possibilities to get a term of the form $\left(\supbracket{\upsilon}{1}\right)^{p_1} \ldots \left(\supbracket{\upsilon}{n}\right)^{p_n}$ is
$l!/(p_1! \ldots p_n!)$, corresponding to the number of distinct permutations with repeated elements.
As a result, Eq.~\ref{SMeq:Delta_f_contraction} is rewritten as 
\begin{align}
    \Delta \fNminus (\epsilon) &= \sum_{n=2}^\infty \epsilon^n \sum_{l=2}^n \frac{1}{l!} \supbracket{A}{l} \sum_{\substack{\{ p_1, \ldots, p_n\} \in \mathbb{N}^n_0 \\ \sum_{k=1}^n p_k = l \\ \sum_{k=1}^n k p_k = n}} \frac{l!}{p_1! \ldots p_n!} \prod_{j=1}^n \left(\frac{\supbracket{\upsilon}{j}}{j!}\right)^{p_j} = \\
    &= \sum_{n=2}^\infty \frac{\epsilon^n}{n!} \sum_{l=2}^n  \supbracket{A}{l} \sum_{\substack{\{ p_1, \ldots, p_n\} \in \mathbb{N}^n_0 \\ \sum_{k=1}^n p_k = l \\ \sum_{k=1}^n k p_k = n}} n!  \prod_{j=1}^n \left[\frac{1}{p_j!} \left(\frac{\supbracket{\upsilon}{j}}{j!}\right)^{p_j} \right].
\end{align}
The last sum exactly corresponds to the partial exponential Bell polynomial $B_{n,l} (\supbracket{\upsilon}{1}, \supbracket{\upsilon}{2}, \ldots)$ and we find
\begin{align}
    \Delta \fNminus (\epsilon) =\sum_{n=2}^\infty \frac{\epsilon^n}{n!} \sum_{l=2}^n  \supbracket{A}{l} B_{n,l} (\supbracket{\upsilon}{1}, \supbracket{\upsilon}{2}, \ldots)=\sum_{n=2}^\infty \frac{\epsilon^n}{n!} \sum_{l=2}^n  \supbracket{A}{l} B_{n,l} (\rhocp \supbracket{\omega}{1}, \rhocp \supbracket{\omega}{2}, \ldots),
    \label{SMeq:expansion_free_energy_Bell}
\end{align}
with the implicit contraction between $\supbracket{A}{l}$ and the Bell polynomial $B_{n,l}$.

To find conditions for the occurrence of (higher-order) critical points, we successively determine the optimal vectors $ \supbracket{\Omega}{m}$ by minimizing the coefficients in front of $\epsilon^n$ (only necessary for even $n$; see below) and setting them to zero one after the other (up to order $2n-1$ for an $n$-th order critical point as discussed before). 

\subsection{Critical point}

For the usual critical point $n=2$, we have to second order
\begin{align}
    \mathcal{O}(\epsilon^2): \hspace{10pt} \minom{1} \left( \supbracket{A}{2} B_{2,2} (\rhocp \supbracket{\omega}{1}) \right) = \minom{1} \left( \supbracket{A}{2}_{ij} \rhocp_i \supbracket{\omega}{1}_i \rhocp_j \supbracket{\omega}{1}_j \right) = \minnu{1} \left( \supbracket{\upsilon}{1}_i \supbracket{A}{2}_{ij} \supbracket{\upsilon}{1}_j \right) \stackrel{!}{=} 0.
\end{align}
It follows that the Hessian $H=\supbracket{A}{2}$ needs to be positive semi-definite and the system is metastable thermodynamically (each critical point lies on the spinodal).
Correspondingly, the tangent to the optimal path $\rhocp \supbracket{\Omega}{1}$ coincides with the direction of the instability:
\begin{align}
    \supbracket{\Upsilon}{1}:= \rhocp \supbracket{\Omega}{1} = \supbracket{e}{1} = \rhocp \supbracket{E}{1} \hspace{10pt} \text{or} \hspace{10pt} \supbracket{\Omega}{1} = \supbracket{E}{1}.
    \label{SMeq:tangent_direction_instability}
\end{align}
Note that, in principle, there could be several directions of instability if the smallest, zero eigenvalue of the Hessian is degenerate.
Here, we focus on the non-degenerate case with a unique direction of instability, but we will briefly comment on the degenerate case below in a separate section.

Using Eq.~\ref{SMeq:tangent_direction_instability} with the optimal $\supbracket{\upsilon}{1} = \supbracket{\Upsilon}{1}$, we find for the third order
\begin{align}
    \mathcal{O}(\epsilon^3): \hspace{5pt} &\minom{2} \left( \supbracket{A}{2} B_{3,2} (\supbracket{\Upsilon}{1}, \rhocp \supbracket{ \omega}{2}) + \supbracket{A}{3} B_{3,3} (\supbracket{\Upsilon}{1}) \right)=\minnu{2} \left( 3 \supbracket{A}{2}_{ij}  \supbracket{\Upsilon}{1}_i  \supbracket{\upsilon}{2}_j + \supbracket{A}{3}_{ijk}  \supbracket{\Upsilon}{1}_i  \supbracket{\Upsilon}{1}_j \supbracket{\Upsilon}{1}_k \right) = \nonumber \\
    &=  \supbracket{A}{3}_{ijk} \rhocp_i \supbracket{\Omega}{1}_i \rhocp_j \supbracket{\Omega}{1}_j \rhocp_k \supbracket{\Omega}{1}_k =- \avgrhoNcp{\left(\supbracket{\Omega}{1} \right)^3} =- \avgrhoNcp{\left(\supbracket{E}{1} \right)^3} \stackrel{!}{=} 0,
    \label{SMeq:third_order_coeff}
\end{align}
where in the second step we used Eq.~\ref{SMeq:tangent_direction_instability} and the definition of the direction of instability, $\supbracket{A}{2}_{ij}  \supbracket{\Upsilon}{1}_i = \supbracket{A}{2}_{ij}  \supbracket{e}{1}_i = H_{ij}  \supbracket{e}{1}_i =0$, and in the third step Eq.~\ref{SMeq:useful_full_contraction}.

Taken together, apart from critical points lying on the spinodal, the third cumulant (or moment) of the direction of instability needs to be zero at a critical point:
\begin{align}
    \avgcp{\left(\supbracket{E}{1} \right)^3} := \avgrhoNcp{\left(\supbracket{E}{1} \right)^3} = 0 \hspace{10pt} \text{(critical \ point \ condition)}.
\end{align}

Before discussing higher-order critical points, we observe that the third order coefficient, Eq.~\ref{SMeq:third_order_coeff}, is independent of any vector $\supbracket{\omega}{i \geq 2}$ not yet determined previously, and therefore does not provide a condition on the next higher vector $\supbracket{\omega}{2}$.
Instead setting it to zero yields a further condition on $\supbracket{\omega}{1}$ without the need for minimization.
As we will see in the next subsection, this pattern that minimizing the coefficient in front of $\epsilon^{2n}$ defines the vector $\supbracket{\omega}{n}$ and that the coefficient of the next higher order $\epsilon^{2n+1}$ only depends on the previously determined vectors $\supbracket{\omega}{m}$, $m\leq n$, is true for all values of $n$.
Correspondingly, increasing the order of the critical point by 1, $n\rightarrow n+1$, imposes two more conditions on the critical point and simultaneously defines one additional vector.

\subsection{Higher-order critical points}

More precisely, for an $n$-th order critical point we find the following recursive equation defining vectors $\supbracket{\Upsilon}{m}, m \leq n-1$:
\begin{align}
    \supbracket{\Upsilon}{1} &= \supbracket{e}{1} \nonumber \\
    \supbracket{\Upsilon}{m} &= - (\supbracket{A}{2}_{\text{pseudo}})^{-1} \sum_{l=3}^{m+1}  \underbrace{\supbracket{A}{l} B_{m,l-1} (\supbracket{\Upsilon}{1}, \ldots, \supbracket{\Upsilon}{m-1})}_{\text{vector \ (single \ index): implicit \ contraction}}  =: - (\supbracket{A}{2}_{\text{pseudo}})^{-1} \supbracket{b}{m} \hspace{10pt} \forall m=2, \ldots, n-1.
    \label{SMeq:recursive_equation_vectors}
\end{align}
Here, $(\supbracket{A}{2}_{\text{pseudo}})^{-1}$ is the pseudo-inverse of the Hessian $H=\supbracket{A}{2}$ (which is singular on the spinodal where all critical points lie).
Specifically, we define/choose it as
\begin{align}
    (\supbracket{A}{2}_{\text{pseudo}})^{-1}_{ij} := \delta_{ij}\rho_i - \rho_i \rho_j + \sum_{\gamma=2}^R \frac{1}{1-\supbracket{\lambda}{\gamma}} \supbracket{e}{\gamma}_i \supbracket{e}{\gamma}_j,
    \label{SMeq:pseudo_inverse}
\end{align}
where -- as noted previously -- we assume a non-degenerate maximal eigenvalue of 1 for the covariance matrix, i.e.\ that $\supbracket{\lambda}{\gamma} < 1 \ \forall \gamma\geq 2$.
The vectors $\supbracket{b}{m}$, $m=2,\ldots,n-1$ satisfy the ``orthogonality" relation
\begin{align}
    \avgcp{\supbracket{\Omega}{1} \supbracket{b}{m}} = 0.
    \label{SMeq:b_orthogonality_Omega1}
\end{align}

The conditions of the $n$-th order critical point, Eqs.~\ref{SMeq:critical_point_opt_path} and~\ref{SMeq:expansion_free_energy_Bell}, can then be rewritten in terms of the vectors $\supbracket{\Upsilon}{m}$ as
\begin{align}
    \sum_{l=2}^m  \supbracket{A}{l} B_{m,l} (\supbracket{\Upsilon}{1},  \supbracket{\Upsilon}{2}, \ldots) &= \supbracket{A}{2} B_{m,2} (\supbracket{\Upsilon}{1},  \supbracket{\Upsilon}{2}, \ldots) +  \sum_{l=3}^m  (-1)^l (l-2)! \avgrhoNcp{B_{m,l} (\supbracket{\Omega}{1},  \supbracket{\Omega}{2}, \ldots)} = \nonumber\\
    &= 0 \hspace{30pt} \forall \ m=2, \ldots, 2n-1,
    \label{SMeq:condition_free_energy_critical_point_vectors}
\end{align}
where we used Eq.~\ref{SMeq:useful_full_contraction} to write the full contraction for $l\geq 3$ in terms of averages.

We will prove the recursion relation, Eq.~\ref{SMeq:recursive_equation_vectors}, together with the orthogonality relation, Eq.~\ref{SMeq:b_orthogonality_Omega1}, by induction. 

\begin{description}
    
\item[Base case] For an ordinary critical point ($n=2$), we already know that $\supbracket{\Upsilon}{1} = \supbracket{e}{1}$ and Eq.~\ref{SMeq:recursive_equation_vectors} is satisfied.
\item[Induction step: $n
\rightarrow n+1$] Suppose now that the recursion, Eq.~\ref{SMeq:recursive_equation_vectors}, holds for a critical point of order $n$.
We will show that it then also holds for a critical point of order $n+1$.
The argument will be done in three steps:
\begin{description}
    \item[1) Orthogonality] We show that if the recursion relation and the orthogonality condition hold for all $m=2, \ldots, n-1$, then the orthogonality condition holds for $n$ as well: $\avgrhoNcp{\supbracket{\Omega}{1} \supbracket{b}{n}} = 0$.
    \item[2) $\epsilon^{2n}$-term: Independence of higher-order vectors and minimization]
    We demonstrate that the orthogonality condition implies that the coefficients in front of the $\epsilon^{2n}$ term, which is the first additional one considered for a critical point of order $n+1$, does not depend on any vector $\supbracket{\upsilon}{i}$ with $i\geq n+1$.
    By minimizing the coefficient for $\epsilon^{2n}$ with respect to $\supbracket{\upsilon}{n}$, we then recover the next step in the recursion, relating $\supbracket{\Upsilon}{n}$ to the previously determined vectors $\supbracket{\Upsilon}{i}, i \leq n-1$.
    \item[3) $\epsilon^{2n+1}$-term: Independence of higher-order vectors] 
    From the recursion relation for $\supbracket{\Upsilon}{n}$ together with the orthogonality condition, it then follows that also the coefficient in front of the $\epsilon^{2n+1}$ term is independent of any vector $\supbracket{\upsilon}{i}$ with $i\geq n+1$.
\end{description}
We will go through them one-by-one.

Suppose that Eqs.~\ref{SMeq:recursive_equation_vectors} and~\ref{SMeq:b_orthogonality_Omega1} hold for a critical point of order $n$.
\begin{description}
    \item[1) Orthogonality] 
    To begin with, we use the definition of $\supbracket{b}{n}$ to write
    \begin{align}
        \avgcp{\supbracket{\Omega}{1} \supbracket{b}{n}} &= \avgcp{\supbracket{\Omega}{1} \sum_{l=3}^{n+1} \supbracket{A}{l} B_{n,l-1} \left( \supbracket{\Upsilon}{1}, \supbracket{\Upsilon}{2}, \ldots  \right)} = \\
        &= \sum_{l=3}^{n+1} (-1)^l (l-2)!\avgcp{\supbracket{\Omega}{1}  B_{n,l-1} \left( \supbracket{\Omega}{1}, \supbracket{\Omega}{2}, \ldots   \right)},
    \end{align}
    where we used Eq.~\ref{SMeq:useful_partial_contraction} and $\avgcp{\supbracket{\Omega}{1}}=0$ to rewrite the partial contraction of $\supbracket{A}{l}$ and $B_{n,l-1}$.
    The partial Bell polynomials satisfy the recursion relation
    \begin{align}
        B_{n+1,l}(\supbracket{x}{1}, \supbracket{x}{2},  \ldots) = \sum_{i=1}^{n+2-l} \binom{n}{i-1} \supbracket{x}{i} B_{n+1-i,l-1}.
        \label{SMeq:Bell_recursion}
    \end{align}
    Using this relation, we rewrite $\avgcp{\supbracket{\Omega}{1} \supbracket{b}{n}}$ as
    \begin{align}
        \avg{\supbracket{\Omega}{1} \supbracket{b}{n}} = \sum_{l=3}^{n+1} ({-}1)^l (l{-}2)! \left( \avg{B_{n+1,l}\left( \supbracket{\Omega}{1}, {...}\right)} - \sum_{i=2}^{n+2-l} \binom{n}{i{-}1} \avg{\supbracket{\Omega}{i} B_{n+1-i,l-1} \left( \supbracket{\Omega}{1}, {...}\right)} \right), \nonumber
    \end{align}
    where, for simplicity, we skip the index $cp$ here and in the following.
    We have $n\geq 2$, or $n+1 \leq 2n-1$, and so 
    \begin{align}
        \sum_{l=3}^{n+1} ({-}1)^l (l{-}2)! \avg{B_{n+1,l}\left( \supbracket{\Omega}{1}, {...}\right)} = - \supbracket{A}{2} B_{n+1,2} (\supbracket{\Upsilon}{1}, \supbracket{\Upsilon}{2}, \ldots) , 
    \end{align}
    according to Eq.~\ref{SMeq:condition_free_energy_critical_point_vectors}.
    Thus,
    \begin{align}
         &\avg{\supbracket{\Omega}{1} \supbracket{b}{n}} = \nonumber\\
         &=- \supbracket{A}{2} B_{n+1,2} (\supbracket{\Upsilon}{1}, \supbracket{\Upsilon}{2}, \ldots)  -\sum_{l=3}^{n+1} ({-}1)^l (l{-}2)! \sum_{i=2}^{n+2-l} \binom{n}{i{-}1} \avg{\supbracket{\Omega}{i} B_{n+1-i,l-1} \left( \supbracket{\Omega}{1}, {...}\right)} = \nonumber \\
         &= -\frac{1}{2} \sum_{k=2}^{n-1} \binom{n+1}{k} \supbracket{A}{2} \left[\rhocp \supbracket{\Omega}{k} \right] \left[\rhocp \supbracket{\Omega}{n+1-k} \right] - \sum_{i=2}^{n-1} \binom{n}{i-1} \avg{\supbracket{\Omega}{i} \supbracket{b}{n+1-i}},
    \end{align}
    where  we used that $\supbracket{A}{2} \supbracket{\Upsilon}{1} = 0$ (spinodal condition) and the definition of
    \begin{align}
        \supbracket{b}{n+1-i} {=} \sum_{l=3}^{n+2-i} ({-}1)^l (l{-}2)! B_{n+1-i,l-1} (\supbracket{\Omega}{1}, \supbracket{\Omega}{2},{...}) {-} \sum_{l=3}^{n+2-i} ({-}1)^l (l{-}2)! B_{n+1-i,l-1} (\supbracket{\Omega}{1}_N \supbracket{\Omega}{2}_N,{...})
        \label{SMeq:b_Bell_N}
    \end{align}
    according to Eq.~\ref{SMeq:useful_partial_contraction}, together with $\avg{\supbracket{\Omega}{i}} =0$ for all $i=2, \ldots, n-1$.
    Furthermore, $\left[ x y\right]$ denotes a vector with indices $\left[ x y\right]_i = x_i y_i$.
    
    For all $k=2, \ldots, n-1$ the recursion relation $\rhocp \supbracket{\Omega}{k} = \supbracket{\Upsilon}{k} = - (\supbracket{A}{2}_{\text{pseudo}})^{-1} \supbracket{b}{k}$ holds by induction and we have
    \begin{align}
        \avg{\supbracket{\Omega}{1} \supbracket{b}{n}} = \frac{1}{2} \sum_{k=2}^{n-1} \binom{n+1}{k} \supbracket{A}{2} \left[(\supbracket{A}{2}_{\text{pseudo}})^{-1} \supbracket{b}{k} \right] \left[\rhocp \supbracket{\Omega}{n+1-k} \right] - \sum_{i=2}^{n-1} \binom{n}{i-1} \avg{\supbracket{\Omega}{i} \supbracket{b}{n+1-i}}.
    \end{align}
    Employing Eq.~\ref{SMeq:prod_pseudoinv} for the product of the Hessian $\supbracket{A}{2}$ with its pseudo-inverse and using that $\supbracket{b}{k}_N=0$, Eq.~\ref{SMeq:b_Bell_N}, we find
    \begin{align}
        \supbracket{A}{2} \left[(\supbracket{A}{2}_{\text{pseudo}})^{-1} \supbracket{b}{k} \right] = \supbracket{b}{k} - \avg{\supbracket{\Omega}{1} \supbracket{b}{k}} \sum_{\gamma=1}^R \supbracket{V}{1}_{\gamma} \supbracket{r}{\gamma} = \supbracket{b}{k},
        \label{SMeq:b_orthogonality_matrices}
    \end{align}
    where we used that $\avg{\supbracket{\Omega}{1} \supbracket{b}{k}}=0 \ \forall k=2, \ldots, n-1$ by induction.
    Thus,
    \begin{align}
        \avg{\supbracket{\Omega}{1} \supbracket{b}{n}} = \frac{1}{2} \sum_{k=2}^{n-1} \left( \binom{n+1}{k} \avg{\supbracket{\Omega}{n+1-k} \supbracket{b}{k}} - 2 \binom{n}{k-1} \avg{\supbracket{\Omega}{k} \supbracket{b}{n+1-k}} \right).
    \end{align}
    
    The recursion relation, Eq.~\ref{SMeq:recursive_equation_vectors} together with the definition of the pseudo-inverse, Eq.~\ref{SMeq:pseudo_inverse}, implies that $\forall i,j = 2, \ldots, n-1$
    \begin{align}
        \avg{\supbracket{\Omega}{i} \supbracket{b}{j}} &= \supbracket{b}{j} [\rhocp \supbracket{\Omega}{i}] = \supbracket{b}{j} \supbracket{\Upsilon}{i} = - \supbracket{b}{j} \left[ (\supbracket{A}{2}_{\text{pseudo}})^{-1} \supbracket{b}{i} \right] = \\
        &= -\avg{\supbracket{b}{i} \supbracket{b}{j}} + \avg{\supbracket{b}{i}} \avg{\supbracket{b}{j}} - \sum_{\gamma=2}^R \frac{1}{1-\supbracket{\lambda}{\gamma}} \avg{\supbracket{b}{i} \supbracket{E}{\gamma}} \avg{\supbracket{b}{j} \supbracket{E}{\gamma}} = \avg{\supbracket{\Omega}{j} \supbracket{b}{i}},
    \end{align}
    where the last step is due to symmetry $i \leftrightarrow j$.
    As a result, we have
    \begin{align}
        \avg{\supbracket{\Omega}{1} \supbracket{b}{n}} &{=} \frac{1}{2} \sum_{k=2}^{n-1} \left( \binom{n{+}1}{k} \avg{\supbracket{\Omega}{n+1-k} \supbracket{b}{k}} {-} \binom{n}{k{-}1} \avg{\supbracket{\Omega}{k} \supbracket{b}{n+1-k}} {-} \binom{n}{k{-}1} \avg{\supbracket{b}{k} \supbracket{\Omega}{n+1-k}} \right) = \nonumber\\
        &{=} \frac{1}{2} \sum_{k=2}^{n-1} \left( \binom{n{+}1}{k} \avg{\supbracket{\Omega}{n+1-k} \supbracket{b}{k}} {-} \binom{n}{n{-}k} \avg{\supbracket{\Omega}{n+1-k} \supbracket{b}{k}} {-} \binom{n}{k{-}1} \avg{\supbracket{b}{k} \supbracket{\Omega}{n+1-k}} \right) = \nonumber \\
        &{=} \frac{1}{2} \sum_{k=2}^{n-1} \left( \binom{n{+}1}{k}  - \binom{n}{k}  - \binom{n}{k{-}1} \right) \avg{\supbracket{b}{k} \supbracket{\Omega}{n+1-k}}.
    \end{align}
    In the second step, we used the variable transformation $k\rightarrow n+1-k$ on the second term.
    Overall, since $\binom{n{+}1}{k}  - \binom{n}{k}  - \binom{n}{k{-}1}=0 \ \forall \ 2\leq k \leq n-1$, we conclude that
    \begin{align}
        \avg{\supbracket{\Omega}{1} \supbracket{b}{n}}=0.
    \end{align}
    
    \vspace{10pt}
    
    \item[2) $\epsilon^{2n}$-term: Independence of higher-order vectors and minimization: next recursion step] 
    
    Consider now the coefficient in front of $\epsilon^{2n}$,
    \begin{align}
    \frac{1}{(2n)!} \sum_{l=2}^{2n}  \supbracket{A}{l} B_{2n,l} ( \supbracket{\upsilon}{1},  \supbracket{\upsilon}{2}, \ldots) 
    = \sum_{l=2}^{2n}  \supbracket{A}{l} \sum_{\substack{\{ p_1, \ldots, p_{2n}\} \in \mathbb{N}^{2n}_0 \\ \sum_{k=1}^{2n} p_k = l \\ \sum_{k=1}^{2n} k p_k = 2n}} \prod_{j=1}^{2n} \left(\frac{1}{p_j!} \left(\frac{\supbracket{\upsilon}{j}}{j!}\right)^{p_j} \right)
    \end{align}
    Collecting all terms containing $\supbracket{\upsilon}{n+1+s}$ for $0 \leq s \leq n-2$ ($B_{2n,l}$ only includes vectors $\supbracket{\upsilon}{i}$ with $i\leq 2n-l+1$ since all its terms contain $l$ factors whose sum of superscripts needs to equal $2n$), we find
    \begin{align}
    \tau_{n+1+s}:= \sum_{l=2}^{2n}  \supbracket{A}{l} \frac{\supbracket{\upsilon}{n+1+s}}{(n+1+s)!} \sum_{\substack{\{ p_1, \ldots, p_{n-1-s}\} \in \mathbb{N}^{n-1-s}_0 \\ \sum_{k=1}^{n-1-s} p_k = l-1 \\ \sum_{k=1}^{n-1-s} k p_k = 2n-(n+1+s)}} \prod_{j=1}^{n-1-s} \left(\frac{1}{p_j!} \left(\frac{\supbracket{\upsilon}{j}}{j!}\right)^{p_j} \right),
    \end{align} 
    where we used that any vector $\supbracket{\upsilon}{n+1+s}$ with $s \geq 0$ cannot be in a product together with any other vector $\supbracket{\upsilon}{n-1-s+t}$ with $t > 0$ (otherwise the sum of the superscripts exceeds $2n$). 
    In particular $\supbracket{\upsilon}{n+1+s}$ with $s \geq 0$ cannot occur with a power $>1$.
    Furthermore, since we factored out $\supbracket{\upsilon}{n+1+s}$, in the last sum the number of factors $l$ gets reduced by 1 and the sum of the superscripts by $n+1+s$ (effectively setting $p_{n+1+s} = 1$).
    We observe that the last sum is exactly $B_{n-1-s,l-1}/(n-1-s)!$, giving 
    \begin{align}
    \tau_{n{+}1{+}s} &{=} \left[\sum_{l=2}^{n-s}  \supbracket{A}{l} B_{n-1-s,l-1}  ( \supbracket{\upsilon}{1},  \supbracket{\upsilon}{2}, \ldots) \right] \frac{\supbracket{\upsilon}{n+1+s}}{(n+1+s)! (n-1-s)!}  = \nonumber \\
    &{=} \left[ \supbracket{A}{2} B_{n{-}1{-}s,1}  ( \supbracket{\upsilon}{1},  \supbracket{\upsilon}{2}, {\ldots})  {+} \sum_{l{=}3}^{n{-}s}  \supbracket{A}{l} B_{n{-}1{-}s,l{-}1}  ( \supbracket{\upsilon}{1},  \supbracket{\upsilon}{2}, {\ldots}) \right] \frac{\supbracket{\upsilon}{n{+}1{+}s}}{(n{+}1{+}s)! (n{-}1{-}s)!}  = \nonumber\\
    &{=} \left[ \supbracket{A}{2}  \supbracket{\upsilon}{n-1-s} + \sum_{l=3}^{n-s}  \supbracket{A}{l} B_{n-1-s,l-1}  ( \supbracket{\upsilon}{1},  \supbracket{\upsilon}{2}, \ldots) \right] \frac{\supbracket{\upsilon}{n+1+s}}{(n+1+s)! (n-1-s)!},
    \end{align}
    where terms in brackets $[]$ are to be interpreted as vectors with single index.
    Plugging in the optimal vectors as determined from the conditions for the $n$-th order critical point, $\supbracket{\upsilon}{i} =  \supbracket{\Upsilon}{i} \ \forall i=1,\ldots,n-1$, we have $\forall \ 0 \leq s\leq n-2$:
    \begin{align}
    \tau_{n+1+s}  =  \left[ \supbracket{A}{2}  \supbracket{\Upsilon}{n-1-s} + \supbracket{b}{n-s-1} \right] \frac{\supbracket{\upsilon}{n{+}1{+}s}}{(n{+}1{+}s)! (n{-}1{-}s)!}.
    \end{align}
    Furthermore, the recursion relation for the $n$-th order critical point: 
    \begin{align}
         \supbracket{\Upsilon}{n-1-s}  = -  (\supbracket{A}{2}_{\text{pseudo}})^{-1} \supbracket{b}{n-1-s}
    \end{align}
    can be rewritten as 
    \begin{align}
        \supbracket{A}{2}  \supbracket{\Upsilon}{n-1-s}  &=  - \supbracket{A}{2} (\supbracket{A}{2}_{\text{pseudo}})^{-1} \supbracket{b}{n-1-s} = \\
        &=- \supbracket{b}{n-1-s} + \left(\sum_{\gamma=1}^R \supbracket{V}{1}_{\gamma} \supbracket{r}{\gamma}\right) \avgrhoNcp{\supbracket{E}{1} \supbracket{b}{n-1-s}} = \\
        &= - \supbracket{b}{n-1-s} + \left(\sum_{\gamma=1}^R \supbracket{V}{1}_{\gamma} \supbracket{r}{\gamma}\right) \avgrhoNcp{\supbracket{\Omega}{1} \supbracket{b}{n-1-s}} =  - \supbracket{b}{n-1-s}
    \end{align}
    using Eq.~\ref{SMeq:prod_pseudoinv} and that $\avgrhoNcp{\supbracket{\Omega}{1} \supbracket{b}{n-1-s}} =0$ for all $0 \leq s \leq n-2$ due to the orthogonality relation for the $n$-th order critical point.
    Thus, 
    \begin{align}
        \tau_{n+1+s}  = 0
    \end{align}
    and the coefficient in front of $\epsilon^{2n}$ is independent of $\supbracket{\upsilon}{n+1+s}$ for $s \geq 0$.
    Correspondingly, the $\epsilon^{2n}$-coefficient is only a function of $\supbracket{\upsilon}{n}$ and the previously determined $\supbracket{\Upsilon}{i}, i=1, \ldots, n-1$, and by minimizing it with respect to $\supbracket{\upsilon}{n}$, the optimal vector $\supbracket{\Upsilon}{n}$ can be determined.
    
    Specifically, the coefficient in front of $\epsilon^{2n}$, $c_{2n}$ contains a quadratic term in $\supbracket{\Upsilon}{n}$, a linear term in $\supbracket{\Upsilon}{n}$ and a constant term:
    \begin{align}
        c_{2n} &=   \frac{1}{2 (2n)!} \binom{2n}{n} \supbracket{A}{2} \supbracket{\upsilon}{n} \supbracket{\upsilon}{n} + \sum_{l=3}^{2n} \supbracket{A}{l} \frac{\supbracket{\upsilon}{n}}{n! n!} B_{n,l-1} (\supbracket{\Upsilon}{1}, \supbracket{\Upsilon}{2}, \ldots)  + \text{const} = \\
        &=\frac{1}{(n!)^2} \left( \frac{1}{2} \supbracket{A}{2} \supbracket{\upsilon}{n} \supbracket{\upsilon}{n} + \left[\sum_{l=3}^{n+1} \supbracket{A}{l}  B_{n,l-1} (\supbracket{\Upsilon}{1}, \supbracket{\Upsilon}{2}, \ldots)\right] \supbracket{\upsilon}{n} \right) + \text{const} =\\
        &=\frac{1}{(n!)^2} \left( \frac{1}{2} \supbracket{A}{2} \supbracket{\upsilon}{n} \supbracket{\upsilon}{n} + \supbracket{b}{n} \supbracket{\upsilon}{n} \right) + \text{const}.
    \end{align}
    Using the orthogonality condition for $\supbracket{b}{n}$ analogously as  in Eq.~\ref{SMeq:b_orthogonality_matrices} for $\supbracket{b}{k}, k=2, \ldots, n-1$, $\supbracket{A}{2} \left[(\supbracket{A}{2}_{\text{pseudo}})^{-1} \supbracket{b}{n} \right] = \supbracket{b}{n}$, we rewrite $\tilde{c}_{2n}:=2 (n!)^2(c_{2n}-\text{const})$ as 
    \begin{align}
        \tilde{c}_{2n} 
        = \left(  \left[\supbracket{\upsilon}{n} + (\supbracket{A}{2}_{\text{pseudo}})^{-1} \supbracket{b}{n}\right]^T \supbracket{A}{2} \left[ \supbracket{\upsilon}{n} + (\supbracket{A}{2}_{\text{pseudo}})^{-1} \supbracket{b}{n}\right] - \left[\supbracket{b}{n}\right]^T (\supbracket{A}{2}_{\text{pseudo}})^{-1} \supbracket{b}{n}  \right),
    \end{align}
    where we used the normal matrix and vector $[]$ notation ($[]^T$ is the transpose).
    
    Since the Hessian $\supbracket{A}{2}$ is a positive semi-definite matrix, $\tilde{c}_{2n}$ (or, equivalently, $c_{2n}$) has a minimum for $\supbracket{\upsilon}{n} = \supbracket{\Upsilon}{n}$ if
    \begin{align}
        \supbracket{\Upsilon}{n} = - (\supbracket{A}{2}_{\text{pseudo}})^{-1} \supbracket{b}{n} + \alpha \supbracket{e}{1}
        \label{SMeq:recursion_alpha_dependence}
    \end{align}
    for any $\alpha \in \mathbb{R}$.
    Here, we choose $\alpha=0$ for simplicity.
    This choice, however, does not influence any of our conditions for critical points as it just corresponds to a rescaling of the path in density space (see below).
    We thus recover the next step of the recursion relation:
    \begin{align}
        \supbracket{\Upsilon}{n} = - (\supbracket{A}{2}_{\text{pseudo}})^{-1} \supbracket{b}{n}.
    \end{align}
    
    The minimized coefficient $c^{\text{min}}_{2n}$ must then equal zero and we find the first additional condition when going from an $n$-th to and $n+1$-th order critical point:
    \begin{align}
        \sum_{l=2}^{2n}  \supbracket{A}{l} B_{2n,l} ( \supbracket{\Upsilon}{1},  \supbracket{\Upsilon}{2}, \ldots) = 0.
    \end{align}
    As discussed before, this expression is a function of $\supbracket{\Upsilon}{1}, \ldots, \supbracket{\Upsilon}{n}$ only.
    All terms with higher-order vectors cancel each other.
    The same is true for the coefficient in front of the $\epsilon^{2n+1}$ term, as we will show next.
    Therefore, for an $(n{+}1)$-th order critical point, only the vectors $\supbracket{\Upsilon}{1}, \ldots, \supbracket{\Upsilon}{n}$ need to be (or can uniquely be) specified.

    \vspace{10pt}
    
    \item[3) $\epsilon^{2n+1}$-term: Independence of higher-order vectors]
    Similarly, we rewrite the coefficient in front of $\epsilon^{2n+1}$:
    \begin{align}
        \frac{1}{(2n+1)!} \sum_{l=2}^{2n+1}  \supbracket{A}{l} B_{2n+1,l} ( \supbracket{\upsilon}{1},  \supbracket{\upsilon}{2}, \ldots) 
        = \sum_{l=2}^{2n+1}  \supbracket{A}{l} \sum_{\substack{\{ p_1, \ldots, p_{2n+1}\} \in \mathbb{N}^{2n+1}_0 \\ \sum_{k=1}^{2n+1} p_k = l \\ \sum_{k=1}^{2n+1} k p_k = 2n+1}} \prod_{j=1}^{2n+1} \left(\frac{1}{p_j!} \left(\frac{\supbracket{\upsilon}{j}}{j!}\right)^{p_j} \right).
    \end{align}
    The terms linear in $\supbracket{\upsilon}{n+1+s}$ for $0 \leq s \leq n-2$ are
    \begin{align}
    \tilde{\tau}_{n+1+s}&:= \sum_{l=2}^{2n+1}  \supbracket{A}{l} \frac{\supbracket{\upsilon}{n+1+s}}{(n+1+s)!} \sum_{\substack{\{ p_1, \ldots, p_{n-s}\} \in \mathbb{N}^{n-s}_0 \\ \sum_{k=1}^{n-s} p_k = l-1 \\ \sum_{k=1}^{n-s} k p_k = 2n+1-(n+1+s)}} \prod_{j=1}^{n-s} \left(\frac{1}{p_j!} \left(\frac{\supbracket{\upsilon}{j}}{j!}\right)^{p_j} \right) = \\
    &= \left[\sum_{l=2}^{2n+1} \supbracket{A}{l} B_{n-s,l-1} ( \supbracket{\upsilon}{1},  \supbracket{\upsilon}{2}, {\ldots})\right] \frac{\supbracket{\upsilon}{n+1+s}}{(n+1+s)!(n-s)!}  =  \\
    &=\left[ \supbracket{A}{2} \supbracket{\upsilon}{n-s} + \sum_{l=3}^{n-s+1} \supbracket{A}{l} B_{n-s,l-1} ( \supbracket{\upsilon}{1},  \supbracket{\upsilon}{2}, {\ldots})\right] \frac{\supbracket{\upsilon}{n+1+s}}{(n+1+s)!(n-s)!}.
    \end{align} 
    Evaluated at $\supbracket{\upsilon}{i} = \supbracket{\Upsilon}{i} \ \forall i=1,\ldots,n-1$, this expression yields
    \begin{align}
    \tilde{\tau}_{n+1+s} = \left[ \supbracket{A}{2} \supbracket{\Upsilon}{n-s} + \supbracket{b}{n-s} \right] \frac{\supbracket{\upsilon}{n+1+s}}{(n+1+s)!(n-s)!} =0,
    \end{align} 
    where we used the recursion relation for $\supbracket{\Upsilon}{i}$ together with the orthogonality condition $\avgrhoNcp{\supbracket{\Omega}{1} \supbracket{b}{i}} =0$ for all $=1, \ldots, n$ (with the ones for $i=n$ as determined by induction in the first and second step, respectively).
    Furthermore, the term linear in $\supbracket{\upsilon}{2n}$ is $\tilde{\tau}_{2n}=\supbracket{A}{2} \supbracket{\Upsilon}{1} \frac{\supbracket{\upsilon}{2n}}{(2n)!}=0$.
    
    The coefficient $c_{2n+1}$ in front of $\epsilon^{2n+1}$ is thus fully determined by the vectors $\supbracket{\Upsilon}{i}, i=1,\ldots, n$ and cannot be minimized by a specific choice of higher-order vectors -- in contrast to $c_{2n}$.
    
\end{description}

\end{description}

Taken together, we find the following to additional conditions for an $n+1$-th order critical point as compared to an $n$-th order critical point:
    \begin{align}
        \sum_{l=2}^{2n}  \supbracket{A}{l} B_{2n,l} ( \supbracket{\Upsilon}{1},  \supbracket{\Upsilon}{2}, \ldots) &= 0 \\
        \sum_{l=2}^{2n+1}  \supbracket{A}{l} B_{2n+1,l} ( \supbracket{\Upsilon}{1},  \supbracket{\Upsilon}{2}, \ldots) &= 0,
    \end{align}
    which both only depend on the vectors $ \supbracket{\Upsilon}{i}, i=1, \ldots, n$ given by the recursion relation
    \begin{align}
    \supbracket{\Upsilon}{1} &= \supbracket{e}{1} \nonumber \\
    \supbracket{\Upsilon}{i} &= - (\supbracket{A}{2}_{\text{pseudo}})^{-1} \sum_{l=3}^{i+1}  \supbracket{A}{l} B_{i,l-1} (\supbracket{\Upsilon}{1}, \ldots, \supbracket{\Upsilon}{i-1}), \hspace{10pt} i\geq 2.
\end{align}

It follows that
\begin{align}
    \supbracket{\Omega}{1} = \supbracket{E}{1}
\end{align}
and for $i=2,\ldots,n$
\begin{align}
    \rhocp \supbracket{\Omega}{i} &= - (\supbracket{A}{2}_{\text{pseudo}})^{-1} \sum_{l=3}^{i+1} (-1)^l (l-2)! \left[ B_{i,l-1} (\supbracket{\Omega}{1}, \supbracket{\Omega}{2}, \ldots) - B_{i,l-1} (\supbracket{\Omega}{1}_N, \supbracket{\Omega}{2}_N, \ldots)\right] = \\
    &= - \rhocp \sum_{l=3}^{i+1} (-1)^l (l-2)! \left[ B_{i,l-1} (\Omega) - \avgcp{B_{i,l-1} (\Omega)} +\sum_{\gamma=2}^R \frac{1}{1-\supbracket{\lambda}{\gamma}} \supbracket{E}{\gamma} \avgcp{\supbracket{E}{\gamma} B_{i,l-1} (\Omega)}\right] =\\
    &=  \rhocp \sum_{l=2}^{i} (-1)^l (l-1)! \left[ B_{i,l} (\Omega) - \avgcp{B_{i,l} (\Omega)} +\sum_{\gamma=2}^R \frac{1}{1-\supbracket{\lambda}{\gamma}} \supbracket{E}{\gamma} \avgcp{\supbracket{E}{\gamma} B_{i,l} (\Omega)}\right],
\end{align}
where we used Eqs.~\ref{SMeq:useful_partial_contraction},~\ref{SMeq:pseudo_inv_vector} and abbreviated $B_{i,l} (\Omega):=B_{i,l} (\supbracket{\Omega}{1}, \supbracket{\Omega}{2}, \ldots)$.
Thus, we find the following vectors for an $n$-th order critical point:
\begin{align}
    \supbracket{\Omega}{1} &= \supbracket{E}{1} \\
    \supbracket{\Omega}{i} &= \sum_{l=2}^i \left[ \tilde{B}_{i,l} (\Omega) - \avgcp{\tilde{B}_{i,l} (\Omega)} +\sum_{\gamma=2}^R \frac{1}{1-\supbracket{\lambda}{\gamma}} \supbracket{E}{\gamma} \avgcp{\supbracket{E}{\gamma} \tilde{B}_{i,l} (\Omega)}\right] \hspace{10pt} \forall i=2,\ldots,n-1, \label{SMeq:vectors_n_order}
\end{align}
with $\tilde{B}_{i,l} (\Omega):= (-1)^l (l-1)! B_{i,l} (\Omega)$, as given in the main text.

The conditions for the $n$-th order critical point are then
\begin{align}
    0&=\sum_{l=2}^m \supbracket{A}{l} B_{m,l} (\Upsilon) = \\
    &=-\frac{1}{2} \sum_{\gamma=1}^R \sum_{k=1}^{m-1} \binom{m}{k} \avgcp{\supbracket{r}{\gamma} \supbracket{\Omega}{k}} \avgcp{\supbracket{r}{\gamma} \supbracket{\Omega}{m-k}} + \sum_{l=2}^m (-1)^l (l-2)! \avgcp{B_{m,l}(\Omega)} \hspace{10pt} \forall m=2,\ldots, 2n-1,
    \label{SMeq:conditions_n_order}
\end{align}
where we used Eqs.~\ref{SMeq:useful_full_contraction_Hessian},~\ref{SMeq:useful_full_contraction} and that $B_{m,2} (\Omega) = \frac{1}{2} \sum_{k=1}^{m-1} \binom{m}{k} \supbracket{\Omega}{k} \supbracket{\Omega}{m-k}$.

Note that choosing a different pseudo-inverse by adding a term like $\beta \supbracket{e}{1}_i \supbracket{e}{1}_j$ to our specific choice of the pseudo-inverse, Eq.~\ref{SMeq:pseudo_inverse}, does not influence our results: $\supbracket{\Omega}{1}$ is orthogonal to $\supbracket{b}{n}$ for any choice of $\beta$.
Furthermore, note that in order to conclusively show that a certain mixture composition corresponds to an $n$-th order critical point, it is also necessary to show that the next-higher-order term in the expansion of the free energy is positive for all possible continuations of the optimal path; see Section~\ref{SMsec:Potts} for a discussion in the context of the suggested mean-field $q{=}3$-states Potts model.

\subsection{Recursion relation in terms of derivatives of a  ``Hamiltonian"-like function}

Instead of expressing the recursion relation in terms of the partial exponential Bell polynomials, they can also be rewritten in terms of derivatives of a ``Hamiltonian"-like function.
To see this equivalence, we define the analytic function $G(\epsilon)$ with $G(0)\equiv 1$ (for convenience, see below) via its derivatives at $\epsilon=0$:
\begin{align}
    \left. \supbracket{G}{m}(\epsilon) \right|_{\epsilon=0} := \left. \frac{\mathrm{d}^m G(\epsilon)}{\mathrm{d} \epsilon^m} \right|_{\epsilon=0} = \supbracket{\Omega}{m} \hspace{10pt} \forall m \geq 1, 
\end{align}
or alternatively, 
\begin{align}
    G(\epsilon) = \sum_{n=0}^{\infty} \frac{\epsilon^n}{n!} \supbracket{G}{n}(0) = \sum_{n=0}^{\infty} \frac{\epsilon^n}{n!} \supbracket{\Omega}{n},
\end{align}
with $\supbracket{G}{0}:=G$ and $\supbracket{\Omega}{0}:=1$.
The function $G$ needs to fulfil
\begin{align}
0 = \avgcp{\supbracket{\Omega}{m}} = \left. \frac{\mathrm{d}^m}{\mathrm{d}\epsilon^m} \avgcp{G} \right|_{\epsilon=0} \hspace{10pt} \forall m \geq 1,
\end{align}
or equivalently,
\begin{align}
    \avgcp{G} = \mathrm{const} = \avgcp{G(0)}=1.
\end{align}
Note that since $\supbracket{\Omega}{m}$ is a vector, $\supbracket{G}{m}$ is also a vector:
$\left. \supbracket{G}{m}_N (\epsilon) \right|_{\epsilon=0} = \supbracket{\Omega}{m}_N$.

The recursion relation for an $n$-th order critical point, Eq.~\ref{SMeq:recursive_equation_vectors}, is then rewritten as
\begin{align}
    \supbracket{\Upsilon}{m} &= - (\supbracket{A}{2}_{\text{pseudo}})^{-1} \left. \sum_{l=2}^m \supbracket{A}{l+1} B_{m,l} (\rhocp \supbracket{G}{1}(\epsilon), \rhocp \supbracket{G}{2} (\epsilon), \ldots) \right|_{\epsilon=0} = \\
    &=- (\supbracket{A}{2}_{\text{pseudo}})^{-1} \left. \sum_{l=2}^m (-1)^{l+1} (l{-}1)! \left[ B_{m,l} (\supbracket{G}{1}(\epsilon), \supbracket{G}{2} (\epsilon), {...}) {-} B_{m,l} (\supbracket{G}{1}_N (\epsilon), \supbracket{G}{2}_N (\epsilon), {...}) \right] \right|_{\epsilon=0} \ \forall m =2, {...}, n-1,
\end{align}
using Eq.~\ref{SMeq:useful_partial_contraction}.
Importantly, the partial Bell polynomials satisfy Faà di Bruno's formula:
\begin{align}
    \frac{\mathrm{d}^n}{\mathrm{d} \epsilon^n} F(G(\epsilon)) = \sum_{k=1}^n \supbracket{F}{k} (G(\epsilon)) B_{n,k} (\supbracket{G}{1}(\epsilon), \supbracket{G}{2}(\epsilon), \ldots).
\end{align}

We use this formula for the function $F(x) = 1-x + \log x$, whose derivatives fulfil
$\left.\supbracket{F}{k}(G(\epsilon))\right|_{\epsilon=0} = \supbracket{F}{k}(1) = (-1)^{k+1} (k-1)! \ \forall k \geq 2$ and $\left.\supbracket{F}{k}(G(\epsilon))\right|_{\epsilon=0}=0$ otherwise.
This allows us to express $\supbracket{\Upsilon}{m}$ as follows:
\begin{align}
    \supbracket{\Upsilon}{m} &= - (\supbracket{A}{2}_{\text{pseudo}})^{-1} \left. \frac{\mathrm{d}^m}{\mathrm{d}\epsilon^m} \left[ F(G(\epsilon)) - F(G_N(\epsilon)) \right] \right|_{\epsilon=0} = (\supbracket{A}{2}_{\text{pseudo}})^{-1}  \left. \frac{\mathrm{d}^m}{\mathrm{d}\epsilon^m} \left[ G-\log G-G_N + \log G_N \right] \right|_{\epsilon=0} =\\
    &= \rho \left. \frac{\mathrm{d}^m}{\mathrm{d}\epsilon^m} \left[ G-\log G -\avgcp{G} + \avgcp{\log G} + \sum_{\gamma=2}^R \frac{1}{1-\supbracket{\lambda}{\gamma}} \supbracket{E}{\gamma} \avgcp{\supbracket{E}{\gamma} \left(G-\log G \right)}\right] \right|_{\epsilon=0} \hspace{10pt} \forall m =2, {...}, n-1,
\end{align}
where in the last step we used Eq.~\ref{SMeq:pseudo_inv_vector}.
Thus,
\begin{align}
    \left. \frac{\mathrm{d}^m}{\mathrm{d}\epsilon^m} G \right|_{\epsilon=0} = \left. \frac{\mathrm{d}^m}{\mathrm{d}\epsilon^m} \left[ G-\log G -\avgcp{G} + \avgcp{\log G} + \sum_{\gamma=2}^R \frac{1}{1-\supbracket{\lambda}{\gamma}} \supbracket{E}{\gamma} \avgcp{\supbracket{E}{\gamma} \left(G-\log G \right)}\right] \right|_{\epsilon=0} \hspace{10pt} \forall m =2, {...}, n-1.
    \label{SMeq:deriv_G} 
\end{align}

We now express 
\begin{align}
    G = \frac{e^h}{\avgcp{e^h}} \hspace{10pt} \mathrm{or} \hspace{10pt} \log G = h - \log \avgcp{e^h}
\end{align}
in terms of a Hamiltonian-like function $h$ with $h(0) =0$.
This specific form satisfies $G(0) = 1$ and $\avgcp{G}=1$ as desired.

Eq.~\ref{SMeq:deriv_G} then requires that the Hamiltonian fulfils
\begin{align}
    0 = \left. \frac{\mathrm{d}^m}{\mathrm{d}\epsilon^m} \left[ -h + \avgcp{h} +\sum_{\gamma=2}^R \frac{1}{1-\supbracket{\lambda}{\gamma}} \supbracket{E}{\gamma} \avgcp{\supbracket{E}{\gamma} \left(\frac{e^h}{\avgcp{e^h}} -h \right)} \right] \right|_{\epsilon=0} \hspace{10pt} \forall m =2, {...}, n-1.
\end{align}
since $\avgcp{\supbracket{E}{\gamma} \log \avgcp{e^h}} = 0$.
Assuming that $h$ is analytic, we thus find
\begin{align}
    h - \avgcp{h} -\sum_{\gamma=2}^R \frac{1}{1-\supbracket{\lambda}{\gamma}} \supbracket{E}{\gamma} \avgcp{\supbracket{E}{\gamma} \left(\frac{e^h}{\avgcp{e^h}} -h \right)} = c_0 + c_1 \epsilon + \mathcal{O}(\epsilon^n).
\end{align}
Evaluating this expression and its derivative at $\epsilon =0$ determines the constants of integration:
\begin{align}
    c_0 &= 0 \\
    c_1 &= \supbracket{\Omega}{1},
\end{align}
where we used that $\supbracket{\Omega}{1} =  \supbracket{G}{1}(\epsilon=0) = \left. \supbracket{h}{1} (\epsilon)-\avgcp{\supbracket{h}{1} (\epsilon)} \right|_{\epsilon=0}$.
Taken together,
\begin{align}
    h - \avgcp{h} -\sum_{\gamma=2}^R \frac{1}{1-\supbracket{\lambda}{\gamma}} \supbracket{E}{\gamma} \avgcp{\supbracket{E}{\gamma} \left(\frac{e^h}{\avgcp{e^h}} -h \right)} =  \epsilon \supbracket{\Omega}{1} + \mathcal{O}(\epsilon^n).
    \label{SMeq:recursion_Hamiltonian}
\end{align}
This form of the recursion relation has to be combined with the conditions for the critical points, Eqs.~\ref{SMeq:condition_free_energy_critical_point_vectors}, which are rewritten as 
\begin{align}
    \frac{\mathrm{d}^k}{\mathrm{d} \epsilon^k} \left.\left[ \frac{\avgcp{e^h h}}{\avgcp{e^h}} - \log \avgcp{e^h} - \frac{1}{2} \sum_{\gamma=1}^R \frac{\avgcp{\supbracket{r}{\gamma}\left(e^h-2 \avgcp{e^h}\right)} \avgcp{\supbracket{r}{\gamma}e^h}}{\avgcp{e^h}^2} \right]\right|_{\epsilon=0} = 0 \hspace{20pt} \forall k=2, \ldots, 2n-1
\end{align}
in terms of the Hamiltonian $h$, using Faà di Bruno's formula with $F(x) = -x+x\log x +1$.

While for general number of features, we did not manage to find an explicit solution to these equations, we did solve them for $R=1$.
We discuss this solution in the next subsection.

\subsection{Single feature $R=1$: Solution of the recursion relation and explicit conditions for higher-order critical points}

For a single feature $R=1$, the recursion relation for the Hamiltonian $h$, Eq.~\ref{SMeq:recursion_Hamiltonian}, reduces to \begin{align}
    h - \avgcp{h}  =  \epsilon \supbracket{\Omega}{1} + \mathcal{O}(\epsilon^n)= \epsilon \left(r-\avgcp{r} \right) + \mathcal{O}(\epsilon^n)
    \label{SMeq:recursion_Hamiltonian_single_feature},
\end{align}
where we abbreviated $\supbracket{r}{1}=r$ and have used that $\supbracket{\Omega}{1} = \supbracket{E}{1} = \sum_{\gamma=1}^1 \supbracket{V}{1}_{\gamma} (\supbracket{r}{\gamma} - \avgcp{\supbracket{r}{\gamma}}) =r - \avgcp{r}$.
This recursion -- indeed for all orders $n$ -- is solved by
\begin{align}
    h=\epsilon r + c = \epsilon r,
    \label{SMeq:recursion_Hamiltonian_single_feature_sol}
\end{align}
where $c$ is a constant in terms of the vector, i.e.\ independent of the vector index.
Without loss of generality, we choose $c=0$ (any constant cancels in $G=e^{h+c}/\avgcp{e^{h+c}} = e^h/\avgcp{e^h}$) and we find
\begin{align}
    G = \frac{e^{\epsilon r}}{\avgcp{e^{\epsilon r}}}.
\end{align}

In terms of $G$ (or the Hamiltonian $h=\epsilon r$), the optimal path is given by
\begin{align}
    \rho_i^{\text{opt}}(\epsilon) = \supbracket{\rho}{cp}_i \left(1+\sum_{m=1}^\infty  \frac{\epsilon^m}{m!} \supbracket{\Omega}{m}_i\right) = \supbracket{\rho}{cp}_i  \sum_{m=0}^\infty \frac{\epsilon^m}{m!} \supbracket{G}{m}_i (0) = \rhocp_i G_i(\epsilon) = \rhocp_i \frac{e^{\epsilon r_i}}{\avgcp{e^{\epsilon r}}}.
\end{align}
Here, it becomes apparent why we chose $G(0)=1$: This choice allows to directly express the optimal path as Taylor series of $G$ in $\epsilon$.

This explicit representation of the optimal path (vectors) in terms of $G$ also enables us to derive explicit conditions for critical points of arbitrary order $n$.
To that end, we rewrite the conditions for higher-order critical points, Eq.~\ref{SMeq:condition_free_energy_critical_point_vectors}, in terms of $G$:
\begin{align}
    0&=\sum_{l=2}^m \supbracket{A}{l} B_{m,l}(\supbracket{\Upsilon}{1}, \supbracket{\Upsilon}{2}, \ldots) =\nonumber\\
    &=-\frac{1}{2} \sum_{k=1}^{m-1} \binom{m}{k} \avgcp{r \supbracket{\Omega}{k}} \avgcp{r \supbracket{\Omega}{m-k}} +  \sum_{l=2}^m  (-1)^l (l-2)! \avgcp{B_{m,l} (\supbracket{\Omega}{1},  \supbracket{\Omega}{2}, \ldots)} 
    \hspace{10pt} \forall m=2, \ldots, 2n-1,
\end{align}
where we used Eq.~\ref{SMeq:useful_full_contraction_Hessian} and that $B_{m,2}(\supbracket{x}{1}, \supbracket{x}{2}, \ldots)= \frac{1}{2} \sum_{k=1}^{m-1} \binom{m}{k} \supbracket{x}{k} \supbracket{x}{m-k}$.
Defining $\tilde{F}(x):=1-x+x\log x$, whose derivatives are $\supbracket{\tilde{F}}{k}(G(0)) = \supbracket{\tilde{F}}{k}(1) = (-1)^k (k-2)! \ \forall k\geq 2$ and $\supbracket{\tilde{F}}{k}(G(0))=0$ otherwise, we use Faà di Bruno's formula again and find
\begin{align}
    0&= \left. \left[-\frac{1}{2} \sum_{k=1}^{m-1} \binom{m}{k} \avgcp{r \supbracket{G}{k} (\epsilon)} \avgcp{r \supbracket{G}{m-k} (\epsilon)} +   \sum_{l=2}^m  (-1)^l (l-2)! \avgcp{B_{m,l} (\supbracket{G}{1}(\epsilon),  \supbracket{G}{2}(\epsilon), \ldots)} \right] \right|_{\epsilon=0} = \\
    &= \left. \left[-\frac{1}{2} \sum_{k=1}^{m-1} \binom{m}{k} \left(\frac{\mathrm{d}^k}{\mathrm{d}\epsilon^k} \avgcp{r G} \right) \left( \frac{\mathrm{d}^{m-k}}{\mathrm{d}\epsilon^{m-k}} \avgcp{r G} \right) +   \frac{\mathrm{d}^m}{\mathrm{d}\epsilon^m} \avgcp{ \tilde{F}(G(\epsilon))} \right] \right|_{\epsilon=0} = \\
    &= \left. \left[-\frac{1}{2} \sum_{k=1}^{m-1} \binom{m}{k} \left(\frac{\mathrm{d}^k}{\mathrm{d}\epsilon^k} \avgcp{r G} \right) \left( \frac{\mathrm{d}^{m-k}}{\mathrm{d}\epsilon^{m-k}} \avgcp{r G} \right) +   \frac{\mathrm{d}^m}{\mathrm{d}\epsilon^m} \avgcp{G \log G} \right] \right|_{\epsilon=0} \hspace{10pt} \forall m=2, \ldots, 2n-1,
    \label{SMeq:condition_higher_order_single_1}
\end{align}
since $\frac{\mathrm{d}^m}{\mathrm{d}\epsilon^m} \avgcp{1-G} =- \avgcp{\supbracket{\Omega}{m}}=0$ for $m\geq 2$.
Importantly, both $\avgcp{r G}$ and $\avgcp{G \log G}$ can be written in terms of the cumulant-generating function 
\begin{align}
    K_r (\epsilon) = \log \avgcp{e^{\epsilon r}}
\end{align}
of $r$:
\begin{align}
    \avgcp{r G} &= \frac{\avgcp{r e^{\epsilon r}}}{\avgcp{e^{\epsilon r}}} = \frac{\mathrm{d}}{\mathrm{d} \epsilon} \log \avgcp{e^{\epsilon r}} = \frac{\mathrm{d}}{\mathrm{d} \epsilon} K_r (\epsilon) \\
    \avgcp{G \log G} &= \frac{\avgcp{ e^{\epsilon r} \left(\epsilon r - \log \avgcp{e^{\epsilon r}} \right)}}{\avgcp{e^{\epsilon r}}} = \epsilon \frac{\mathrm{d}}{\mathrm{d} \epsilon} \log \avgcp{e^{\epsilon r}} - \log \avgcp{e^{\epsilon r}}= \epsilon \frac{\mathrm{d}}{\mathrm{d} \epsilon} K_r (\epsilon) - K_r (\epsilon).
\end{align}
Using that 
\begin{align}
    \frac{\mathrm{d}^m}{\mathrm{d} \epsilon^m} \left(\epsilon \frac{\mathrm{d}}{\mathrm{d} \epsilon} K_r (\epsilon) - K_r (\epsilon) \right) = (m-1) \frac{\mathrm{d}^m K_r}{\mathrm{d}\epsilon^m} + \epsilon \frac{\mathrm{d}^{m+1}
    K_r}{\mathrm{d}\epsilon^{m+1}} \hspace{10pt} \forall m\geq 1,
\end{align}
which can be shown by induction, the conditions for an $n$-th order critical point, Eq.~\ref{SMeq:condition_higher_order_single_1}, are given in terms of the cumulants of $r$:
\begin{align}
    0&=\left. \left[-\frac{1}{2} \sum_{k=1}^{m-1} \binom{m}{k} \left(\frac{\mathrm{d}^{k+1}}{\mathrm{d}\epsilon^{k+1}} K_r (\epsilon) \right) \left( \frac{\mathrm{d}^{m-k+1}}{\mathrm{d}\epsilon^{m-k+1}} K_r (\epsilon) \right) +   (m-1) \frac{\mathrm{d}^m }{\mathrm{d}\epsilon^m} K_r (\epsilon) \right]  \right|_{\epsilon=0} \hspace{10pt} =\\
    &= -\frac{1}{2} \sum_{k=1}^{m-1} \binom{m}{k} \kappa_{k+1} \kappa_{m-k+1} +   (m-1) \kappa_m \hspace{10pt} \forall m=2, \ldots, 2n-1.
    \label{SMeq:condition_higher_order_single_2}
\end{align}
Here, $\kappa_i$ is the $i$-th cumulant of $r$.
Solving these equations recursively yields:
\begin{align}
    \kappa_2 &= 1 \\
    \kappa_i &= 0 \hspace{10pt} \forall i=3, \ldots, 2n-1,
\end{align}
which we briefly demonstrate by induction:
\begin{description}
    
\item[Base case $n=2$]
    Eq.~\ref{SMeq:condition_higher_order_single_2} gives:
    \begin{align}
      m=2: \hspace{10pt} 0 &=-\frac{1}{2} \sum_{k=1}^{1} \binom{2}{k} \kappa_{k+1} \kappa_{3-k} + \kappa_2 =  \kappa_2 (1-\kappa_2) \\
      m=3: \hspace{10pt} 0 &=-\frac{1}{2} \sum_{k=1}^{2} \binom{3}{k} \kappa_{k+1} \kappa_{4-k} + 2 \kappa_3 =  -3 \kappa_2 \kappa_3 + 2 \kappa_3.
    \end{align}
    Since $\kappa_2$ corresponds to the variance of $r=\sqrt{2C} s$ and since $\kappa_2 =0$ can be excluded as this case would correspond to all component types being equal, we conclude that $\kappa_2=1$ and $\kappa_3=0$.

\item[Induction step $n \rightarrow n+1$]
    Assuming that $\kappa_2=1$ and $\kappa_i = 0 \ \forall i=3, \ldots, 2n-1$,
    Eq.~\ref{SMeq:condition_higher_order_single_2} yields
    \begin{align}
        m=2n: \hspace{10pt} 0 &=-\frac{1}{2} \sum_{k=1}^{2n-1} \binom{2n}{k} \kappa_{k+1} \kappa_{2n-k+1} +   (2n-1) \kappa_{2n} = \\
        &=-\frac{1}{2} \sum_{k=2}^{2n-2} \binom{2n}{k} \kappa_{k+1} \kappa_{2n-k+1} - 2n \kappa_2 \kappa_{2n} +  (2n-1) \kappa_{2n} = \\
        &=-\frac{1}{2} \sum_{k=2}^{2n-2} \binom{2n}{k} \kappa_{k+1} \kappa_{2n-k+1} -  \kappa_{2n} \hspace{30pt} \text{and} \\
        m=2n+1: \hspace{10pt} 0 &=-\frac{1}{2} \sum_{k=1}^{2n} \binom{2n+1}{k} \kappa_{k+1} \kappa_{2n+2-k} +   2n \kappa_{2n+1} = \\
        &=-\frac{1}{2} \sum_{k=2}^{2n-1} \binom{2n+1}{k} \kappa_{k+1} \kappa_{2n+2-k} - (2n+1) \kappa_2 \kappa_{2n+1} +  2n \kappa_{2n+1} = \\
        &=-\frac{1}{2} \sum_{k=2}^{2n-1} \binom{2n+1}{k} \kappa_{k+1} \kappa_{2n+2-k} -  \kappa_{2n+1} \hspace{30pt}.
    \end{align}
    By induction, we know that $\kappa_i = 0 \ \forall i=3, \ldots, 2n-1$, and we conclude that $\kappa_{2n}=\kappa_{2n+1}=0$.

\end{description}

Overall, we thus find the following conditions for an $n$-th order critical point:
\begin{align}
    \kappa_2 &= 1 \\
    \kappa_m &= 0 \hspace{20pt} \forall m=3, \ldots, 2n-1,
\end{align}
depending solely on the cumulants $\kappa_i$ of the rescaled spin value $r=\sqrt{2C} s$.
Rewriting them in terms of the cumulants $\supbracket{\kappa}{s}_i$ of the spin $s$ gives
\begin{align}
    \supbracket{\kappa}{s}_2 &= \frac{1}{2C} = \frac{k_B T}{z J} \\
    \supbracket{\kappa}{s}_m &= 0 \hspace{20pt} \forall m=3, \ldots, 2n-1.
\end{align}
In the limit $n\rightarrow \infty$, the spins $s_i$ need to be distributed according to a Gaussian with variance $1/(2C)$, implicitly requiring an infinite number of component types $N \rightarrow \infty$.
    
\subsection{Choice of $\alpha$ in the recursion relation}

In Eq.~\ref{SMeq:recursion_alpha_dependence}, we have seen that in principle there are several solutions for the recursion relation, namely all recursions of the form
\begin{align}
	\supbracket{\Upsilon}{1} &= \supbracket{\alpha}{1} \supbracket{e}{1} \\
	\supbracket{\Upsilon}{n} &= - (\supbracket{A}{2}_{\text{pseudo}})^{-1} \supbracket{b}{n} (\supbracket{\Upsilon}{1}, \supbracket{\Upsilon}{2}, \ldots) + \supbracket{\alpha}{n} \supbracket{e}{1},
\end{align}
where $\supbracket{\alpha}{n}$ is a constant and $\supbracket{\alpha}{1}\neq 0$ (otherwise, all $\supbracket{\Upsilon}{n}=0$).
So far, we have chosen $\supbracket{\alpha}{n} = \delta_{n1}$, and one may wonder whether other choices lead to the same conditions for the spinodal and critical points.

In this section, we show that the choice of $\supbracket{\alpha}{n}$ (with $\supbracket{\alpha}{1}\neq 0$) does not change our results. 
Instead it just corresponds to an effective rescaling of $\epsilon$.
To see this, we perform the following steps:

\begin{description}
\item[1] We show that if $\supbracket{Z}{i} = \sum_{j=1}^i \supbracket{Y}{j} B_{i,j} (\supbracket{\alpha}{1}, \supbracket{\alpha}{2}, \ldots)$ for all $i=1,\ldots, n-1$, then
\begin{align}
B_{k,m} (\supbracket{Z}{1}, \supbracket{Z}{2}, \ldots)  = \sum_{j=m}^k B_{k,j} (\supbracket{\alpha}{1}, \supbracket{\alpha}{2}, \ldots) B_{j,m} ( \supbracket{Y}{1}, \supbracket{Y}{2}, \ldots) \hspace{10pt} \forall k=2, \ldots, n \ \forall m= 2, \ldots, k.
\end{align}

\item[2] 
Using the previous equation, we then demonstrate that if $Y$ solves the recursion relation 
\begin{align}
	\supbracket{Y}{1} &= \supbracket{e}{1} \\
	\supbracket{Y}{m} &= - (\supbracket{A}{2}_{\text{pseudo}})^{-1} \supbracket{b}{m} (\supbracket{Y}{1}, \ldots) \hspace{10pt} \forall m=2, \ldots, n-1
	\label{SMeq:recursionY}
\end{align}
and $Z$ solves 
\begin{align}
	\supbracket{Z}{1} &= \supbracket{\alpha}{1} \supbracket{e}{1} \\
	\supbracket{Z}{m} &= - (\supbracket{A}{2}_{\text{pseudo}})^{-1} \supbracket{b}{m} (\supbracket{Z}{1}, \ldots) + \supbracket{\alpha}{m} \supbracket{e}{1}  \hspace{10pt} \forall m=2, \ldots, n-1,
	\label{SMeq:recursionZ}
\end{align}
then 
\begin{align}
\supbracket{Z}{i} = \sum_{j=1}^i \supbracket{Y}{j} B_{i,j} (\supbracket{\alpha}{1}, \supbracket{\alpha}{2}, \ldots) \hspace{10pt} \forall i=1, \ldots, n-1.
\end{align}

\item[3] We show that the conditions for a critical point of order $n$ determined previously, $\sum_{l=2}^m \supbracket{A}{l} B_{m,l} =0 \forall m=2, \ldots, 2n-1$, are equivalent for $\supbracket{Y}{n}$ and $\supbracket{Z}{n}$:
If $\sum_{l=2}^m \supbracket{A}{l} B_{m,l} ( \supbracket{Y}{1}, \supbracket{Y}{2}, \ldots) =0 \forall m=2, \ldots, 2n-1$ then $\sum_{l=2}^m \supbracket{A}{l} B_{m,l} ( \supbracket{Z}{1}, \supbracket{Z}{2}, \ldots) =0 \forall m=2, \ldots, 2n-1$ and vice versa.

\end{description}

\vspace{10pt}

The partial exponential Bell polynomials can be defined via the following series expansion:
\begin{align}
\sum_{n=k}^{\infty} B_{n,k} (\supbracket{x}{1}, \supbracket{x}{2}, \ldots) \frac{t^n}{n!} = \frac{1}{k!} \left(\sum_{j=1}^{\infty} \supbracket{x}{j} \frac{t^j}{j!} \right)^k.
\end{align}
Thus, $B_{n,k} (\supbracket{x}{1}, \supbracket{x}{2}, \ldots)$ is given by
\begin{align}
	B_{n,k} (\supbracket{x}{1}, \supbracket{x}{2}, \ldots) =\frac{1}{k!} \frac{\partial^n}{\partial t^n} \left. \left[  \left( \sum_{m=1}^{\infty} \supbracket{x}{m} \frac{t^m}{m!} \right)^k\right] \right|_{t=0}.
\end{align}
Using this expression and abbreviating $B_{n,k} (\supbracket{x}{1}, \supbracket{x}{2}, \ldots) =: B_{n,k} (x)$, we rewrite
\begin{align}
	\sum_{j=m}^k B_{k,j} (\alpha) B_{j,m}(Y) &= \sum_{j=m}^k \frac{1}{j!m!} \frac{\partial^k}{\partial t^k} \left.\left[\left( \sum_{r=1}^{\infty} \supbracket{\alpha}{r} \frac{t^r}{r!} \right)^j \right]\right|_{t=0} \frac{\partial^j}{\partial s^j} \left.\left[\left( \sum_{n=1}^{\infty} \supbracket{Y}{n} \frac{s^n}{n!} \right)^m \right]\right|_{s=0} = \\
	&= \frac{1}{m!} \frac{\partial^k}{\partial t^k} \left.\left[ \sum_{j=m}^k \frac{1}{j!} \left( \sum_{r=1}^{\infty} \supbracket{\alpha}{r} \frac{t^r}{r!} \right)^j \frac{\partial^j}{\partial s^j}  \left( \sum_{n=1}^{\infty} \supbracket{Y}{n} \frac{s^n}{n!} \right)^m \right] \right|_{t,s=0}.
\end{align}
As a next step, we observe that the boundaries for $j$ can be extended to $0$ and $\infty$, respectively.
For $j\geq k+1$, the term $\left( \sum_{r=1}^{\infty} \supbracket{\alpha}{r} \frac{t^r}{r!} \right)^j$ is of order $\mathcal{O}(t^{k+1})$ since $r \geq 1$, and yields a zero contribution when the $k$-th derivative $\partial^k/\partial t^k$ is evaluated at $t=0$.
Similarly, for $j \leq m-1$, the term $\left( \sum_{n=1}^{\infty} \supbracket{Y}{n} \frac{s^n}{n!} \right)^m$ is of order $\mathcal{O}(s^{j+1})$ and is zero when the derivative $\frac{\partial^j}{\partial s^j} $ is evaluated at $s=0$.
We thus find
\begin{align}
	\sum_{j=m}^k B_{k,j} (\alpha) B_{j,m}(Y) &= \frac{1}{m!} \frac{\partial^k}{\partial t^k} \left.\left[ \sum_{j=0}^{\infty} \frac{1}{j!} \left( \sum_{r=1}^{\infty} \supbracket{\alpha}{r} \frac{t^r}{r!} \right)^j \left.\frac{\partial^j}{\partial s^j}  \left( \sum_{n=1}^{\infty} \supbracket{Y}{n} \frac{s^n}{n!} \right)^m \right|_{s=0} \right] \right|_{t=0} = \\
	&= \frac{1}{m!} \left. \frac{\partial^k}{\partial t^k} \left[ \left( \sum_{n=1}^{\infty} \supbracket{Y}{n} \frac{1}{n!} \left(\sum_{r=1}^{\infty} \supbracket{\alpha}{r} \frac{t^r}{r!}\right)^n \right)^m \right] \right|_{t=0}.
\end{align}
Finally, we express $\sum_{n=1}^{\infty} \supbracket{Y}{n} \frac{1}{n!} \left( \sum_{r=1}^{\infty} \supbracket{\alpha}{r} \frac{t^r}{r!}\right)^n$ in terms of Bell polynomials:
\begin{align}
	 \sum_{n=1}^{\infty} \supbracket{Y}{n} \frac{1}{n!} \left( \sum_{r=1}^{\infty} \supbracket{\alpha}{r} \frac{t^r}{r!}\right)^n &=\sum_{n=1}^{\infty} \supbracket{Y}{n}  \frac{1}{n!} \sum_{s=n}^{\infty} t^s \sum_{\substack{\{ p_1, \ldots, p_n\} \in \mathbb{N}^n \\ \sum_{k=1}^n p_k = s}} \prod_{j=1}^n \frac{\supbracket{\alpha}{p_j}}{p_j!} = \\
	 &= \sum_{n=1}^{\infty} \supbracket{Y}{n} \frac{1}{n!} \sum_{s=n}^{\infty} t^s \sum_{\substack{\{ q_1, \ldots, q_s\} \in \mathbb{N}^n_0 \\ \sum_{k=1}^s k q_k = s \\ \sum_{k=1}^s q_k = n}}  \frac{n!}{q_1! \ldots q_s!} \prod_{r=1}^s \left(\frac{\supbracket{\alpha}{r}}{r!}\right)^{q_r} = \\
	 &= \sum_{n=1}^{\infty} \supbracket{Y}{n} \frac{1}{n!} \sum_{s=n}^{\infty} t^s \frac{n!}{s!} B_{s,n} (\alpha) = \sum_{s=1}^{\infty} \frac{t^s}{s!} \sum_{n=1}^s \supbracket{Y}{n} B_{s,n}(\alpha).
\end{align}
Taken together,
\begin{align}
	\sum_{j=m}^k B_{k,j} (\alpha) B_{j,m}(Y)
	&= \frac{1}{m!} \left. \frac{\partial^k}{\partial t^k} \left[ \left(  \sum_{s=1}^{\infty} \frac{t^s}{s!} \sum_{n=1}^s \supbracket{Y}{n} B_{s,n}(\alpha) \right)^m \right] \right|_{t=0} = \frac{1}{m!} \left. \frac{\partial^k}{\partial t^k} \left[ \left(  \sum_{s=1}^{k-1} \frac{t^s}{s!} \sum_{n=1}^s \supbracket{Y}{n} B_{s,n}(\alpha) \right)^m \right] \right|_{t=0} \\
	&= \frac{1}{m!} \left. \frac{\partial^k}{\partial t^k} \left[ \left(  \sum_{s=1}^{k-1} \frac{t^s}{s!}  \supbracket{Z}{s}  \right)^m \right] \right|_{t=0} = B_{k,m}(Z) \hspace{10pt} \forall k=2,\ldots, n \ \forall m=2, \ldots, k,
\end{align}
where we used the definition of $\supbracket{Z}{i}$ for $i=1, \ldots, n-1$ and that, since $m\geq 2$, if $s\geq k$ in one factor, the terms are of order $\mathcal{O}(t^{k+1})$ and yield zero after taking the derivative and setting $t=0$,
thus concluding part 1. 

Suppose $Y$ and $Z$ satisfy the recursion relations, Eqs.~\ref{SMeq:recursionY},~\ref{SMeq:recursionZ}, respectively. 
We will show by induction that then $\supbracket{Z}{i} = \sum_{j=1}^i \supbracket{Y}{j} B_{i,j} (\alpha) \ \forall i=1, \ldots, n-1$.

\begin{description}
\item[Base case $n=2$] 
\begin{align}
\supbracket{Z}{1} = \supbracket{\alpha}{1} \supbracket{e}{1} = \supbracket{\alpha}{1} \supbracket{Y}{1} = \sum_{j=1}^1\supbracket{Y}{j} B_{1,j} (\alpha).
\end{align}
\item[Induction step: $n \rightarrow n+1$] Suppose $\supbracket{Z}{i} = \sum_{j=1}^i \supbracket{Y}{j} B_{i,j} (\alpha) \ \forall i=1, \ldots, n-1$, then since $Z$ satisfies recursion~\ref{SMeq:recursionZ},
\begin{align}
	\supbracket{Z}{n} -  \supbracket{\alpha}{n} \supbracket{Y}{1} &= - (\supbracket{A}{2}_{\text{pseudo}})^{-1} \sum_{l=3}^{n+1} \supbracket{A}{l} B_{n,l-1} (Z)  = - (\supbracket{A}{2}_{\text{pseudo}})^{-1} \sum_{l=3}^{n+1} \supbracket{A}{l} \sum_{j=l-1}^n B_{n,j} (\alpha) B_{j,l-1}(Y) =\\
	&= - (\supbracket{A}{2}_{\text{pseudo}})^{-1} \sum_{j=2}^n B_{n,j} (\alpha)  \sum_{l=3}^{j+1} \supbracket{A}{l} B_{j,l-1}(Y) = \sum_{j=2}^n B_{n,j} (\alpha) \left( - (\supbracket{A}{2}_{\text{pseudo}})^{-1}  \sum_{l=3}^{j+1} \supbracket{A}{l} B_{j,l-1}(Y)\right) = \\
	&=  \sum_{j=2}^n B_{n,j} (\alpha) \supbracket{Y}{j},
\end{align}
or, equivalently, since $B_{n,1} (\alpha) = \supbracket{\alpha}{n}$,
\begin{align}
	\supbracket{Z}{n} =   \supbracket{\alpha}{n} \supbracket{Y}{1} + \sum_{j=2}^n B_{n,j} (\alpha) \supbracket{Y}{j} = \sum_{j=1}^n B_{n,j} (\alpha) \supbracket{Y}{j},
\end{align}
concluding our proof by induction and thereby part 2.
\end{description}

As a last step, we show that $Y$ and $Z$ yield the same conditions for critical points and the spinodal.
Suppose that 
\begin{align}
	\sum_{l=2}^m \supbracket{A}{l} B_{m,l} (Z) = 0 \hspace{10pt} \forall m=2, \ldots, n.
\end{align}
Then 
\begin{align}
	\sum_{l=2}^m \supbracket{A}{l} B_{m,l} (Y) = 0 \hspace{10pt} \forall m=2, \ldots,n,
\end{align}
which is proven by a simple induction.
\begin{description}
	\item[Base case $n=2$] If $ \supbracket{A}{2} B_{2,2} (Z) = 0$, then $0= \supbracket{A}{2} (\supbracket{Z}{1})^2 = (\supbracket{\alpha}{1})^2 \supbracket{A}{2} (\supbracket{Y}{1})^2$ and since $\supbracket{\alpha}{1}\neq 0$, we conclude that $\supbracket{A}{2} (\supbracket{Y}{1})^2 =  \supbracket{A}{2} B_{2,2}(Y)=0$.
	\item[Induction step $n\rightarrow n+1$] If $\sum_{l=2}^m \supbracket{A}{l} B_{m,l} (Z) = 0 \ \forall m=2, \ldots, n+1$ and suppose that $\sum_{l=2}^m \supbracket{A}{l} B_{m,l} (Y) = 0 \ \forall m=2, \ldots,n$, then
	\begin{align}
		0 &= \sum_{l=2}^{n+1} \supbracket{A}{l} B_{n+1,l} (Z) = \sum_{l=2}^{n+1} \supbracket{A}{l} \sum_{j=l}^{n+1} B_{n+1,j}(\alpha) B_{j,l} (Y) =  \sum_{j=2}^{n+1} B_{n+1,j}(\alpha) \sum_{l=2}^{j} \supbracket{A}{l}  B_{j,l} (Y) = \label{SMeq:induction_rescaling}\\
		& =  B_{n+1,n+1}(\alpha) \sum_{l=2}^{n+1} \supbracket{A}{l}  B_{n+1,l} (Y)  =  (\supbracket{\alpha}{1})^{n+1} \sum_{l=2}^{n+1} \supbracket{A}{l}  B_{n+1,l} (Y).
	\end{align}
	Since $\supbracket{\alpha}{1}\neq 0$, we conclude that $\sum_{l=2}^{n+1} \supbracket{A}{l}  B_{n+1,l} (Y)=0$, thus concluding our proof.
\end{description}

\vspace{20pt}

Conversely, if 
\begin{align}
	\sum_{l=2}^m \supbracket{A}{l} B_{m,l} (Y) = 0 \hspace{10pt} \forall m=2, \ldots, n.
\end{align}
Then 
\begin{align}
	\sum_{l=2}^{m} \supbracket{A}{l} B_{m,l} (Z) = \sum_{l=2}^{m} \supbracket{A}{l}  \sum_{j=l}^m B_{m,j} (\alpha) B_{j,l} (Y)= \sum_{j=2}^{m} B_{m,j}(\alpha) \sum_{l=2}^{j} \supbracket{A}{l}  B_{j,l} (Y) =0.
\end{align}

Overall, we thus find that irrespective of the choice of $\supbracket{\alpha}{n}$ ($\supbracket{\alpha}{1} \neq 0$) in the recursion relation, our analysis yields the same conditions.
Without loss of generality, we can thus set $\supbracket{\alpha}{n} = \delta_{n,1}$.

Note that in Eq.~\ref{SMeq:induction_rescaling} we implicitly used that the conditions only depend on the vectors $\supbracket{Y}{i}$ or $\supbracket{Z}{i}$ determined up to this point: $i\leq m-1$ for an $m$-th order critical point.

\vspace{20pt}

In terms of the optimal path in component space, the choice of $\supbracket{\alpha}{n}$ just corresponds to a certain parameterization of the curve:
\begin{align}
	\sum_{m=1}^{\infty} \frac{\epsilon^m}{m!} \supbracket{Z}{m} &= \sum_{m=1}^{\infty} \frac{\epsilon^m}{m!} \sum_{j=1}^m \supbracket{Y}{j} B_{m,j} (\alpha) = \sum_{j=1}^{\infty} \supbracket{Y}{j} \sum_{m=j}^{\infty} \frac{\epsilon^m}{m!} B_{m,j} (\alpha) = \\
	&=  \sum_{j=1}^{\infty} \supbracket{Y}{j} \sum_{m=j}^{\infty} \frac{\epsilon^m}{m!}  \left. \frac{\partial^m}{\partial t^m} \left[ \frac{1}{j!} \left(\sum_{p=1}^{\infty} \supbracket{\alpha}{p} \frac{t^p}{p!} \right)^j\right] \right|_{t=0} = \\
	&=  \sum_{j=1}^{\infty} \supbracket{Y}{j} \sum_{m=0}^{\infty} \frac{\epsilon^m}{m!}  \left. \frac{\partial^m}{\partial t^m} \left[ \frac{1}{j!} \left(\sum_{p=1}^{\infty} \supbracket{\alpha}{p} \frac{t^p}{p!} \right)^j\right] \right|_{t=0} = \\
	&=  \sum_{j=1}^{\infty} \supbracket{Y}{j} \frac{1}{j!} \left(\sum_{p=1}^{\infty} \supbracket{\alpha}{p} \frac{\epsilon^p}{p!} \right)^j =: \sum_{j=1}^{\infty}  \frac{\tilde{\epsilon}^j}{j!} \supbracket{Y}{j}.
\end{align}
Thus, going from $Y$ to $Z$ leads to a reparameterization of $\epsilon$ as 
\begin{align}
 	\sum_{p=1}^{\infty} \supbracket{\alpha}{p} \frac{\epsilon^p}{p!}.
\end{align}

\section{Components of different sizes}
\label{SMsec:diff_sizes}

In Eq.~\ref{SMeq:free_energy_Nminus}, the densities $\rho_i$ can be interpreted as either the volume or the number fraction of component $i$ in the mixture.
These are equivalent if all components have the same size and occupy the same number of lattice sites.
In realistic mixtures of lipids, proteins and DNA, this assumption might not be satisfied.
So how does the size of components modify the phase and critical behavior of the mixture?
Addressing this question requires modifying Eq.~\ref{SMeq:free_energy_Nminus} to include the sizes/lengths $l_i$ of components $i$.
Here we do this in a purely mean-field way that neglects correlations between lattice sites arising due to the finite length of the components, as done in Flory's and Huggins' pioneering work for two-component systems~\cite{Huggins1941, Flory1941}:
\begin{align}
\tilde{f}^{(l)} =  \sum_{i=1}^N \frac{\rho_i}{l_i} \log \rho_i -\sum_{i,j=1}^N \rho_i \chi_{ij} \rho_j, \label{eq:free_energy_different_sizes}
\end{align}
where $\rho_i$ is now the volume fraction of component $i$.
The number fraction is given by $(\rho_i/l_i) /\sum_{j} (\rho_j/l_j)$.  

Expressing the free energy in terms of the volume fractions $\rho_i, i=1,\ldots,N-1$ (using that $\rho_N=1-\sum_{i=1}^{N-1} \rho_i$ due to the incompressibility) and differentiating the free energy twice yields the Hessian
\begin{align}
\tilde{H}_{ij} = \supbracket{\tilde{A}}{2}_{ij} = \partial_i \partial_j \supbracket{\tilde{f}}{l} = \frac{\delta_{ij}}{\rho_i l_i} + \frac{1}{l_N (1-\Rho)} - \sum_{\gamma=1}^R \supbracket{r}{\gamma}_i \supbracket{r}{\gamma}_j=: L_{ij} - \sum_{\gamma=1}^R \supbracket{r}{\gamma}_i \supbracket{r}{\gamma}_j \hspace{20pt} i,j=1,\ldots,N-1.
\end{align}
Using Woodbury's matrix identity~\cite{woodbury1950} on $L$, we find for its inverse:
\begin{align}
(L^{-1})_{ij} = \delta_{ij} \rho_i l_i - \frac{1}{\bar{l}} \rho_i l_i \rho_j l_j,
\end{align}
where $\bar{l} = \sum_{i=1}^N \rho_i l_i$ is the average length.
Using similar arguments as before, the Hessian is invertible if and only if $\mathbf{1} - U^T L^{-1} U$ is invertible.
This matrix can be rewritten as
\begin{align}
	\delta_{\alpha \beta} - U_{n\alpha} (L^{-1})_{nm} U_{m\beta} &= \delta_{\alpha\beta} - \supbracket{r}{\alpha}_n (\delta_{nm} l_n \rho_n - \frac{1}{\bar{l}} l_n \rho_n l_m \rho_m) \supbracket{r}{\beta}_m = \\
	&= \delta_{\alpha \beta} - \rho_m l_m \supbracket{r}{\alpha}_m \supbracket{r}{\beta}_m + \frac{1}{\bar{l}} (\rho_n l_n \supbracket{r}{\alpha}_n)(\rho_m l_m \supbracket{r}{\beta}_m).
\end{align}
Due to the factors of $l$ appearing here, the second part is not as straightforwardly written as a covariance matrix, at least not with respect to $\rho$.
However, we can define a new probability measure $\phi$ by
\begin{align}
\phi_m = \frac{\rho_m l_m}{\bar{l}},
\end{align}
which satisfies $\phi_m \geq 0$ and $\sum_{m=1}^N \phi_m  =(\sum_{m=1}^N \rho_m l_m)/\bar{l} =1$, as desired.
In terms of $\phi$, the matrix $\mathbf{1} - U^T L^{-1} U$ is 
\begin{align}
	\delta_{\alpha \beta} - U_{n\alpha} (L^{-1})_{nm} U_{m\beta} = \delta_{\alpha \beta} - \bar{l} \phi_m \supbracket{r}{\alpha}_m \supbracket{r}{\beta}_m + \bar{l} (\phi_n \supbracket{r}{\alpha}_n)(\phi_m \supbracket{r}{\beta}_m),
\end{align}
and so 
\begin{align}
 \mathbf{1} - U^T L^{-1} U = \mathbf{1} - \bar{l} \supbracket{\text{Cov}}{\phi},
\end{align}
where $\supbracket{\text{Cov}}{\phi}$ is the covariance matrix of the rescaled and shifted features $\supbracket{r}{\gamma}$ with respect to the new probability measure $\phi$:
\begin{align}
\supbracket{\text{Cov}}{\phi}_{\alpha \beta} = \avgphi{ \supbracket{r}{\alpha} \supbracket{r}{\beta}} - \avgphi{\supbracket{r}{\alpha}} \avgphi{\supbracket{r}{\beta}}.
\end{align}
Here, all averages are taken with respect to the weighted volume fractions of the different components: $\avgphi{X}:= \sum_{i=1}^N \phi_i X_i = \sum_{i=1}^N \rho_i X_i l_i/\bar{l}$.

Overall, we  find that the Hessian matrix is invertible if and only if $\mathbf{1} - \bar{l} \supbracket{\text{Cov}}{\phi}$ is invertible.
By analogous arguments as before we thus conclude that the condition for the spinodal is
\begin{align}
\supbracket{\tilde{\lambda}}{1} = \frac{1}{\bar{l}},
\end{align}
where $\supbracket{\tilde{\lambda}}{1}$ is the variance of the first principal component of $\supbracket{\text{Cov}}{\phi}$.

The direction of instability is determined by looking at the inverse of the Hessian matrix
$\tilde{H}^{-1} = L^{-1} + L^{-1} U (\mathbf{1} - \bar{l} \supbracket{\text{Cov}}{\phi})^{-1} U^T L^{-1}$.
We rewrite its components as
\begin{align}
(\tilde{H}^{-1})_{ij} &= \delta_{ij} \rho_i l_i - \frac{1}{\bar{l}} \rho_i l_i \rho_j l_j + (\delta_{im} \rho_i l_i - \frac{1}{\bar{l}} \rho_i l_i \rho_m l_m) \supbracket{r}{\alpha}_m (\mathbf{1} - \bar{l} \supbracket{\text{Cov}}{\phi})^{-1}_{\alpha \beta} \supbracket{r}{\beta}_k (\delta_{kj} \rho_j l_j - \frac{1}{\bar{l}} \rho_k l_k \rho_j l_j) = \\
&= \delta_{ij} \rho_i l_i - \frac{1}{\bar{l}} \rho_i l_i \rho_j l_j + \rho_i l_i \rho_j l_j \left( \supbracket{r}{\alpha}_i - \avgphi{\supbracket{r}{\alpha}} \right) (\mathbf{1} - \bar{l} \supbracket{\text{Cov}}{\phi})^{-1}_{\alpha \beta}\left( \supbracket{r}{\beta}_j - \avgphi{\supbracket{r}{\beta}} \right). 
\end{align}
The covariance matrix $\supbracket{\text{Cov}}{\phi}$ is a real and symmetric matrix and therefore can be decomposed into an orthonormal set of eigenvectors $\supbracket{\tilde{V}}{\gamma}$ corresponding to the descending eigenvalues $\supbracket{\tilde{\lambda}}{\gamma}$:
\begin{align}
\supbracket{\text{Cov}}{\phi}_{\alpha \beta} = \sum_{\gamma=1}^R  \supbracket{\tilde{\lambda}}{\gamma} \supbracket{\tilde{V}}{\gamma}_{\alpha} \supbracket{\tilde{V}}{\gamma}_{\beta}.
\end{align}
Using this decomposition the inverse of the Hessian is 
\begin{align}
(\tilde{H}^{-1})_{ij}  &= \delta_{ij} \rho_i l_i - \frac{1}{\bar{l}} \rho_i l_i \rho_j l_j + \rho_i l_i \rho_j l_j \sum_{\gamma=1}^R \frac{1}{1-\bar{l} \ \supbracket{\tilde{\lambda}}{\gamma} } \supbracket{\tilde{V}}{\gamma}_{\alpha} (\supbracket{r}{\alpha}_i - \avgphi{\supbracket{r}{\alpha}}) \supbracket{\tilde{V}}{\gamma}_{\beta}  (\supbracket{r}{\beta}_j - \avgphi{\supbracket{r}{\beta}}) =: \\
&=:  \delta_{ij} \rho_i l_i - \frac{1}{\bar{l}} \rho_i l_i \rho_j l_j + \rho_i l_i \rho_j l_j \sum_{\gamma=1}^R \frac{1}{1-\bar{l} \ \supbracket{\tilde{\lambda}}{\gamma} } \supbracket{\tilde{E}}{\gamma}_i \supbracket{\tilde{E}}{\gamma}_j,
\end{align}
where
\begin{align}
\supbracket{\tilde{E}}{\gamma}_i  = \sum_{\alpha=1}^R \supbracket{\tilde{V}}{\gamma}_{\alpha} (\supbracket{r}{\alpha}_i - \avgphi{\supbracket{r}{\alpha}}).
\end{align}
On the spinodal, where $\supbracket{\tilde{\lambda}}{\gamma} = 1/\bar{l}$, the inverse of the Hessian is dominated by the term $ \frac{1}{1-\bar{l} \ \supbracket{\tilde{\lambda}}{1} } (\rho_i l_i \supbracket{\tilde{E}}{1}_i) ( \rho_j l_j  \supbracket{\tilde{E}}{1}_j)$, suggesting that the direction of instability is given by
\begin{align}
\supbracket{\tilde{e}}{1}_i:= \phi_i \supbracket{\tilde{E}}{1}_i \sim \rho_i l_i \supbracket{\tilde{E}}{1}_i.
\end{align}
This can also be verified by explicitly calculating
\begin{align}
\tilde{H}_{ij} \supbracket{\tilde{e}}{1}_j = \sum_{\gamma=1}^R \supbracket{\tilde{V}}{1}_{\alpha} \supbracket{r}{\alpha}_i (\frac{1}{\bar{l}} - \supbracket{\tilde{\lambda}}{1}),
\end{align}
which is zero on the spinodal, where $ \supbracket{\tilde{\lambda}}{1} = 1/\bar{l}$.

Looking at the (tilted) free energy change along a path $\rho(\epsilon) = \rhocp + \delta \rho(\epsilon) = \rhocp + \sum_{n=1}^{\infty} \frac{\epsilon^n}{n!} \supbracket{\tilde{\Upsilon}}{n}$ centered on a point $\rhocp$ on the spinodal, we find up to second order in $\epsilon$
\begin{align}
\tilde{f}(\rho(\epsilon)) - \tilde{f}(\rhocp) =\left. \left(\epsilon^2 \frac{1}{2}  \supbracket{\tilde{A}}{2}_{ij}  \supbracket{\tilde{\Upsilon}}{1}_i \supbracket{\tilde{\Upsilon}}{1}_j + \epsilon^3 \left[ \frac{1}{2} \supbracket{\tilde{A}}{2}_{ij}  \supbracket{\tilde{\Upsilon}}{1}_i \supbracket{\tilde{\Upsilon}}{2}_j   + \frac{1}{3!}  \supbracket{\tilde{A}}{3}_{ijk}  \supbracket{\tilde{\Upsilon}}{1}_i \supbracket{\tilde{\Upsilon}}{1}_j \supbracket{\tilde{\Upsilon}}{1}_k \right] + \mathcal{O} (\epsilon^4) \right)\right|_{cp},
\end{align}
where 
\begin{align}
\supbracket{\tilde{A}}{3}_{ijk} = \partial_i \partial_j \partial_k \supbracket{\tilde{f}}{l} = -\frac{\delta_{ijk}}{l_i \rho_i^2}  + \frac{1}{l_N (1-\Rho)^2}
\end{align}
and all terms are evaluated at $\rhocp$.

Analogously to before, if we require that two minima and one maximum merge along the path, we identify $\supbracket{\tilde{\Upsilon}}{1} = \supbracket{\tilde{e}}{1}$.
To determine the condition for the critical point, we then set the third order term in $\epsilon$ to zero, giving
\begin{align}
0 &= \left. \supbracket{\tilde{A}}{3}_{ijk}  \supbracket{\tilde{\Upsilon}}{1}_i \supbracket{\tilde{\Upsilon}}{1}_j \supbracket{\tilde{\Upsilon}}{1}_k  \right|_{cp}= \left. \left(-\sum_{i=1}^{N-1} \frac{(\supbracket{\tilde{\Upsilon}}{1}_i)^3}{l_i \rho_i^2} + \frac{(\sum_{i=1}^{N-1} \supbracket{\tilde{\Upsilon}}{1}_i)^3}{l_N (1-\Rho)^2} \right) \right|_{cp} = - \left. \sum_{i=1}^N \frac{(\supbracket{\tilde{\Upsilon}}{1}_i)^3}{l_i \rho_i^2} \right|_{cp},
\end{align}
where we used that $\sum_{i=1}^N \supbracket{\tilde{\Upsilon}}{1}_i  = \sum_{i=1}^N \supbracket{\tilde{e}}{1}= \sum_{i=1}^N \phi_i \sum_{\alpha=1}^R \supbracket{\tilde{V}}{1}_{\alpha} (\supbracket{r}{\alpha}_i - \avgphi{\supbracket{r}{\alpha}}) = 0$.
Using $\supbracket{\tilde{\Upsilon}}{1}_i  = \phi_i \supbracket{\tilde{E}}{1}_i$ together with $l_i \rho_i^2 = \bar{l}^2 \phi_i^2/l_i$ we thus find
\begin{align}
0 = \bar{l}^2\left. \sum_{i=1}^N \frac{(\supbracket{\tilde{\Upsilon}}{1}_i)^3}{l_i \rho_i^2} \right|_{cp} = \bar{l}^2 \left. \sum_{i=1}^N \frac{\phi_i^3 (\supbracket{\tilde{E}}{1}_i)^3 l_i}{\bar{l}^2 \phi_i^2} \right|_{cp} = \left. \sum_{i=1}^N \phi_i (\supbracket{\tilde{E}}{1}_i)^3 l_i \right|_{cp} = \avgphi{l \left(\supbracket{\tilde{E}}{1}\right)^3}
\end{align}
as condition for the critical point.

\section{Negative interaction strengths}
\label{SMsec:neg_interaction_strengths}

So far, we have implicitly assumed that the rescaled features $\supbracket{r}{\gamma}$ are all real and that $\text{Cov}$ is a true covariance matrix, which is symmetric and positive semi-definite and has an orthonormal set of eigenvectors.
However, these assumptions are only true if the interaction strengths $J^{(\gamma)}$ between the features are all positive.
Our previous analysis is thus restricted to the case where all eigenvalues of the interaction matrix $\sim \supbracket{J}{\gamma} \left| \bm{s}^{(\gamma)}\right|^2$ are positive and where  the pairwise interactions satisfy $2\chi_{ij} - (\chi_{ii} + \chi_{jj}) = - \sum_{\gamma=1}^R C^{(\gamma)} [s^{(\gamma)}_i -s^{(\gamma)}_j]^2 < 0 \ \forall \ i,j$; interactions between alike components are always energetically preferred as compared to interactions between dislike components.

To resolve this limitation, we now discuss the general case with a combination of positive and negative interaction strengths.
Without loss of generality, we assume that the first $R^+$ ``positive" features are attractive and the remaining $R^-=R-R^+$ ``negative" ones are repulsive:
\begin{align}
J^{(\gamma)} > 0 \ \forall \gamma = 1, \ldots, R^+ \hspace{10pt} \& \hspace{10pt}
J^{(\gamma)} < 0 \ \forall \gamma = R^+ +1, \ldots, R. \label{eq:pos_neg_interaction_strengths}
\end{align}
The rescaled and shifted features can thus be written in terms of real numbers $\tilde{r}^{(\gamma)} \in \mathbb{R} \ \forall \ \gamma=1,{\ldots},R$ as
\begin{align}
r^{(\gamma)} &=: \tilde{r}^{(\gamma)} \ \forall \gamma = 1, \ldots, R^+  \nonumber \\
r^{(\gamma)} &=: i \tilde{r}^{(\gamma)} \ \forall \gamma = R^+ +1, \ldots, R, \label{eq:complex_conj_r}
\end{align}
and the ``covariance" matrix $\text{Cov}$ (which, in the general case, is in fact a pseudo-covariance matrix) as
\begin{align}
\label{eq:cov_matrix_neg_interaction_strengths}
\text{Cov} = \begin{pmatrix} \cpp & i \cpm \\ i \cmp & -\cmm \end{pmatrix},
\end{align}
with the real matrices
\begin{align}
\label{eq:cov_matrix_entries_neg_interaction_strengths}
\cpp_{\alpha \beta} &= \text{Cov}^{(\tilde{r})}_{\alpha, \beta} \hspace{5pt} &\alpha, \beta {=} 1, {\ldots}, R^+ \nonumber \\
\cpm_{\alpha \beta} &= \text{Cov}^{(\tilde{r})}_{\alpha, \beta+R^+} \hspace{5pt} &\alpha {=} 1, {\ldots}, R^+, \hspace{5pt} \beta {=} 1, {\ldots}, R^- \nonumber \\
\cmp_{\alpha \beta} &= \text{Cov}^{(\tilde{r})}_{\alpha+R^+, \beta}  = \cpm_{\beta \alpha}  \hspace{5pt} &\alpha {=} 1, {\ldots}, R^-, \hspace{5pt} \beta {=} 1, {\ldots}, R^+\nonumber \\
\cmm_{\alpha \beta} &= \text{Cov}^{(\tilde{r})}_{\alpha+R^+, \beta+R^+} \hspace{5pt}  &\alpha, \beta {=} 1, \ldots, R^-.
\end{align}
These matrices quantify the covariances between the subsets of positive ($+$) and negative ($-$) features, respectively: $\text{Cov}^{(\tilde{r})}_{\alpha, \beta} = \avgrho{ \tilde{r}^{(\alpha)}  \tilde{r}^{(\beta)}} {-} \avgrho{ \tilde{r}^{(\alpha)} } \avgrho{ \tilde{r}^{(\beta)}}$.
In particular, $\cpp$ and $\cmm$ are true covariance matrices restricted to $\tilde{r}^{(\gamma)}$ for $\gamma=1,{\ldots},R^+$ and $\gamma=R^+ +1,{\ldots},R$, respectively, and are positive semi-definite: $\cpp, \cmm \succeq 0$.
The matrix $\mathbf{1}-\text{Cov}$ occurring in the condition for the Hessian to be invertible is thus a $2 \times 2$ block matrix whose lower diagonal element is $\mathbf{1}+\cmm \succeq \mathbf{1}$ and hence invertible.
Therefore, $\mathbf{1}-\text{Cov}$ is invertible if its Schur complement with respect to $\mathbf{1}+\cmm$, denoted by $(\mathbf{1}-\text{Cov})/(\mathbf{1}+\cmm) =:\mathbf{1}-\bar{C}$, is invertible and has no zero eigenvalue.
Here $\bar{C}$ is given by
\begin{align}
\bar{C} =  \cpp - \cpm (\mathbf{1}+\cmm)^{-1} \cmp,
\label{eq:Schur_complement}
\end{align}
which is a real and symmetric matrix since $\cpp$ and $\cmm$ are symmetric and $(\cpm)^T = \cmp$.

\paragraph{Spinodal}
For small absolute values of the interaction strengths, for which the system should not phase separate, the matrices $\supbracket{C}{\pm\pm/\pm\mp}$ are small and, consequently, $\mathbf{1}-\bar{C}$ is close to $\mathbf{1}$ and has only positive eigenvalues.
The condition for the spinodal is thus that the smallest eigenvalue of $\mathbf{1}-\bar{C}$ is 0, or, equivalently, that the largest eigenvalue of $\bar{C}$ is 1:
\begin{align}
\supbracket{\lambda}{1}_{\bar{C}} \overset{!}{=} 1 \hspace{10pt} \text{\textit{(spinodal criterion)}},
\end{align}
where $\{\supbracket{\lambda}{\alpha}_{\bar{C}}\}_{\alpha}$ denote the eigenvalues of $\bar{C}$ in descending order.
Note that the matrix dimension of the Schur complement, $\mathbf{1}-\bar{C}$, Eq.~\ref{eq:Schur_complement}, is $R^+ \times R^+$.
This illustrates that phase separation relies on features with positive interaction strengths (positive eigenvalues of the interaction matrix).
Then again, features with negative interaction strengths modify the propensity to phase separate.
More specifically, since $(\cpm)^T = \cmp$ and  $\mathbf{1}+\cmm$ (or, equivalently, $(\mathbf{1}+\cmm)^{-1}$) are positive semi-definite, $ \cpm (\mathbf{1}+\cmm)^{-1} \cmp$ is positive semi-definite: $v^T \cpm (\mathbf{1}+\cmm)^{-1} \cmp v = (\cmp v)^T (\mathbf{1}+\cmm)^{-1} (\cmp v) \geq  0$ for any real vector $v$.
For two real, symmetric, positive semi-definite matrices $A$ and $B$, the largest eigenvalue of $A-B$ is smaller than the largest eigenvalue of $A$: Let $v$ be the normalized eigenvector corresponding to the largest eigenvalue $\supbracket{\lambda}{A-B}_{\text{max}}$ of $A-B$.
Then  $\supbracket{\lambda}{A-B}_{\text{max}} = v^T (A-B) v = v^T A v - v^T B v \leq \supbracket{\lambda}{A}_{\text{max}}$ since $v^T B v \geq 0$ and $v^T A v \leq \supbracket{\lambda}{A}_{\text{max}}$.
Thus, the negative features always tend to lower the eigenvalue of $\bar{C}$ and thereby prevent phase separation.
The extent to which they influence the phase behavior depends on the relative correlations between all features:
Suppose $\supbracket{v}{{+}{+}}$ is the normalized eigenvector of $\cpp$ corresponding to the largest eigenvalue.
Or, in other words, assume that the dominant combination of positive features that drives phase separation is $\supbracket{R}{{+}{+}} := \sum_{\alpha=1}^{R^+} \supbracket{v}{{+}{+}}_{\alpha} \supbracket{r}{\alpha}$.
Then we find $\sum_{\alpha=1}^{R^+} \cmp_{\beta \alpha} \supbracket{v}{{+}{+}}_{\alpha} = \sum_{\alpha=1}^{R^+} \left[\avgrho{\supbracket{\tilde{r}}{\beta+R^+} \supbracket{r}{\alpha}} - \avgrho{\supbracket{\tilde{r}}{\beta+R^+}}\avgrho{ \supbracket{r}{\alpha}} \right] \supbracket{v}{{+}{+}}_{\alpha} = \avgrho{\supbracket{\tilde{r}}{\beta+R^+} \supbracket{R}{{+}{+}}} - \avgrho{\supbracket{\tilde{r}}{\beta+R^+}}\avgrho{ \supbracket{R}{{+}{+}}}$.
As a result, if the dominant combinations of positive features driving phase separation correlate only weakly with the negative features, $\avgrho{\supbracket{\tilde{r}}{\beta+R^+} \supbracket{R}{{+}{+}}} - \avgrho{\supbracket{\tilde{r}}{\beta+R^+}}\avgrho{ \supbracket{R}{{+}{+}}} \approx 0 \ \forall \beta$, the thermodynamic stability is barely modified by the presence of the latter:  $\bar{C}\supbracket{v}{{+}{+}} = \supbracket{C}{{+}{+}}\supbracket{v}{{+}{+}} - \cpm (\mathbf{1}+\cmm)^{-1} \cmp \supbracket{v}{{+}{+}} \approx \cpp \supbracket{v}{{+}{+}}$.
Conversely, if for each dominant combination of positive features, there is a highly correlated negative feature, the two effects counteract each other and, depending on the relative strengths and correlations, the effective interactions between the components may be small -- the mixture will not phase separate.
In a similar spirit, depending on the correlation structure, the addition of repulsive features may shift the relevance of the different attractive features with respect to their role for phase separation. 
Taken together, $\bar{C}$ can be interpreted as representing a multicomponent system with $R^+$ positive features with effective, reduced interactions integrating the interplay of the positive and negative features in the original system.

\paragraph{Direction of instability}

If the Schur complement $\mathbf{1}-\bar{C}$ of $\mathbf{1}-\text{Cov}$ with respect to $\mathbf{1}+\cmm$ is invertible, the inverse of $\mathbf{1}-\text{Cov}$ exists and is given by
\begin{align}
	\left(\mathbf{1} {-}\text{Cov} \right)^{-1} = 
	\begin{pmatrix} 
	\left( \mathbf{1} {-} \bar{C} \right)^{-1} &  i \left( \mathbf{1} {-} \bar{C} \right)^{-1} \cpm \left( \mathbf{1} {+} \cmm  \right)^{-1} \\
	 i \left( \mathbf{1} {+} \cmm \right)^{-1}\cmp\left( \mathbf{1} {-} \bar{C} \right)^{-1}  & \left( \mathbf{1} {+} \cmm \right)^{-1}  \left[\mathbf{1} {-}   \cmp    \left( \mathbf{1} {-} \bar{C} \right)^{-1}  \cpm \left( \mathbf{1} {+} \cmm \right)^{-1} \right]
	\end{pmatrix}. 
\end{align}
For our purposes, this expression is conveniently rewritten as
\begin{align}
	\left(\mathbf{1} {-}\text{Cov} \right)^{-1} &= 
	\begin{pmatrix} 
		0 & 0 \\
		0 & \left( \mathbf{1} {+}\cmm \right)^{-1} 
	\end{pmatrix} + \\
	&+
	\begin{pmatrix}
		\mathbf{1}_{R^+ {\times} R^+} & 0 \\
		0 & i \left( \mathbf{1} {+}\cmm \right)^{-1} \cmp
	\end{pmatrix}
	\begin{pmatrix}
		\left( \mathbf{1} {-} \bar{C} \right)^{-1}& \left( \mathbf{1} {-} \bar{C} \right)^{-1} \\
		\left( \mathbf{1} {-} \bar{C} \right)^{-1} & \left( \mathbf{1} {-} \bar{C} \right)^{-1}
	\end{pmatrix}
	\begin{pmatrix}
		\mathbf{1}_{R^+ {\times} R^+} & 0 \\
		0 & i \cpm \left( \mathbf{1} {+}\cmm \right)^{-1} 
	\end{pmatrix}. \nonumber
\end{align}
The inverse of the Hessian matrix then exists and is given by $H^{-1} = K^{-1} + K^{-1} U \left(\mathbf{1} - \text{Cov} \right)^{-1} U^T K^{-1}$ or
\begin{align}
\left( H^{-1} \right)_{ij} &= \delta_{ij} \rho_i - \rho_i \rho_j +( \delta_{ik} \rho_i - \rho_i \rho_k) \supbracket{r}{\alpha}_k (\mathbf{1}-\text{Cov})^{-1}_{\alpha \beta} \supbracket{r}{\beta}_m (\delta_{mj} \rho_j - \rho_m \rho_j) = \\
&= \delta_{ij} \rho_i - \rho_i \rho_j + \rho_i \rho_j \left( \supbracket{r}{\alpha}_i - \avgrho{\supbracket{r}{\alpha}}\right) \left(\mathbf{1}-\text{Cov} \right)^{-1}_{\alpha \beta} \left( \supbracket{r}{\beta}_j - \avgrho{\supbracket{r}{\beta}}\right) = \nonumber\\
&=  \delta_{ij} \rho_i - \rho_i \rho_j + \rho_i \rho_j  \left[ -\vec{\nu}_i^{\ T} \left( \mathbf{1} {+}\cmm \right)^{-1}  \vec{\nu}_j + \vec{z}_i^{\ T} \left( \mathbf{1} {-} \bar{C} \right)^{-1} \vec{z}_j \right]
\nonumber,
\end{align}
where for all $i,j=1,\ldots,N-1$ we have defined the deviations of the rescaled features from the mean and their weighted sum 
\begin{align}
\vec{\pi}_i := 
\begin{pmatrix}
\supbracket{\tilde{r}}{1}_i - \avgrho{\supbracket{\tilde{r}}{1}} \\
\vdots \\
\supbracket{\tilde{r}}{R^+}_i - \avgrho{\supbracket{\tilde{r}}{R^+}}
\end{pmatrix} &= 
\begin{pmatrix}
\supbracket{\pi}{1}_i \\
\vdots \\
\supbracket{\pi}{R^+}_i
\end{pmatrix}, 
\hspace{15pt}
\vec{\nu}_i := 
\begin{pmatrix}
\supbracket{\tilde{r}}{R^+ +1}_i - \avgrho{\supbracket{\tilde{r}}{R^+ +1}} \\
\vdots \\
\supbracket{\tilde{r}}{R}_i - \avgrho{\supbracket{\tilde{r}}{R}}
\end{pmatrix} =
\begin{pmatrix}
\supbracket{\nu}{1}_i \\
\vdots \\
\supbracket{\nu}{R^-}_i
\end{pmatrix}, \\
\vec{z}_i &:= \vec{\pi}_i - \cpm \left( \mathbf{1} {+}\cmm \right)^{-1} \vec{\nu}_i.
\end{align}

Since $\cpp$ and $\cmm$ are symmetric, $\cmp  = (\cpm)^T$ and all matrices are real, $\bar{C}$ is real and symmetric and thus has an orthonomal set of $R^+$ eigenvectors $\supbracket{\vec{\phi}}{\gamma}$ corresponding to the descending eigenvalues $\supbracket{\bar{\lambda}}{\gamma}$:
\begin{align}
\bar{C}_{\alpha \beta} = \sum_{\gamma=1}^{R^+} \supbracket{\bar{\lambda}}{\gamma}  \supbracket{\phi}{\gamma}_{\alpha} \supbracket{\phi}{\gamma}_{\beta}.
\end{align}
If $\supbracket{\bar{\lambda}}{\gamma} \neq 1 \forall \gamma=1,\ldots,R^+$, then $\left(\mathbf{1}-\bar{C}\right)^{-1}_{\alpha\beta} =  \sum_{\gamma=1}^{R^+} (1-\supbracket{\bar{\lambda}}{\gamma})^{-1} \supbracket{\phi}{\gamma}_{\alpha} \supbracket{\phi}{\gamma}_{\beta}$ and we find
\begin{align}
\left( H^{-1} \right)_{ij} = \delta_{ij} \rho_i - \rho_i \rho_j - \rho_i \vec{\nu}_i^{\ T} \left( \mathbf{1} {+}\cmm \right)^{-1}  \rho_j \vec{\nu}_j + \sum_{\gamma=1}^{R^+} \frac{1}{1-\supbracket{\bar{\lambda}}{\gamma}} \supbracket{\bar{e}}{\gamma}_i \supbracket{\bar{e}}{\gamma}_j,
\end{align}
where
\begin{align}
	\supbracket{\bar{e}}{\gamma}_i  = \rho_i \vec{z}^{\ T}_i \supbracket{\vec{\phi}}{\gamma} = \rho_i \sum_{\alpha=1}^{R^+} \supbracket{\phi}{\gamma}_{\alpha} \left[ \supbracket{\pi}{\alpha}_i - \sum_{\beta,\delta=1}^{R^-} \cpm_{\alpha \beta} \left( ( \mathbf{1}+\cmm)^{-1} \right)_{\beta \delta} \supbracket{\nu}{\delta}_i \right].
	\label{eq:rank_corrections_inverse_Hessian_neg}
\end{align}
Close to the spinodal, the inverse of the Hessian is dominated by the term $\frac{1}{1-\supbracket{\bar{\lambda}}{1}} \supbracket{\bar{e}}{1}_i \supbracket{\bar{e}}{1}_j$.
Similarly to before, we thus identify $\supbracket{\bar{e}}{1}$ with the eigenvector of the Hessian to eigenvalue 0 at the spinodal:
\begin{align}
    \left. H \bar{e}^{(1)} \right|_{\text{spinodal}} = 0 \hspace{10pt}   \text{\textit{(direction of instability)}}.
    \label{eq:direction_instability_neg}
\end{align}

\paragraph{Critical point}

Expanding the free energy along a path $\rho(\epsilon)$ up to third order in $\epsilon$, we find an analogous condition for the critical point as for the original model:
\begin{align}
    \avgcp{ \left(\bar{E}^{(1)}\right)^3}  \overset{!}{=}  0 \hspace{10pt} \text{\textit{(condition critical point)}} \label{SMeq:condition_critical},
\end{align}
where $\supbracket{\bar{e}}{1}_i =: \rho_i  \supbracket{\bar{E}}{1}_i$.

\section{Linearly dependent features}
\label{SMsec:linearly_dependent_features}

In this section, we show that adding linearly dependent features does not change the results of our analysis.
We restrict our discussion to the case of only positive interaction strengths and find that there is an additional zero eigenvalue $\supbracket{\lambda}{\Gamma}=0$ (all others remain the same) but the corresponding additional eigenvector $\supbracket{V}{\Gamma}$ leads to a zero vector $\supbracket{E}{\Gamma}=0$, thus not changing our predictions.

To simplify notation, we assume without loss of generality that $2 \supbracket{C}{\gamma}= 1$ (absorbing the interaction strengths into the spin values) and that $\supbracket{s}{\gamma}_N =0$ (shifting the spin values; see also Section~\ref{SMsec:global_rotation_translation}).
Thus, $\supbracket{r}{\gamma}=\supbracket{s}{\gamma}$.

Suppose now that the $R$-th feature can be written as a linear combination of the other features:
\begin{align}
    \supbracket{s}{R} = \sum_{\alpha=1}^{R-1} \nu_{\alpha} \supbracket{s}{\alpha}
    \label{SMeq:linear_dep_features}
\end{align}
with $\nu \in \mathbb{R}^{R-1}$ and  $\supbracket{s}{\gamma} = (\supbracket{s}{\gamma}_1, \ldots, \supbracket{s}{\gamma}_N)$.
Then we can rewrite the interaction matrix $\chi$ as
\begin{align}
    \chi_{ij} = \sum_{\gamma=1}^R \supbracket{s}{\gamma}_i \supbracket{s}{\gamma}_j = \sum_{\gamma=1}^{R-1} \supbracket{s}{\gamma}_i \supbracket{s}{\gamma}_j + \sum_{\alpha,\beta=1}^{R-1} \nu_{\alpha} \nu_{\beta} \supbracket{s}{\alpha}_i \supbracket{s}{\beta}_j = \sum_{\alpha,\beta=1}^{R-1} \left(\delta_{\alpha \beta} + \nu_{\alpha} \nu_{\beta} \right) \supbracket{s}{\alpha}_i \supbracket{s}{\beta}_j = (\tilde{S}^T M \tilde{S})_{ij},
\end{align}
where we defined the matrices $\tilde{S}_{\alpha i} = \supbracket{s}{\alpha}_i$ for $i=1,\ldots, N$, $\alpha=1,\ldots,R-1$ and $M_{\alpha \beta} = \delta_{\alpha \beta} + \nu_{\alpha} \nu_{\beta}$ for $\alpha, \beta = 1, \ldots, R-1$.
The latter matrix is real and symmetric and can thus be written in terms of a diagonal matrix $D$ and an orthogonal matrix $Q$:
\begin{align}
    M=QDQ^T \hspace{20pt} \text{with} \hspace{20pt} QQ^T=Q^TQ=\mathbf{1}.
\end{align}
Further, since $M$ is a rank-1 correction of the identity matrix, the eigenvalues of $M$ are $1+||\nu||^2$ and 1 ($R{-}2$ times degenerate):
\begin{align}
    M_{\alpha \beta} \nu_{\beta} = \nu_{\alpha} (1+||\nu||^2) \hspace{20pt} \text{and} \hspace{20pt} M_{\alpha \beta} w_{\beta} = w_{\alpha} + \nu_{\alpha} (\bm{\nu} \cdot \bm{w}) = w_{\alpha} \ \ \forall \bm{w} \perp \bm{\nu}. 
\end{align}
The corresponding normalized eigenvectors are thus $\bm{\nu}/||\nu||=: \supbracket{w}{1}$ and a set of $R{-}2$ normal vectors $\supbracket{w}{\gamma}, \gamma=2,...,R-1$ with $\supbracket{w}{\gamma} \perp \supbracket{w}{\delta}$ for all $\gamma \neq \delta$.
In terms of these, the matrices are
\begin{align}
    D=\begin{pmatrix}
        1+||\nu||^2 & & & \\
        & 1 & & \\
        & & \ddots & \\
        & & & 1
    \end{pmatrix}
    \hspace{40pt}
    Q=(\supbracket{w}{1}, \ldots, \supbracket{w}{R-1}).
\end{align}
Moreover, we can express the interaction matrix as
\begin{align}
    \chi_{ij} = (\tilde{S}^T QDQ^T \tilde{S})_{ij} = (P^T P),
\end{align}
where 
\begin{align}
    P = \sqrt{D} Q^T \tilde{S}
\end{align}
and the square of the diagonal matrix is 
\begin{align}
    \sqrt{D}=\begin{pmatrix}
        \sqrt{1+||\nu||^2} & & & \\
        & 1 & & \\
        & & \ddots & \\
        & & & 1
    \end{pmatrix}.
\end{align}
We thus have two representations of $\chi$, one in terms of $R$ linearly dependent features and the other one where we have integrated out the last feature:
\begin{align}
    \sum_{\gamma=1}^R \supbracket{s}{\gamma}_i \supbracket{s}{\gamma}_j = (S^T S)_{ij} = \chi_{ij} = (P^T P) = \sum_{\delta=1}^{R-1} \supbracket{p}{\delta}_i \supbracket{p}{\delta}_j,
\end{align}
where $S_{\gamma i} = \supbracket{s}{\gamma}_i$ has dimensions $R \times N$ and $P_{\delta i} =: \supbracket{p}{\delta}_i$ has $(R{-}1)\times N$.
Do these two representations yield the same conditions for the spinodal and the critical points?

To answer these questions, we consider the respective covariance matrices $\supbracket{\text{Cov}}{s}$ and $\supbracket{\text{Cov}}{p}$ and their eigenvalues and eigenvectors next.

\begin{align}
    \supbracket{\text{Cov}}{p}_{\alpha \beta} &= \avg{\supbracket{p}{\alpha}\supbracket{p}{\beta}} - \avg{\supbracket{p}{\alpha}}\avg{\supbracket{p}{\beta}} = \sum_{i=1}^N \rho_i \supbracket{p}{\alpha}_i \supbracket{p}{\beta}_i - \sum_{i,j=1}^N \rho_i \supbracket{p}{\alpha}_i \rho_j \supbracket{p}{\beta}_j = \sum_{i=1}^N \rho_i P_{\alpha i} P_{\beta i} - \sum_{i,j=1}^N \rho_i P_{\alpha i} \rho_j P_{\beta j} =\nonumber \\
    &= \sum_{\gamma, \delta, \phi, \epsilon=1}^{R-1} \sum_{i=1}^N \rho_i \sqrt{D}_{\alpha \gamma} Q^T_{\gamma\delta} S_{\delta i} \sqrt{D}_{\beta \epsilon} Q^T_{\epsilon \phi} S_{\phi i} - \sum_{\gamma, \delta, \phi, \epsilon=1}^{R-1} \sum_{i,j=1}^N \rho_i \rho_j \sqrt{D}_{\alpha \gamma} Q^T_{\gamma\delta} S_{\delta i} \sqrt{D}_{\beta \epsilon} Q^T_{\epsilon \phi} S_{\phi j} = \\
    &=\sum_{\gamma, \delta, \phi, \epsilon=1}^{R-1} \sqrt{D}_{\alpha \gamma} Q^T_{\gamma \delta} \supbracket{\text{Cov}}{s}_{\delta \phi} Q_{\phi \epsilon} \sqrt{D}_{\epsilon \beta},
\end{align}
where we used that 
\begin{align}
    \sum_{i=1}^N \rho_i 
    S_{\delta i} S_{\phi i} - \sum_{i,j=1}^N \rho_i S_{\delta i} \rho_j S_{\phi j} = 
    \sum_{i=1}^N \rho_i \supbracket{s}{\delta}_i \supbracket{s}{\phi}_i - \sum_{i,j=1}^N \rho_i \supbracket{s}{\delta}_i \rho_j \supbracket{s}{\phi}_j = \supbracket{\text{Cov}}{s}_{\delta \phi}.
\end{align}
Thus,
\begin{align}
    \supbracket{\text{Cov}}{p} = \sqrt{D} Q^T \supbracket{\text{Cov}}{s|R-1} Q \sqrt{D},
    \label{SMeq:relation_Covs}
\end{align}
where $\supbracket{\text{Cov}}{s|R-1}$ is the covariance matrix $\supbracket{\text{Cov}}{s}$ restricted to the first $R-1$ features/dimensions.
More specifically, since
\begin{align}
    \supbracket{\text{Cov}}{s}_{\alpha R} &= \avg{\supbracket{s}{\alpha} \supbracket{s}{R}} - \avg{\supbracket{s}{\alpha}}\avg{ \supbracket{s}{R}} = \sum_{\beta=1}^{R-1} \nu_{\beta} \left( \avg{\supbracket{s}{\alpha} \supbracket{s}{\beta}} - \avg{\supbracket{s}{\alpha}}\avg{ \supbracket{s}{\beta}}\right) =  \sum_{\beta=1}^{R-1} \nu_{\beta} \supbracket{\text{Cov}}{s}_{\alpha \beta} \\
    \supbracket{\text{Cov}}{s}_{R R} &= \sum_{\alpha,\beta=1}^{R-1} \nu_{\alpha} \nu_{\beta} \supbracket{\text{Cov}}{s}_{\alpha \beta},
\end{align}
the covariance matrix $\supbracket{\text{Cov}}{s}$ is expressed in terms of the restricted covariance matrix $\supbracket{\text{Cov}}{s|R-1}$ as
\begin{align}
    \supbracket{\text{Cov}}{s} = \begin{pmatrix}
    \supbracket{\text{Cov}}{s|R-1} & \supbracket{\text{Cov}}{s|R-1} \bm{\nu} \\
    \bm{\nu}^T \supbracket{\text{Cov}}{s|R-1} & \bm{\nu}^T \supbracket{\text{Cov}}{s|R-1} \bm{\nu}
    \end{pmatrix} = 
    \begin{pmatrix}
    \mathbf{1}_{R-1} \\
    \bm{\nu}^T
    \end{pmatrix}
    \supbracket{\text{Cov}}{s|R-1}
    \begin{pmatrix}
    \mathbf{1}_{R-1} & \bm{\nu}
    \end{pmatrix},
\end{align}
where $\mathbf{1}_{R-1}$ is the identity matrix of dimensions $(R-1)\times (R-1)$ and $\bm{\nu} = (\nu_1, \ldots, \nu_{R-1})^T$.
This representation demonstrates that
\begin{align}
    \supbracket{V}{R}:= \begin{pmatrix}
        \bm{\nu} \\ -1
    \end{pmatrix}
\end{align}
is an eigenvector of $\supbracket{\text{Cov}}{s}$ to eigenvalues $\supbracket{\lambda}{R}=0$.
The corresponding vector $\supbracket{E}{R}$ is
\begin{align}
    \supbracket{E}{R} = \sum_{\alpha=1}^R \supbracket{V}{R}_{\alpha} \left(\supbracket{s}{\alpha}-\avg{\supbracket{s}{\alpha}} \right) = \left( \sum_{\alpha=1}^{R-1} \nu_{\alpha} \left(\supbracket{s}{\alpha}-\avg{\supbracket{s}{\alpha}}\right)\right) - \left(\supbracket{s}{R}-\avg{\supbracket{s}{R}} \right)=0,
\end{align}
and hence does not add any contribution to e.g.\ the pseudoinverse of the Hessian.

Since $D$ is diagonal and $Q$ orthogonal, the relation Eq.~\ref{SMeq:relation_Covs} between $\supbracket{\text{Cov}}{p}$ and $\supbracket{\text{Cov}}{s|R-1}$ is easily inverted as
\begin{align}
    \supbracket{\text{Cov}}{s|R-1} = Q \sqrt{D}^{-1} \supbracket{\text{Cov}}{p} \sqrt{D}^{-1} Q^T.
\end{align}
As a result, 
\begin{align}
    \supbracket{\text{Cov}}{s} = \begin{pmatrix}
    \mathbf{1}_{R-1} \\
    \bm{\nu}^T
    \end{pmatrix}
    Q \sqrt{D}^{-1} \supbracket{\text{Cov}}{p} \sqrt{D}^{-1} Q^T
    \begin{pmatrix}
    \mathbf{1}_{R-1} & \bm{\nu}
    \end{pmatrix}=: \tilde{Q} \supbracket{\text{Cov}}{p} \tilde{Q}^T.
    \label{SMeq:rel_Covs_Q}
\end{align}
The explicit expression of $\tilde{Q}^T$ is calculated as
\begin{align}
    \tilde{Q}^T = \sqrt{D}^{-1} Q^T
    \begin{pmatrix}
    \mathbf{1}_{R-1} & \bm{\nu}
    \end{pmatrix} = 
    \begin{pmatrix}
    \frac{1}{\sqrt{1+||\nu||^2}} (\supbracket{w}{1})^T & \frac{||\nu||}{\sqrt{1+||\nu||^2}} \\
    (\supbracket{w}{2})^T & 0 \\
    \vdots & \vdots \\
    (\supbracket{w}{R-1})^T & 0
    \end{pmatrix},
\end{align}
where we used that $(\supbracket{w}{\alpha})^T \bm{\nu} = (\supbracket{w}{\alpha})^T \supbracket{w}{1} ||\nu|| = \delta_{\alpha 1} ||\nu||$.
Importantly,
\begin{align}
    \tilde{Q}^T \tilde{Q} &= 
    \begin{pmatrix}
    \frac{1}{\sqrt{1+||\nu||^2}} (\supbracket{w}{1})^T & \frac{||\nu||}{\sqrt{1+||\nu||^2}} \\
    (\supbracket{w}{2})^T & 0 \\
    \vdots & \vdots \\
    (\supbracket{w}{R-1})^T & 0
    \end{pmatrix}
    \begin{pmatrix}
    \frac{1}{\sqrt{1+||\nu||^2}} \supbracket{w}{1} & \supbracket{w}{2} & \ldots & \supbracket{w}{R-1} \\ \frac{||\nu||}{\sqrt{1+||\nu||^2}} & 0 & \ldots &0
    \end{pmatrix}= \\
    &=\begin{pmatrix}
    \frac{1}{1+||\nu||^2} + \frac{||\nu||^2}{1+||\nu||^2} &  & & & \\
    & 1 & & & \\
    & & \ddots & & \\
    & & &  1 & \\
    & & & &  1
    \end{pmatrix} = 
    \begin{pmatrix}
    1 &  & & & \\
    & 1 & & & \\
    & & \ddots & & \\
    & & &  1 & \\
    & & & &  1
    \end{pmatrix} 
    =\mathbf{1}_{R-1},
\end{align}
allowing Eq.~\ref{SMeq:rel_Covs_Q} to be rewritten as
\begin{align}
    \tilde{Q}^T \supbracket{\text{Cov}}{s} = \supbracket{\text{Cov}}{p} \tilde{Q}^T.
\end{align}
Thus, if $\supbracket{\text{Cov}}{s}$ has eigenvectors $\supbracket{V}{\alpha}, \alpha=1, \ldots, R$ corresponding to eigenvalues $\supbracket{\lambda}{\alpha}$,
\begin{align}
    \supbracket{\text{Cov}}{s} \supbracket{V}{\alpha} = \supbracket{\lambda}{\alpha} \supbracket{V}{\alpha},
\end{align}
then
\begin{align}
    \supbracket{\text{Cov}}{p} (\tilde{Q}^T \supbracket{V}{\alpha}) = \tilde{Q}^T \supbracket{\text{Cov}}{s} \supbracket{V}{\alpha} =
    \supbracket{\lambda}{\alpha} (\tilde{Q}^T  \supbracket{V}{\alpha}) =: \supbracket{\lambda}{\alpha}  \supbracket{\tilde{V}}{\alpha}.
\end{align}
Note, however, that 
\begin{align}
    \supbracket{\tilde{V}}{R} = \tilde{Q}^T  \supbracket{V}{R} = 0,
\end{align}
and so we conclude that  $\supbracket{\text{Cov}}{p}$ has the same eigenvalues $\supbracket{\lambda}{\alpha}, \alpha=1, \ldots, R-1$ as $\supbracket{\text{Cov}}{s}$, except for $\supbracket{\lambda}{R}=0$.
The corresponding eigenvectors are $\supbracket{\tilde{V}}{\alpha} = \tilde{Q}^T  \supbracket{V}{\alpha}, \alpha=1, \ldots, R-1$.

To conclude, we will check that these different sets of eigenvectors give the same vectors $\supbracket{E}{\gamma}$ in the space of components.
To this end, we observe that for $\gamma=1,\ldots, R-1$
\begin{align}
    \supbracket{\tilde{E}}{\gamma} := \sum_{\alpha=1}^{R-1} \supbracket{\tilde{V}}{\gamma}_{\alpha} \left( \supbracket{p}{\alpha} - \avg{\supbracket{p}{\alpha}} \right) = \sum_{\alpha=1}^{R-1} \sum_{\beta=1}^R \tilde{Q}^T_{\alpha \beta} \supbracket{V}{\beta} (\supbracket{p}{\alpha}-\avg{\supbracket{p}{\alpha}}) = \sum_{\beta=1}^R \supbracket{V}{\beta} \sum_{\alpha=1}^{R-1} \tilde{Q}_{\beta \alpha}  (\supbracket{p}{\alpha}-\avg{\supbracket{p}{\alpha}}).
    \label{SMeq:tildeE}
\end{align}
Using the definitions of $\tilde{Q}$ and $P$ and that $Q$ is an orthogonal matrix, we find
\begin{align}
 \sum_{\alpha=1}^{R-1} \tilde{Q}_{\beta \alpha}  \supbracket{p}{\alpha}_i = (\tilde{Q} P)_{\beta i} = \left(\begin{pmatrix}
 \mathbf{1}_{R-1} \\ 
 \bm{\nu}^T
 \end{pmatrix}
 Q \sqrt{D}^{-1} \sqrt{D} Q^T \tilde{S}\right)_{\beta i} = 
 \left(\begin{pmatrix}
 \mathbf{1}_{R-1} \\ 
 \bm{\nu}^T
 \end{pmatrix} \tilde{S} \right)_{\beta i} = 
 \begin{pmatrix}
 \tilde{S} \\ 
 \bm{\nu}^T \tilde{S}
 \end{pmatrix}_{\beta i}.
\end{align}
For $\beta=1, \ldots, R-1$, this expression simplifies to
$
    \sum_{\alpha=1}^{R-1} \tilde{Q}_{\beta \alpha}  \supbracket{p}{\alpha}_i = \tilde{S}_{\beta i} = \supbracket{s}{\beta}_i
$
and, similarly, for $\beta=R$, we have
$
    \sum_{\alpha=1}^{R-1} \tilde{Q}_{R \alpha}  \supbracket{p}{\alpha}_i = (\bm{\nu}^T \tilde{S})_{i} = \sum_{\delta=1}^{R-1} \nu_{\delta} \supbracket{s}{\delta}_i = \supbracket{s}{R}_i
$.
Overall, we conclude that 
\begin{align}
    \sum_{\alpha=1}^{R-1} \tilde{Q}_{\beta \alpha} \supbracket{p}{\alpha}_i = \supbracket{s}{\beta}_i  \hspace{20pt} \forall \beta=1, \ldots, R.
\end{align}
Combining this with Eq.~\ref{SMeq:tildeE}, we find
\begin{align}
    \supbracket{\tilde{E}}{\gamma} = \sum_{\beta=1}^R \supbracket{V}{\beta} \left( \supbracket{s}{\beta} - \avg{\supbracket{s}{\beta}} \right) = \supbracket{E}{\gamma},
\end{align}
showing that indeed both representations lead to the same vectors in the space of components.

Taken together, we find that the addition of linearly dependent features does not change the eigenvalues of the covariance matrix, nor the resulting vectors in component space.
Our results for the spinodal and the critical points thus do not depend on whether or not the features are chosen as linearly independent.

\bigskip

Finally, we observe that if the covariance matrix has a zero eigenvalue $\supbracket{\lambda}{R}=0$, then the corresponding vector $\supbracket{E}{R}=0$.
Furthermore, either one features is deterministic, $\supbracket{r}{\gamma}_i = \supbracket{r}{\gamma}_1 \forall i$, and therefore does not lead to any effective interaction between the components, or the features are linearly dependent.
To see this, we note that since the covariance matrix is real and symmetric, we can write it as
\begin{align}
    \mathrm{Cov} = T \Delta T^T,
\end{align}
where $T$ is orthogonal $T T^T = T^T T = \mathbf{1}$ with $T_{\alpha \delta} = \supbracket{V}{\delta}_{\alpha}$ and $\Delta$ is a diagonal matrix with $\Delta_{RR}=\supbracket{\lambda}{R}=0$.
As a result,
\begin{align}
   0 &=\Delta_{RR} = \left(T^T \mathrm{Cov}T\right)_{RR} = \sum_{\gamma,\delta=1}^R T_{\gamma R} \mathrm{Cov}_{\gamma \delta} T_{\delta R} = \\
   &=  \avgrho{\sum_{\gamma=1}^R T_{\gamma R} \supbracket{r}{\gamma}\sum_{\delta=1}^R T_{\delta R} \supbracket{r}{\delta}}- \avgrho{\sum_{\gamma=1}^R T_{\gamma R} \supbracket{r}{\gamma}} \avgrho{\sum_{\delta=1}^R T_{\delta R} \supbracket{r}{\delta}} = \avgrho{z^2} - \avgrho{z}^2 = \mathrm{Var}_{\rho} (z),
\end{align}
where 
\begin{align}
    z=\sum_{\gamma=1}^R T_{\gamma R} \supbracket{r}{\gamma} = \sum_{\gamma=1}^R \supbracket{V}{R}_{\gamma} \supbracket{r}{\gamma}.
\end{align}
Since the variance of $z$ can only be zero if $z$ is deterministic/constant, we have
\begin{align}
    z_i = c \hspace{20pt} \forall i.
\end{align}
For the vector $\supbracket{E}{R}$ we then find
\begin{align}
    \supbracket{E}{R} = \sum_{\gamma=1}^R \supbracket{V}{R}_{\gamma} \left(\supbracket{r}{\gamma}-\avgrho{\supbracket{r}{\gamma}} \right) = z-\avgrho{z} =0.
\end{align}
Since $\supbracket{V}{R}$ is an eigenvector of $\mathrm{Cov}$, it is not equal to the zero vector and, unless $\supbracket{r}{\delta} - \avgrho{\supbracket{r}{\delta}}=0$ for some $\delta$, we conclude that the (shifted) features $\{ \supbracket{r}{\gamma} - \avgrho{\supbracket{r}{\gamma}} \}_{\gamma=1,\ldots, R}$ are linearly dependent.

Note that, in general, the space spanned by the feature vectors of a mixture with $N$ components is at most $N-1$-dimensional: 
$N$ points always lie in an $N-1$-dimensional Euclidean subspace.
Thus, the covariance matrix has at most $N-1$ non-zero eigenvalues, in accordance with the Hessian matrix having dimensions $(N-1) \times (N-1)$.

\section{Invariance under global rotations and translations of the rescaled feature vectors}
\label{SMsec:global_rotation_translation}

In this section, we show that the system is invariant under global rotations or translations of the rescaled feature vectors.
We proceed in two steps:
First, we demonstrate that the tilted free energy only changes by a constant and thus describes the same physical system.
We then show that, accordingly, the conditions for the spinodal and (higher-order) critical points do not change under global rotations or translations of the rescaled feature vectors.

To simplify notation, without loss of generality we consider the case when the interaction strength is absorbed into the spin values and we have $2\supbracket{C}{\gamma}=1 \forall \gamma$.
In this case, the rescaled and shifted feature vector just corresponds to
\begin{align}
    \supbracket{r}{\alpha} = \supbracket{s}{\alpha}-\supbracket{s}{\alpha}_N,
\end{align}
and the system is invariant to global rotations and translations of the rescaled feature vector:
\begin{align}
    \supbracket{s}{\alpha}_i = O_{\alpha \beta} \supbracket{\tilde{s}}{\beta}_i + \tau_{\alpha},
\end{align}
where $O$ is an orthogonal rotation matrix, $O O^T = O^T O = \mathbf{1}$ and $\mathrm{det} (O) = 1$, and $\vec{\tau}$ is a constant shift~\footnote{If $\supbracket{C}{\alpha} \neq \supbracket{C}{\beta}$ for some $\alpha,\beta$, then the system is invariant to $\vec{s}_i \rightarrow \vec{\tilde{s}}_i$ if
\begin{align}
    \sqrt{\supbracket{C}{\alpha}} \supbracket{s}{\alpha}_i = O_{\alpha \beta} \sqrt{\supbracket{C}{\beta}} \supbracket{\tilde{s}}{\beta}_i + \tau_{\alpha}.
\end{align}
}.
To see this, note that the interaction matrix is
\begin{align}
    \chi_{ij} &= \frac{1}{2} \sum_{\alpha=1}^R \supbracket{s}{\alpha}_i \supbracket{s}{\alpha}_j = \frac{1}{2} \sum_{\alpha=1}^R \left( \sum_{\gamma=1}^R O_{\alpha \gamma} \supbracket{\tilde{s}}{\gamma}_i + \tau_{\alpha}\right) \left( \sum_{\delta=1}^R O_{\alpha \delta} \supbracket{\tilde{s}}{\delta}_j + \tau_{\alpha}\right) = \\
    &=\frac{1}{2} \left( \sum_{\gamma=1}^R \supbracket{\tilde{s}}{\gamma}_i \supbracket{\tilde{s}}{\gamma}_j +
    \sum_{\alpha,\gamma=1}^R \tau_{\alpha} O_{\alpha \gamma} (\supbracket{\tilde{s}}{\gamma}_i+\supbracket{\tilde{s}}{\gamma}_j) + \sum_{\alpha=1}^R \tau_{\alpha}^2\right),
\end{align}
where we used that $\sum_{\alpha=1}^R O_{\alpha \gamma} O_{\alpha \delta} = \sum_{\alpha=1}^R O^T_{\gamma\alpha } O_{\alpha \delta} = \delta_{\gamma\delta}$.
Since $\sum_i \rho_i = 1$, in the free energy (in $-\sum_{i,j} \chi_{ij} \rho_i \rho_j$), the second term is linear in $\rho$ and the third term corresponds to a constant.
Neither term thus contributes to the tilted free energy and the thermodynamic behavior of a system with feature vectors $\vec{s}_i$ is the same as for a system with feature vectors $\vec{\tilde{s}}_i$.

Correspondingly, the conditions we derive are the same in both cases:
The covariance matrices are related via the rotation matrix (the translation cancels since a constant shift in the mean does not change the covariance):
\begin{align}
    \mathrm{Cov}_{\alpha \beta} = \avgrho{\supbracket{r}{\alpha} \supbracket{r}{\beta}} - \avgrho{\supbracket{r}{\alpha}}\avgrho{ \supbracket{r}{\beta}} = O_{\alpha \gamma} \tilde{\mathrm{Cov}}_{\gamma \delta} O^T_{\delta \beta}.
\end{align}
Since the rotation matrix is orthogonal, this implies that the covariance matrices have the same eigenvalues and their eigenvectors are related via
\begin{align}
    \supbracket{V}{\gamma}_{\alpha} = O_{\alpha \beta} \supbracket{\tilde{V}}{\gamma}_{\beta}.
\end{align}
As a result, the relative enrichment is the same:
\begin{align}
    \supbracket{E}{\gamma} = \sum_{\alpha=1}^R \supbracket{V}{\gamma}_{\alpha} \left( \supbracket{r}{\alpha} - \avgrho{\supbracket{r}{\alpha}} \right) = \sum_{\alpha,\beta=1}^R O_{\alpha \beta} \supbracket{\tilde{V}}{\gamma}_{\beta} \left( \supbracket{r}{\alpha} - \avgrho{\supbracket{r}{\alpha}} \right) = \sum_{\beta=1}^R  \supbracket{\tilde{V}}{\gamma}_{\beta} \left( \supbracket{\tilde{r}}{\beta} - \avgrho{\supbracket{\tilde{r}}{\beta}} \right)  = \supbracket{\tilde{E}}{\gamma},
\end{align}
where we used that $\supbracket{r}{\alpha} = O_{\alpha \beta} \supbracket{\tilde{r}}{\beta}$ and so $O_{\alpha \gamma} \supbracket{r}{\alpha} = O_{\alpha \gamma} O_{\alpha \beta} \supbracket{\tilde{r}}{\beta} = \delta_{\gamma \beta} \supbracket{\tilde{r}}{\beta} = \supbracket{\tilde{r}}{\gamma}$ (implicit sum convention).
Since, furthermore, 
\begin{align}
    \sum_{\alpha=1}^R \avgrho{\supbracket{r}{\alpha} \supbracket{\Omega}{k}} \avgrho{\supbracket{r}{\alpha} \supbracket{\Omega}{m-k}} = \sum_{\alpha, \beta,\gamma=1}^R O_{\alpha \beta} O_{\alpha \gamma} \avgrho{\supbracket{\tilde{r}}{\beta} \supbracket{\Omega}{k}} \avgrho{\supbracket{\tilde{r}}{\gamma} \supbracket{\Omega}{m-k}} = \sum_{\gamma=1}^R  \avgrho{\supbracket{\tilde{r}}{\gamma} \supbracket{\Omega}{k}} \avgrho{\supbracket{\tilde{r}}{\gamma} \supbracket{\Omega}{m-k}},
\end{align}
Eq.~\ref{SMeq:vectors_n_order} implies by induction that $\supbracket{\Omega}{k} = \supbracket{\tilde{\Omega}}{k} \forall k$ and, thus, according to Eq.~\ref{SMeq:conditions_n_order} all conditions are the same.
Note that the same is true for reflections $\supbracket{s}{\alpha}_i \rightarrow -\supbracket{s}{\alpha}_i \forall i$ of any subset of features $\alpha \in \mathcal{S} \subseteq \{1,\ldots,R \}$.

\section{Several directions of instability}
\label{SMsec:several_directions_instability}

In this section, we briefly comment on a second series of higher-order critical points, arising from several directions of instability, i.e., in the case when on the spinodal the largest eigenvalue of the covariance matrix is degenerate.

\subsection{Codimension of the critical submanifold}

Suppose that the covariance matrix has a degenerate maximal eigenvalue of 1:
\begin{align}
    \supbracket{\lambda}{\gamma}&=1 \hspace{20pt} \forall \gamma=1, \ldots, D \\
    \supbracket{\lambda}{\gamma}&<1 \hspace{20pt}  \forall \gamma>D.
\end{align}
Then, according to Eq.~\ref{SMeq:eigenvectorsHessian}, we have
\begin{align}
    H_{ij} \supbracket{e}{\gamma}_j = 0 \hspace{20pt} \forall \gamma=1,\ldots,D,
\end{align}
with the directions of instability $\supbracket{e}{\gamma} = \rho \sum_{\delta=1}^R \supbracket{V}{\gamma}_\delta \left( \supbracket{r}{\delta} - \avgrho{\supbracket{r}{\delta}}\right)$ as defined previously in terms of the orthonormal set of eigenvectors $\supbracket{V}{\gamma}$ of the covariance matrix.
Note that since the covariance matrix is real and symmetric, eigenvectors corresponding to degenerate eigenvalue can still be chosen orthogonal, which we assume in the following.

The Hessian is a linear operator and so we have
\begin{align}
    H_{ij} (\sum_{\gamma=1}^D \mu_{\gamma} \supbracket{e}{\gamma})_j = 0
\end{align}
for any linear combination $\sum_{\gamma=1}^D \mu_{\gamma} \supbracket{e}{\gamma}$, $\mu_{\gamma} \in \mathbf{R}$, of the directions of instability.
Furthermore, the directions of instability are linearly independent:
Suppose that 
\begin{align}
    \sum_{\gamma=1}^D \mu_{\gamma} \supbracket{e}{\gamma}_i = 0 \hspace{20pt} \forall i.
\end{align}
Multiplying by $\supbracket{E}{\beta}_i$ for $\beta=1,\ldots,D$ and summing over all $i=1,\ldots, N$ yields
\begin{align}
    0=\avgrho{\supbracket{E}{\beta} \sum_{\gamma=1}^D \mu_{\gamma} \supbracket{E}{\gamma}} = \sum_{\gamma=1}^D \mu_{\gamma} \sum_{\delta,\epsilon=1}^R \supbracket{V}{\beta}_{\delta} \supbracket{V}{\gamma}_{\epsilon} \mathrm{Cov}_{\delta \epsilon} = \sum_{\gamma=1}^D \mu_{\gamma} \delta_{\beta \gamma} \supbracket{\lambda}{\gamma} = \mu_{\beta},
\end{align}
demonstrating that the directions of instability are linearly independent.
They thus span a $D$-dimensional subspace of the component density space along which the second derivative of the free energy is zero.

What are the conditions that $m\leq D+1$ local minima of the free energy merge along this $D$-dimensional subspace?
One possibility is that simultaneously along each of $m-1$ linearly independent directions of instability, two minima merge, i.e.\ that a usual critical point arises simultaneously along $m-1$ linearly independent linear combinations $\sum_{\gamma=1}^D \supbracket{\mu}{i}_{\gamma} \supbracket{e}{\gamma}, i=1,\ldots, m-1$ of the directions of instability.
In analogy to our previous analysis, where we considered the optimal path in the density space, this requires that the third cumulant of these linear combinations has to be zero:
\begin{align}
    \avgcp{\left(\sum_{\gamma=1}^D \supbracket{\mu}{i}_{\gamma} \supbracket{E}{\gamma}\right)^3} =0 \hspace{20pt} \forall i=1, \ldots, m-1.
\end{align}
However, since it should not be possible to minimize the free energy along any other direction (in the $D$-dimensional subspace where the Hessian is zero), the third cumulant needs to be zero for all linear combinations of the $D$ directions of instability:
\begin{align}
    \avgcp{\left(\sum_{\gamma=1}^D \mu_{\gamma} \supbracket{E}{\gamma}\right)^3} =0 \hspace{20pt} \forall \vec{\mu} \in \mathbb{R}^D,
\end{align}
and therefore there will be simultaneous ordinary (or higher-order) critical points along all $D$ directions of instability, thus making the critical point order $D+1$ (or higher).
The above conditions are only fulfilled if
\begin{align}
    \avgcp{\supbracket{E}{\gamma_1} \supbracket{E}{\gamma_2} \supbracket{E}{\gamma_3}} = 0
\end{align}
for all possible combinations of $\gamma_1,\gamma_2, \gamma_3=1,\ldots, D$.
Since the distinct combinations correspond to drawing a number in $1,\ldots, D$ three times without considering the ordering, there are $\binom{3+D-1}{3}$ conditions that need to be satisfied.

What is the codimension resulting from $D$ eigenvalues being 1?
As discussed in \cite{Keller2008}, for real symmetric matrices (such as the covariance matrix or the Hessian), the set of matrices with a $D$-fold degenerate eigenvalue has codimension $(D+2)(D-1)/2$.
Since in our case, these eigenvalues have to be equal to 1, this gives a codimension of $(D+2)(D-1)/2+1=D(D+1)/2=\binom{D+1}{2}$.

Taken together, for this series of $D+1$-th order critical points, this counting suggests a codimension of $\binom{D+1}{2}+\binom{3+D-1}{3}$:  $\binom{D+1}{2}$ from the $D$ largest eigenvalues being 1 and $\binom{D+2}{3}$ from the third cumulants being 0~\footnote{Note that here we posit that these conditions are all independent.
The basis for this argument is that the conditions on the eigenvalues correspond to conditions on the second moments/cumulants whereas the other ones are about the third moments/cumulants. 
However, we cannot entirely exclude that there is a mutually dependent set and that the actual codimension is smaller.}.
Note that for $D$ directions of instability (at least) $D+1$ phases will become indistinguishable at a critical point: either all third cumulants in the $D$-dimensional subspace of directions of instability are zero (and at least $D+1$ phases merge) or there is a direction along which the free energy can be lowered and the point is not thermodynamically stable.

Of course, one could also imagine combining this series of higher-order critical points arising from merging of minima along several directions of instability with the series as discussed before, where several minima merge along a single direction of instability.
This combination will, however, not be discussed here.

To conclude this section, we will briefly touch upon one idea for an effective mean-field theory for the two-dimensional $q{=}3$-states Potts model with two direction of instability.

\subsection{\label{SMsec:Potts}A mean-field version of the $q{=}3$-states Potts model}

The regular mean-field version of the Potts model can be expressed in terms of our framework by a $N=3$-component system with the following $R=3$ features:
\begin{align}
    \supbracket{s}{\gamma}_i = a \delta_{i \gamma},
\end{align}
where $i,\gamma=1,2,3$.
Here and in the following, we directly incorporate the interaction strengths $\supbracket{J}{\gamma}$ (and other constant factors) into the spin values and only keep the dependency on the temperature $T$, effectively setting $z \supbracket{J}{\gamma}/k_B=1$.
Then, the covariance matrix is given by
\begin{align}
    \mathrm{Cov}_{\alpha\beta} = \frac{1}{T} \left( \avgrho{\supbracket{s}{\alpha} \supbracket{s}{\beta}} - \avgrho{\supbracket{s}{\alpha}} \avgrho{ \supbracket{s}{\beta}} \right).
\end{align}
At the symmetry point $\rho_1=\rho_2=\rho_3=1/3$, we have
\begin{align}
\mathrm{Cov} = \frac{a^2}{9 T}
    \begin{pmatrix}
        2 & -1 & -1 \\
        -1 & 2 & -1 \\
        -1 & -1 & 2
    \end{pmatrix}
\end{align}
with eigenvalues $\supbracket{\lambda}{1} = \supbracket{\lambda}{2} = a^2/3/T$ and $\supbracket{\lambda}{3}=0$.
The spinodal is thus located at $a^2 = 3T$ and the system exhibits two directions of instability with corresponding (orthonormal) eigenvectors of $\mathrm{Cov}$
\begin{align}
    \supbracket{V}{1} = \frac{1}{\sqrt{2}} \begin{pmatrix}
        -1 \\
        0 \\
        1
    \end{pmatrix}
    \hspace{20pt} \text{and} \hspace{20pt}
    \supbracket{V}{2} = \frac{1}{\sqrt{6}} \begin{pmatrix}
        -1 \\
        2 \\
        -1
    \end{pmatrix}.
\end{align}
Importantly, not all combinations of third cumulants of these directions of instability are zero.
While two are indeed zero, the other two are equal to $\pm a^3/3/\sqrt{6}$, thus leading to thermodynamic instability along certain directions in the 2-dimensional subspace of directions of instability.
Hence, as noted previously~\cite{Straley1973, Mittag1974}, the symmetric state does not exhibit critical behavior, in contrast to the two-dimensional $q{=}3$-states Potts model~\cite{Baxter1973}.
Ultimately, this lack of a continuous transition is due to the fact that for $D=2$, seven conditions need tuning while any $N=3$, $R=3$ model only exhibits five degrees of freedom, two ratios of densities and three interaction parameters: 
The three feature vectors necessarily lie in a plane, thus making the effective number of features ${=}2$ (a rotation of the feature vectors onto e.g.\ the $x$-$y$-plane does not change the system) and the number of parameters ${=}6$.
However, each feature is invariant under constant shifts and the system is invariant under rotations of the features, thus reducing the number of degrees of freedom to $6{-}2{-1}=3$.

\vspace{20pt}

While this regular mean-field version of the $q{=}3$-states Potts model does not exhibit a critical point, here, we propose a slightly more complex version with $N=6$ and $R=2$, which preserves the $\mathbf{S}_3$ symmetry group of the $q{=}3$-states Potts model but represents each ``lattice component type" by two ``effective component types".

\begin{figure}[t]
\begin{center}
    \includegraphics[width=0.5\columnwidth]{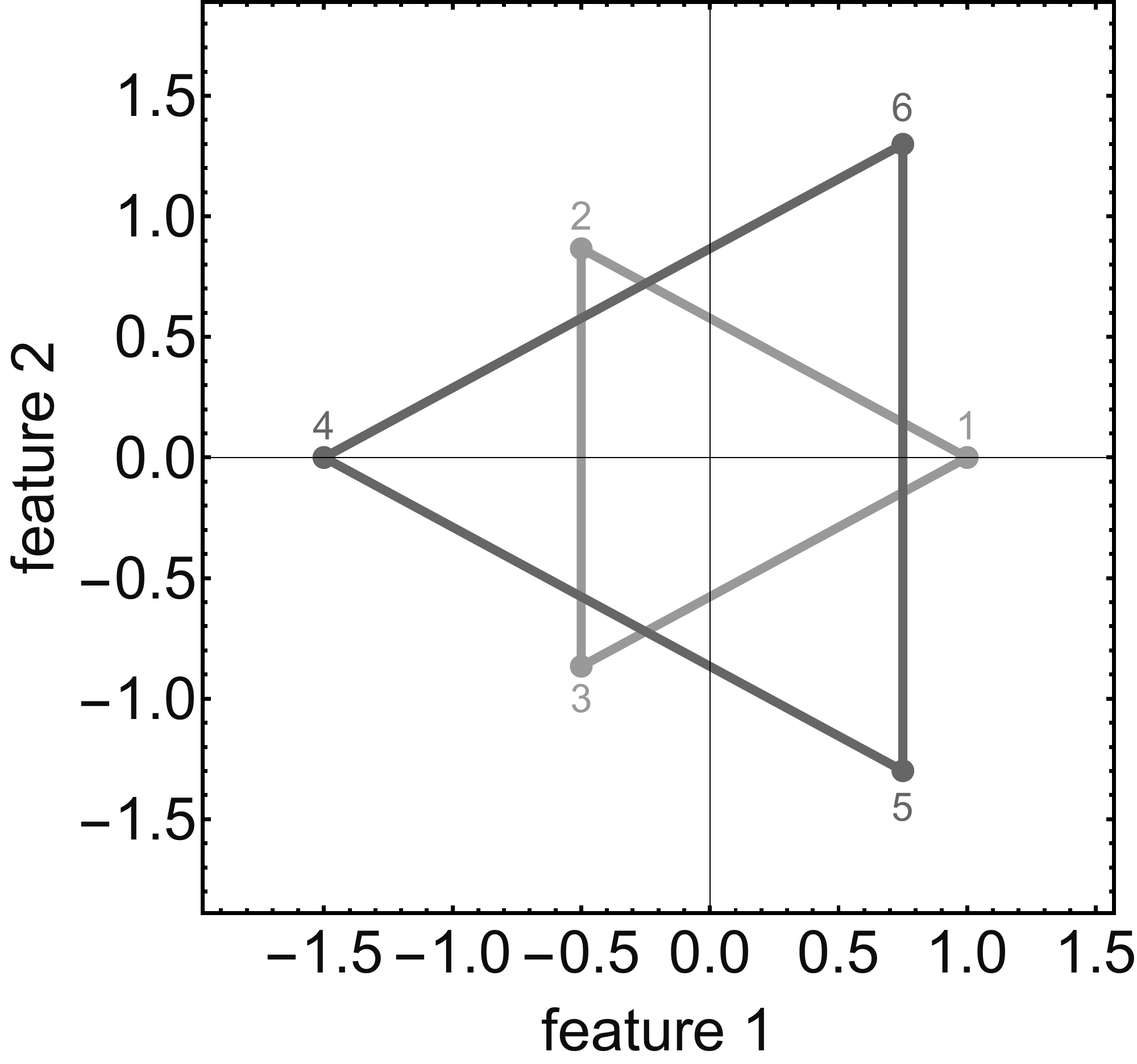}%
\end{center}
\caption{\label{SMfig:Potts} Possible choice of feature vectors for the mean-field Potts model; here $a=1$ and $b=-1.5$.
The component types preserve the permutation symmetry and their feature vectors lie on two equilateral triangles, centered at the origin.
We focus on the symmetric state where the component types belonging to the same triangle are present with equal densities: $\rho_1=\rho_2=\rho_3=\frac{x}{3}$ and $\rho_4=\rho_5=\rho_6=\frac{1-x}{3}$.
}
\end{figure}

More specifically, we consider a mixture with the following feature vectors:
\begin{align}
    \vec{s}_1 = 
    a\begin{pmatrix}
        1\\0
    \end{pmatrix}
    \hspace{10pt}
    \vec{s}_2 = 
    a\begin{pmatrix}
        -1/2\\ \sqrt{3}/2
    \end{pmatrix}
    \hspace{10pt}
    \vec{s}_3 = 
    a \begin{pmatrix}
        -1/2\\ - \sqrt{3}/2
    \end{pmatrix}
    \hspace{10pt}
    \vec{s}_4 = 
    b\begin{pmatrix}
        1\\0
    \end{pmatrix}
    \hspace{10pt}
    \vec{s}_5 = 
    b\begin{pmatrix}
        -1/2\\ \sqrt{3}/2
    \end{pmatrix}
    \hspace{10pt}
    \vec{s}_6 = 
    b \begin{pmatrix}
        -1/2\\ - \sqrt{3}/2
    \end{pmatrix},
\end{align}
corresponding to two equilateral triangles centered around the origin (see Fig.~\ref{SMfig:Potts}).
For simplicity, we absorb all the interaction strengths and constants into the spin values, and assume that $2 \supbracket{C}{1}=2 \supbracket{C}{1}=1/T$, retaining only the dependency on temperature.
Furthermore, we focus on the symmetric state where all component types belonging to the same equilateral triangle have the same densities:
\begin{align}
    \rho_1=\rho_2=\rho_3=\frac{x}{3} \hspace{20pt} \text{and} \hspace{20pt} \rho_4=\rho_5=\rho_6=\frac{1-x}{3}.
\end{align}
For this symmetric case, the average of all spins vanishes, $\avg{\supbracket{s}{1}} = \avg{\supbracket{s}{2}}=0$, and the covariance matrix is given by
\begin{align}
    \mathrm{Cov} = \frac{1}{2 T} \left( x a^2 + (1-x) b^2\right)
    \begin{pmatrix}
        1 & 0 \\
        0 & 1
    \end{pmatrix}.
\end{align}
The eigenvalues can be read off as 
\begin{align}
    \supbracket{\lambda}{1} = \supbracket{\lambda}{2} = \frac{1}{2 T} \left( x a^2 + (1-x) b^2\right),
\end{align}
with corresponding eigenvectors
\begin{align}
    \supbracket{\vec{V}}{1} = 
    \begin{pmatrix}
        1\\0
    \end{pmatrix}
    \hspace{20pt}
    \supbracket{\vec{V}}{2} = 
    \begin{pmatrix}
        0\\1
    \end{pmatrix}.
\end{align}
The largest eigenvalue is thus two-fold degenerate and on the spinodal, 
\begin{align}
    x a^2 + (1-x) b^2 = 2T,
\end{align}
there are two directions of instability.
According to our previous analysis, we thus need the third cumulant of any linear combination of
\begin{align}
    \supbracket{E}{1}_i &= \sum_{\alpha=1}^2 \supbracket{V}{1}_{\alpha} \supbracket{s}{\alpha}_i = \supbracket{s}{1}_i \hspace{20pt} \text{and}\\
    \supbracket{E}{2}_i &= \supbracket{s}{2}_i
\end{align}
to be zero in order for the system to be at its critical point (where then the 3 symmetric phases merge into a single phase):
\begin{align}
    \avg{\left( \alpha_1 \supbracket{E}{1} + \alpha_2 \supbracket{E}{2} \right)^3} = \frac{1}{4} \alpha_1 \left(\alpha_1^2 - 3 \alpha_2^2 \right) \left(x a^3 + (1-x) b^3 \right) = 0 \hspace{10pt} \forall \alpha_1, \alpha_2.
\end{align}
In addition to the condition for the spinodal, we thus find that
\begin{align}
    &x a^3 + (1-x) b^3 = 0 \hspace{20pt} \text{or} \\
    &x = \frac{b^3}{b^3-a^3} \label{SMeq:Potts_constraint_density}
\end{align}
at the critical point.
Since $x$ and $1-x$ denote the combined densities of component types $1,2,3$ and $4,5,6$, respectively, (and since we have seen previously that a three component system is not enough,) we need $x \in (0,1)$, yielding the additional constraint
\begin{align}
    ab<0.
\end{align}
Choosing the density as in Eq.~\ref{SMeq:Potts_constraint_density}, the spinodal condition is
\begin{align}
    \frac{a^2 b^2 (b-a)}{b^3-a^3} = b^2 \frac{(1-r)r^2}{1-r^3} = 2T, \label{SMeq:Potts_spinodal}
\end{align}
where $r=a/b$ is the ratio between the linear dimensions of the two ``feature triangles".

In principle, equality~\ref{SMeq:Potts_spinodal} is satisfied by a one-dimensional submanifold of $(b,r)$-space.
We do, however, need to ensure that the symmetric state is thermodynamically stable (at least to local fluctuations) and that up to the next order in the free energy expansion (the fourth order), the free energy change is positive for all possible paths in density space.
The fourth order term in the free energy expansion  is (see Eq.~\ref{SMeq:expansion_free_energy_Bell})
\begin{align}
    F:=\supbracket{A}{4} B_{4,4} \left(\rho (\alpha_1 \supbracket{E}{1} {+} \alpha_2 \supbracket{E}{2}), \rho \supbracket{\omega}{2} \right)  + \supbracket{A}{3} B_{4,3} \left(\rho (\alpha_1 \supbracket{E}{1} {+} \alpha_2 \supbracket{E}{2}), \rho \supbracket{\omega}{2} \right) + \supbracket{A}{2} B_{4,2} \left(\rho (\alpha_1 \supbracket{E}{1} {+} \alpha_2 \supbracket{E}{2}), \rho \supbracket{\omega}{2} \right).
\end{align}
Here, we have used that the paths of minimal change in free energy have a tangent within the plane of instabilities: $\supbracket{\Omega}{1} = \alpha_1 \supbracket{E}{1} + \alpha_2 \supbracket{E}{2}=: E$.
Thus $\supbracket{A}{2} (\rho E) =0$ and the fourth order term is rewritten as
\begin{align}
    F=6 \left( \frac{1}{2} \supbracket{A}{2} (\rho \supbracket{\omega}{2})^2 + (\rho \supbracket{\omega}{2}) \underbrace{\left[ \supbracket{A}{3} (\rho E)^2\right]}_{=: b} + \frac{1}{6} \supbracket{A}{4} (\rho E)^4 \right).
\end{align}
This quadratic form exhibits a minimal value although $\supbracket{A}{2}$ is only positive semi-definite:
Analogously to the case with a single direction of instability we can define the pseudoinverse of $\supbracket{A}{2}$ by 
\begin{align}
    (\supbracket{A}{2}_{\text{pseudo}})_{ij}^{-1} = \delta_{ij} \rho_i - \rho_i \rho_j.
\end{align}
It satisfies
\begin{align}
    \left(\supbracket{A}{2} (\supbracket{A}{2}_{\text{pseudo}})^{-1}\right)_{ik} &= \left( \delta_{ij} \frac{1}{\rho_i} + \frac{1}{1-\Rho} -\supbracket{r}{1}_i \supbracket{r}{1}_j - \supbracket{r}{2}_i \supbracket{r}{2}_j \right) \left( \delta_{jk} \rho_j - \rho_j \rho_k\right) = \delta_{ik} - \sum_{\gamma=1}^2 \supbracket{r}{\gamma}_i \supbracket{r}{\gamma}_j  \left( \delta_{jk} \rho_j - \rho_j \rho_k\right) = \nonumber\\
    &= \delta_{ik} - \rho_k \sum_{\gamma=1}^2 \supbracket{r}{\gamma}_i \left[ \supbracket{r}{\gamma}_k - \avg{\supbracket{r}{\gamma}} \right] = \delta_{ik} - \rho_k \sum_{\gamma=1}^2 \supbracket{r}{\gamma}_i \sum_{\alpha=1}^2 \underbrace{\delta_{\alpha\gamma}}_{=\sum_{\delta=1}^2 \supbracket{V}{\delta}_{\alpha}\supbracket{V}{\delta}_{\gamma}} \left[ \supbracket{r}{\alpha}_k - \avg{\supbracket{r}{\alpha}} \right]  =\\
    &= \delta_{ik} - \rho_k \sum_{\delta=1}^2 \supbracket{E}{\delta}_k  \sum_{\gamma=1}^2 \supbracket{V}{\delta}_{\gamma} \supbracket{r}{\gamma}_i.
\end{align}
In particular, 
\begin{align}
    \left(\supbracket{A}{2} (\supbracket{A}{2}_{\text{pseudo}})^{-1}\right)b = b - \sum_{\delta=1}^2 \left(\sum_{\gamma=1}^2 \supbracket{V}{\delta}_{\gamma} \supbracket{r}{\gamma}\right) \supbracket{A}{3} (\rho E)^2 \rho \supbracket{E}{\delta} = b + \sum_{\delta=1}^2 \left(\sum_{\gamma=1}^2 \supbracket{V}{\delta}_{\gamma} \supbracket{r}{\gamma}\right) \avg{ ( E)^2  \supbracket{E}{\delta}} = b,
\end{align}
where we used Eq.~\ref{SMeq:useful_full_contraction} and that the third cumulants $\avg{ ( E)^2  \supbracket{E}{\delta}}=0 \ \forall \delta$.
As a result, $F$ has a minimum, which is given by
\begin{align}
    F_{\mathrm{min}} &= 6 \left( -\frac{1}{2} b^T (\supbracket{A}{2}_{\text{pseudo}})^{-1} b + \frac{1}{6} \supbracket{A}{4} \left( \rho E \right)^4 \right) = 6 \left( -\frac{1}{2} b_i (\delta_{ij} \rho_i - \rho_i \rho_j) b_j + \frac{1}{6} 2 \avg{ E^4} \right) = \\
    &= 3 \avg{b}^2 - 3 \avg{b^2} + 2 \avg{E^4} = 3 \avg{E^2}^2 - 3 \avg{E^4} + 2 \avg{E^4} = - \left( \avg{E^4} - 3 \avg{E^2}^2 \right) = - \supbracket{\kappa}{E}_4.
\end{align}
Here $\supbracket{\kappa}{E}_4$ is the fourth cumulant of $E$ and we used Eq.~\ref{SMeq:useful_full_contraction} together with~\ref{SMeq:useful_partial_contraction}.
Overall, we thus find that the fourth order term is positive for all optimal paths only if $\supbracket{\kappa}{E}_4<0$.

Rewriting the fourth cumulant gives
\begin{align}
    \supbracket{\kappa}{E}_4 &= \avg{E^4} - 3 \avg{E^2}^2 = \avg{\left(\alpha_1 \supbracket{E}{1} + \alpha_2 \supbracket{E}{2}  \right)^4} - 3 \avg{\left( \alpha_1 \supbracket{E}{1} + \alpha_2 \supbracket{E}{2}  \right)^2}^2 = \\
    &=-\frac{3}{8} \left(\alpha_1^2 + \alpha_2^2 \right)^2\frac{a^3 b^3 \left(a^2+3ab+b^2 \right)}{\left( a^2 + ab + b^2\right)^2}.
\end{align}
Since $ab<0$, this expression is negative only if 
\begin{align}
    a^2+3ab+b^2 = b^2 (r^2 + 3r + 1)<0. \label{SMeq:Potts_condition_fourth_cumulant}
\end{align}
We thus find the following condition for the ratio $r$ between $a$ and $b$:
\begin{align}
    r \in (\frac{-3-\sqrt{5}}{2},\frac{-3+\sqrt{5}}{2}) \approx (-2.62,-0.38).
\end{align} 
We note that this range is invariant under $r {\rightarrow} 1/r$, as expected from relabelling $a {\leftrightarrow} b$.

Taken together, this $N{=}6$/$R{=}2$ model with threefold symmetry exhibits a critical point where three phases merge if $b$ and $r=a/b$ are chosen as
\begin{align}
    2 T = b^2 \frac{(1-r)r^2}{1-r^3}  \hspace{10pt} \text{for some} \hspace{10pt}
    r \in \frac{1}{2} (-3-\sqrt{5},-3+\sqrt{5}).
\end{align}
We suggest that any such model might therefore be a candidate for a mean-field version of the two-dimensional $q{=}3$-states Potts model which retains a continuous phase transition.

Interestingly, we first tried a version with $R=3$ features and $N=6$ component types whose spin vectors don't lie in one plane but instead in two parallel planes:
\begin{align}
    \vec{s}_1 = 
    a\begin{pmatrix}
        1\\0 \\0
    \end{pmatrix}
    \hspace{10pt}
    \vec{s}_2 = 
    a\begin{pmatrix}
        0\\1\\0
    \end{pmatrix}
    \hspace{10pt}
    \vec{s}_3 = 
    a \begin{pmatrix}
        0\\0\\1
    \end{pmatrix}
    \hspace{10pt}
    \vec{s}_4 = 
    b\begin{pmatrix}
        1\\0\\0
    \end{pmatrix}
    \hspace{10pt}
    \vec{s}_5 = 
    b\begin{pmatrix}
        0\\1\\0
    \end{pmatrix}
    \hspace{10pt}
    \vec{s}_6 = 
    b \begin{pmatrix}
        0\\0\\1
    \end{pmatrix}.
\end{align}
In this case, it is not possible to simultaneously satisfy the spinodal and third cumulants condition and to have a positive fourth order term for all possible paths.
Ultimately, this is due to the fact that in this case where the features space is three- instead of two-dimensional, the free energy can be lowered by choosing a vector of second derivatives $\supbracket{\omega}{2}$ that leaves the plane of instability.

Looking ahead, it will be interesting to see whether a related model with $R=3$ and three directions of instability could serve as a mean-field version of the two-dimensional $q{=}4$-Potts model, which also exhibits a continuous transition in two dimensions but not in regular mean-field theory.
One possibility might be to consider feature vectors spanning a regular tetrahedron and rotations thereof.
Furthermore, it would be enlightening to understand whether such types of models are not possible for $D>4$ (for the two-dimensional $q{>}4$-states Potts models, the transitions are indeed discontinuous).

\section{Coarse-graining a multicomponent mixtures as a binary mixture}

In this section, we discuss one idea how to coarse-grain a multicomponent mixture as a binary mixture of two components with effective features, while conserving two local thermodynamic properties, the spinodal line and the line of (ordinary) critical points.
As we have seen before, the conditions for the spinodal and ordinary critical points only depend on the first principal component of the covariance matrix (note that we restrict our discussion to the case of a non-degenerate maximal eigenvalue of the covariance matrix).
Hence, if only those two characteristics are to be preserved, only the distribution of features along the first principal component is of relevance and it might be enough to consider a binary system with the same first principal component.
We therefore suggest to coarse-grain the multicomponent mixture (denoted by regular symbols) as a binary mixture (denoted by tildes) whose components $i=1,2$ exhibit feature vectors along the 1st principal component of the original system:
\begin{align}
    \sqrt{2 \supbracket{\tilde{C}}{\gamma}} \supbracket{\tilde{s}}{\gamma}_i = x_i \supbracket{V}{1}_{\gamma}
\end{align}
where $x_i$ is the distance of component $i$ along the 1st principal component and $\gamma=1, \ldots, R$ denotes the different features (note that it would equally be possible to do this coarse-graining procedure with a single feature, corresponding to a rotation of this feature vector onto the first axis).
This binary mixture has a covariance matrix 
\begin{align}
    \widetilde{\mathrm{Cov}}_{\alpha \beta}&=\avgvarphi{\supbracket{\tilde{r}}{\alpha} \supbracket{\tilde{r}}{\beta}} - \avgvarphi{\supbracket{\tilde{r}}{\alpha}} \avgvarphi{\supbracket{\tilde{r}}{\beta}} = 2\sqrt{\supbracket{\tilde{C}}{\alpha} \supbracket{\tilde{C}}{\beta}} \left( \avgvarphi{\supbracket{\tilde{s}}{\alpha} \supbracket{\tilde{s}}{\beta}} -  \avgvarphi{\supbracket{\tilde{s}}{\alpha}} \avgvarphi{\supbracket{\tilde{s}}{\beta}} \right) = \\
    &=\varphi_1 x_1^2 \supbracket{V}{1}_{\alpha} \supbracket{V}{1}_{\beta} + \varphi_2 x_2^2 \supbracket{V}{1}_{\alpha} \supbracket{V}{1}_{\beta} - \left(\varphi_1 x_1 \supbracket{V}{1}_{\alpha} + \varphi_2 x_2 \supbracket{V}{1}_{\alpha} \right)  \left(\varphi_1 x_1 \supbracket{V}{1}_{\beta} + \varphi_2 x_2 \supbracket{V}{1}_{\beta} \right) = \\
    &= \ldots = \supbracket{V}{1}_{\alpha} \supbracket{V}{1}_{\beta} \varphi_1 \left(1-\varphi_1 \right) \left(x_1 - x_2 \right)^2,
\end{align}
where $\varphi_i$ is the density of component type $i=1,2$ of the binary mixture: $\varphi_2 = 1-\varphi_1$.
We observe that $\widetilde{\mathrm{Cov}}$ is a rank-1 matrix and, as desired, the first principal component of the original multicomponent mixture, $\supbracket{V}{1}$, is also an eigenvector of $\widetilde{\mathrm{Cov}}$ (corresponding to the only non-zero eigenvalue, $\varphi_1 \left(1-\varphi_1 \right) \left(x_1 - x_2 \right)^2$).

Similarly, for the third cumulant of the direction in component space corresponding to this eigenvector we find 
\begin{align}
    \avgvarphi{\left(\sum_{\alpha=1}^R \supbracket{V}{1}_{\alpha} \left( \supbracket{\tilde{r}}{\alpha}-\avgvarphi{\supbracket{\tilde{r}}{\alpha}}\right) \right)^3} &=  \avgvarphi{\left(\sum_{\alpha=1}^R \supbracket{V}{1}_{\alpha} \left(\sqrt{2 \supbracket{C}{\alpha}} \supbracket{\tilde{s}}{\alpha}- \sqrt{2 \supbracket{C}{\alpha}}\avgvarphi{\supbracket{\tilde{s}}{\alpha}}\right) \right)^3} = \\
    &=\avgvarphi{ \left( x -\avgvarphi{x}\right)^3 } \left( \sum_{\alpha=1}^R \left(\supbracket{V}{1}_{\alpha}\right)^2 \right)^3 = \avgvarphi{ \left( x -\avgvarphi{x}\right)^3 }
    ,
\end{align}
where we used that $\supbracket{\tilde{r}}{\alpha}_i = \sqrt{2 \supbracket{C}{\alpha}}\supbracket{\tilde{s}}{\alpha}_i - \sqrt{2 \supbracket{C}{\alpha}}\supbracket{\tilde{s}}{\alpha}_2 = (x_i-x_2) \supbracket{V}{1}_{\alpha}$ and so  
\begin{align}
 \avgvarphi{\left(\supbracket{\tilde{r}}{\alpha} - \avgvarphi{\supbracket{\tilde{r}}{\alpha}} \right) \left(\supbracket{\tilde{r}}{\beta} - \avgvarphi{\supbracket{\tilde{r}}{\beta}} \right)\left(\supbracket{\tilde{r}}{\gamma} - \avgvarphi{\supbracket{\tilde{r}}{\gamma}} \right)} = \supbracket{V}{1}_{\alpha} \supbracket{V}{1}_{\beta} \supbracket{V}{1}_{\gamma} \avgvarphi{ \left( x -\avgvarphi{x}\right)^3 }.
\end{align}
Using that $\avgvarphi{x} = \varphi_1 x_1 + \varphi_2 x_2 = x_2 + \varphi_1 (x_1-x_2)$, the third cumulant of $x$ (and thus of the potential direction of instability) is rewritten as
\begin{align}
    \avgvarphi{\left(\sum_{\alpha=1}^R \supbracket{V}{1}_{\alpha} \left( \supbracket{\tilde{r}}{\alpha}-\avgvarphi{\supbracket{\tilde{r}}{\alpha}}\right) \right)^3} &= \varphi_1 \left(\left( 1-\varphi_1 \right) \left(x_1-x_2 \right)\right)^3 - \left(1-\varphi_1\right) \left(\varphi_1 \left(x_1-x_2 \right)\right)^3 = \\
    &= \varphi_1 \left( 1-\varphi_1 \right) \left(1-2 \varphi_1 \right) \left(x_1 - x_2 \right)^3.
\end{align}

One possibility to maintain the location of the  spinodal and critical line while coarse-graining is to preserve the second and third moment along the first principal component, i.e. to require that
\begin{align}
    \varphi_1 \left(1-\varphi_1 \right) \left( x_1 -x_2 \right)^2 
    \overset{!}{=} \supbracket{\lambda}{1} = \avgrho{\left(\supbracket{E}{1} \right)^2} &=: \mu_2 \\
    \varphi_1 \left(1-\varphi_1 \right) \left(1-2\varphi_1 \right) \left( x_1 -x_2 \right)^3 
    \overset{!}{=}  \avgrho{\left(\supbracket{E}{1} \right)^3} &=: \mu_3,
\end{align}
where $\avgrho{\ldots}$ denotes the average in the original space of densities of the multicomponent mixture.
As a result,
\begin{align}
    \frac{(\mu_2)^3}{(\mu_3)^2} \overset{!}{=} \frac{\varphi_1^3 \left(1-\varphi_1 \right)^3 \left( x_1 -x_2 \right)^6}{\varphi_1^2 \left(1-\varphi_1 \right)^2 \left(1-2\varphi_1 \right)^2 \left( x_1 -x_2 \right)^6} = \frac{\varphi_1 \left(1-\varphi_1 \right)}{ \left(1-2\varphi_1 \right)^2},
\end{align}
which is solved by
\begin{align}
    \varphi_1 = \frac{1}{2} \left( 1 \pm \sqrt{\frac{(\mu_3)^2}{(\mu_3)^2 + 4 (\mu_2)^3}}\right) = \frac{1}{2} \left( 1 \pm \sqrt{\frac{1}{1+\tau}}\right),
\end{align}
with
\begin{align}
    \tau = \frac{4 (\mu_2)^3}{(\mu_3)^2}.
\end{align}
For $x_1-x_2$ we then find
\begin{align}
    \left( x_1 - x_2\right)^2 \overset{!}{=} \frac{\mu_2}{\frac{1}{4} \left( 1 + \sqrt{\frac{1}{1+\tau}}\right) \left( 1 - \sqrt{\frac{1}{1+\tau}}\right)} = 4 \mu_2 \left(1+\frac{1}{\tau} \right).
\end{align}
Taken together, we find the following mapping
\begin{align}
    \varphi_1 &\overset{!}{=} \frac{1}{2} \left( 1+\frac{1}{\sqrt{1+\tau}}\right) \\
    \left|x_1 - x_2 \right| &\overset{!}{=} 2 \sqrt{\mu_2 \left(1+\frac{1}{\tau} \right)}.
\end{align}
For the density we chose the positive sign without loss of generality.

To proceed we note that $x_1 -x_2$ is related to the Flory-Huggins interaction strength $\bar{\chi}$:
\begin{align}
    \bar{\chi} := \tilde{\chi}_{11} + \tilde{\chi}_{22} - 2 \tilde{\chi}_{12} = \sum_{\gamma=1}^R \supbracket{\tilde{C}}{\gamma} \left(\supbracket{\tilde{s}}{\gamma}_1 - \supbracket{\tilde{s}}{\gamma}_2 \right)^2 = \frac{1}{2} \sum_{\gamma=1}^R \left( x_1-x_2 \right)^2 \left(\supbracket{V}{1}_{\gamma}\right)^2 = \frac{1}{2}\left( x_1-x_2 \right)^2,
\end{align}
which leads to the following conditions for the composition $\varphi:=\varphi_1$ and the interaction strength:
\begin{align}
    \varphi &\overset{!}{=} \frac{1}{2} \left(1+\frac{1}{\sqrt{1+\tau}} \right) \\
    \bar{\chi} &\overset{!}{=} 2 \mu_2 \left(1+\frac{1}{\tau} \right).
\end{align}
At the critical point, where $\mu_3 = 0$ and so $\tau \rightarrow \infty$, we have
\begin{align}
    \varphi_c &= \frac{1}{2} \\
    \bar{\chi}_c &= 2 \mu_2.
\end{align}
We hence find for the distance of the interaction strength from its critical value
\begin{align}
    \Delta \bar{\chi}:=\frac{\bar{\chi}-\bar{\chi}_c}{\bar{\chi}_c} = \frac{1}{\tau},
\end{align}
and for density difference $\Delta \varphi := \varphi_1 - \varphi_2 = 2 \varphi -1$
\begin{align}
    \Delta \varphi =\sqrt{ \frac{1}{1+\tau}} = \sqrt{\frac{\frac{1}{\tau}}{1+\frac{1}{\tau}}} = \sqrt{\frac{\Delta \bar{\chi}}{1+\Delta \bar{\chi}}} \approx \sqrt{\Delta \bar{\chi}} = \sqrt{\frac{1}{\tau}},
\end{align}
where the approximation is valid close to the critical point.

Taken together, coarse-graining procedures based on such (or similar) arguments can  conserve local properties of the free energy like the spinodal and critical points.
Whether (and in what cases) the same can be said for global characteristics like the binodal is, however, not clear a priori and is an interesting question for future research.

\section{Generation of the multicomponent mixtures illustrated in Fig. 2 of the main text}

To illustrate the role of the skewness of the distribution of the relative enrichments $\delta \rho_i/\rho_i$ for the occurrence of critical behavior, we generated the multicomponent mixtures as displayed in Fig. 2 of the main text from random distributions with zero and non-zero skewness, respectively.
More specifically, we followed the procedure as outline next (for simplicity, we assume $2\supbracket{C}{\gamma}=1 \ \forall \gamma$).
\subsection{Mixture 1: zero skewness}
We first drew $N{=}1000$ two-dimensional feature vectors (component types) from a multivariate Gaussian distribution with zero mean and covariance matrix $\begin{pmatrix} \sigma_1^2 &0 \\ 0 &\sigma_2^2 \end{pmatrix}$ with $\sigma_1=1, \sigma_2=0.3$.
We then randomly generated the corresponding densities of the component types by drawing $N{=}1000$ exponentially distributed random variables (with rate $\lambda=100$) and by normalizing the vector so that its sum (1-norm) equals 1~\footnote{This leads to a uniform distribution of the mixture composition on the $N{-}1$-simplex.}.
In order to enforce a variance of 1 along the first principal component and to illustrate the case when the first principal component is not aligned with a feature axis, we proceeded as follows:
\begin{enumerate}
    \item We rotated all feature vectors in a way that the first principal component aligns with the $x$-axis (by multiplying them by the matrix of eigenvectors of the covariance matrix).
    \item We rescaled the first component of the feature vectors by $\sigma_1/\sigma_{\mathrm{PC}1}$ and the second one by $\sigma_2/\sigma_{\mathrm{PC}2}$.
    Here $\sigma^2_{\mathrm{PC}1/2}$ is the variance of the features along the first and second principal component, corresponding to the variance along the $x$- and $y$-axis after step 1.
    (Steps 1 and 2 together thus (re)align the first and second principal component with the $x$- and $y$-axis and fix the variances as $\sigma_{\mathrm{PC}1/2} = \sigma_{1/2}$, respectively.)
    \item We rotated all feature vectors (and thereby the first and second principal component) counterclockwise by an angle $\vartheta=\pi/6$, thus aligning the first principal component $\supbracket{V}{1}$ with $\begin{pmatrix}
        \cos(\vartheta) \\ \sin(\vartheta)
    \end{pmatrix}$.
\end{enumerate}
The relative enrichment $\delta \rho_i/\rho_i$ of component type $i$ was then determined by taking the scalar product of the difference of the respective feature vector from the mean of the feature vectors (weighted by the densities) with $\begin{pmatrix}
        \cos(\vartheta) \\ \sin(\vartheta)
    \end{pmatrix}$.
In mixture 1 displayed in Fig. 2 of the main text, the skewness is $\mu_3\approx 0.022$, thus very close to zero (it is non-zero due to the finite size of the sample).
Note that for the discussion there, we treat it as being equal to zero.

\subsection{Mixture 2: non-zero skewness}
For mixture 2, the densities of the component types were generated as for mixture 1 from an exponential distribution with rate $\lambda=100$ and proper normalization, corresponding to a uniform distribution of the mixture composition on the $N{-}1$-simplex.
In order to achieve a non-zero skewness for mixture 2, we randomly generated the feature vectors as follows:
\begin{enumerate}
    \item Feature 1 (the first component of the feature vectors) was generated as a sum of two independent random variables, a Poisson distribution with variance $\sigma_{\mathrm{Poisson}}=1.5$ and a Gaussian distribution with zero mean and variance $\sigma_{\mathrm{Gaussian1}}=0.1$.
    \item Feature 2 was generated independently of feature 1 from a Gaussian distribution with zero mean and variance $\sigma_{\mathrm{Gaussian2}}=0.03$.
    \item To ensure a variance of 1 along the first principal component $\supbracket{V}{1}$ and to align it with the $x$-axis, we then rotated and rescaled them as described for mixture 1: We multiplied them by the matrix of eigenvectors of the covariance matrix (obtained after steps 1 and 2) and then rescaled the first and second components of the feature vectors by $\sigma_1/\sigma_{\mathrm{PC}1}$ and $\sigma_2/\sigma_{\mathrm{PC}2}$, respectively (again $\sigma_1=1$ and $\sigma_2=0.3$).
\end{enumerate} 
The relative enrichment $\delta \rho_i/\rho_i$ of component type $i$ was then determined analogously to mixture 1 by taking the scalar product of the difference of the respective feature vector from the mean of the feature vectors (weighted by the densities) with $\begin{pmatrix}
\cos(\vartheta) \\ \sin(\vartheta)
\end{pmatrix}$, where $\vartheta=0$ for mixture 2.
Mixture 2 displayed in Fig. 2 of the main text exhibits a skewness of $\mu_3\approx 0.99$.

\subsection{Invariance under constant shifts and rotations}
Note that all the physical results are invariant under constant shifts or rotations of the set of rescaled feature vectors, see Section~\ref{SMsec:global_rotation_translation}.
We tried to illustrate this invariance by choosing a principal component not aligned with any feature axis for mixture 1 and by generating a mixture with non-zero (weighted) mean for the feature vectors for mixture 2.
\subsection{Binary system}
The binary systems for both mixtures were determined as described in the previous section with $\mu_2=1$ and $\mu_3$ as determined from the empirical distributions of the relative enrichments (weighted by the densities of the component types).
In particular, we chose the feature vectors as \begin{align}
   \supbracket{\tilde{r}}{\gamma}_1 = (\sqrt{\mu_2 (1+\frac{1}{\tau})} + \mu_1)  \supbracket{V}{1}_{\gamma} \\
   \supbracket{\tilde{r}}{\gamma}_2 = (-\sqrt{\mu_2 (1+\frac{1}{\tau})} + \mu_1)  \supbracket{V}{1}_{\gamma},
\end{align}
where $\tau = 4 (\mu_2)^3/(\mu_3)^2$. 
Furthermore, $\mu_1$ is the (weighted) mean of the projections of the feature vectors of the original $N{=}1000$ mixture along the first principal component; $\mu_1$ is non-zero for mixture 2.
Note that, as noted previously, the constant shift by $\mu_1 \supbracket{V}{1}$ leaves the physical characteristics of the binary system invariant.
Finally, the densities were chosen as 
\begin{align}
   \phi_1 = \frac{1}{2} \left(1+\frac{1}{\sqrt{1+\tau}} \right) \\
   \phi_2 = \frac{1}{2} \left(1-\frac{1}{\sqrt{1+\tau}} \right).
\end{align}
\subsection{Histograms}
For both histograms, we chose a bin width of $0.1$.

\clearpage

\section{Useful formula}
\label{SMsec:useful_formula}

\subsection{Properties of the Hessian}

\begin{description}
    \item[Hessian acting on a vector] 
    For the Hessian $\supbracket{A}{2}$, Eq.~\ref{SMeq:Hessian}, and a vector $Z$ with $\avgrhoN{Z}=0$ we find
    \begin{align}
        ( \supbracket{A}{2} )_{ij} \rho_j Z_j=  Z_i - Z_N -\sum_{\gamma=1}^R \supbracket{r}{\gamma}_i \avgrhoN{\supbracket{r}{\gamma} Z}.
        \label{SMeq:Hessian_vector}
    \end{align}
    
    \textit{Proof:}
    \begin{align}
    	( \supbracket{A}{2} )_{ij} \rho_j Z_j &= \left( \frac{\delta_{ij}}{\rho_i} + \frac{1_{ij}}{1-\Rho} - \sum_{\gamma=1}^R \supbracket{r}{\gamma}_i \supbracket{r}{\gamma}_j \right) \rho_j Z_j = Z_i + \frac{1}{1-\Rho} \sum_{j=1}^{N-1} \rho_j Z_j - \sum_{\gamma=1}^R \supbracket{r}{\gamma}_i \sum_{j=1}^{N-1} \rho_j \supbracket{r}{\gamma}_j Z_j \\
	&= Z_i + \frac{1}{1-\Rho} \left(\avgrho{Z} - (1-\Rho) Z_N\right)- \sum_{\gamma=1}^R \supbracket{r}{\gamma}_i \sum_{j=1}^{N} \rho_j \supbracket{r}{\gamma}_j Z_j = Z_i - Z_N - \sum_{\gamma=1}^R \supbracket{r}{\gamma}_i \avg {\supbracket{r}{\gamma} Z},
    \end{align}
    since $\supbracket{r}{\gamma}_N=0$.
    
    \item[Full contraction of the Hessian] 
    
    If $\avgrhoN{v}=\avgrhoN{w}=0$, then
    \begin{align}
       \supbracket{A}{2}_{i j} \rho_{i} v_i \rho_j w_j  = \avgrhoN{v w}-\sum_{\gamma=1}^R \avgrhoN{v \supbracket{r}{\gamma}} \avgrhoN{w \supbracket{r}{\gamma}}
      \label{SMeq:useful_full_contraction_Hessian},
    \end{align}
    implicitly summing over all repeated indices on the left-hand side.
    
    \textit{Proof:}
    Using Eq.~\ref{SMeq:Hessian_vector} we have
    \begin{align}
       \supbracket{A}{2}_{i j} \rho_{i} v_i \rho_j w_j  &= \rho_i v_i (w_i - w_N - \sum_{\gamma=1}^R \supbracket{r}{\gamma}_i \avg {\supbracket{r}{\gamma} w}) = \\
       &=\avgrho{vw} - \left( 1-\Rho \right) v_N w_N - \left(\avgrho{v} - \left( 1-\Rho \right) v_N \right) w_N - \sum_{\gamma=1}^R \avgrho{\supbracket{r}{\gamma} v} \avgrho{\supbracket{r}{\gamma} w} = \\
       &=\avgrho{vw}-\sum_{\gamma=1}^R \avgrho{\supbracket{r}{\gamma} v} \avgrho{\supbracket{r}{\gamma} w}.
    \end{align}
    
\end{description}

\subsection{(Partial) Contraction of higher-order derivatives of the free energy}

\begin{description}
    \item[Partial contraction] 
    If $n\geq 3$ and $\avgrhoN{\supbracket{Z}{k}}=0 \ \forall k=1,\ldots,n-1$, then
    \begin{align}
       \supbracket{A}{n}_{i_1 i_2 {\ldots} i_{n-1} j} \prod_{k=1}^{n-1} \left(\rho_{i_k} \supbracket{Z}{k}_{i_k}\right) = (-1)^n (n-2)! \left[ \prod_{k=1}^{n-1}\supbracket{Z}{k}_j  - \prod_{k=1}^{n-1} \supbracket{Z}{k}_N  \right]
      \label{SMeq:useful_partial_contraction},
    \end{align}
    implicitly summing over all repeated indices on the left-hand side.
    
    \textit{Proof:}
    \begin{align*}
        \supbracket{A}{n}_{i_1 i_2 {\ldots} i_{n-1} j} \prod_{k=1}^{n-1} \left(\rho_{i_k} \supbracket{Z}{k}_{i_k}\right) &= (n-2)! \left[ \delta_{i_1 \ldots i_{n-1} j} (-1)^n \frac{1}{\rho_{i_1}^{n-1}} + 1_{i_1 \ldots i_{n-1} j} \frac{1}{(1-\Rho)^{n-1}} \right] \prod_{k=1}^{n-1} \left(\rho_{i_k} \supbracket{Z}{k}_{i_k}\right) = \\
        &= (n-2)! \left[ (-1)^n \prod_{k=1}^{n-1} \supbracket{Z}{k}_j + \frac{1}{(1-\Rho)^{n-1}}  \prod_{k=1}^{n-1} \left( \underbrace{\sum_{i_k=1}^{N-1} \rho_{i_k} \supbracket{Z}{k}_{i_k}}_{= \avgrhoN{\supbracket{Z}{k}} - \rho_N \supbracket{Z}{k}_N} \right) \right] =\\
        &= (n-2)! (-1)^n \left[ \prod_{k=1}^{n-1}\supbracket{Z}{k}_j  - \prod_{k=1}^{n-1} \supbracket{Z}{k}_N \right],
    \end{align*}
    where we used that $\rho_N = 1-\Rho$ and $\avgrhoN{\supbracket{Z}{k}}=0$.
    
    \item[Full contraction] 
    If $n\geq 3$ and $\avgrhoN{\supbracket{Z}{k}}=0 \ \forall k=1,\ldots,n$, then
    \begin{align}
       \supbracket{A}{n}_{i_1 i_2 {\ldots} i_n} \prod_{k=1}^{n} \left( \rho_{i_k} \supbracket{Z}{k}_{i_k} \right)  = (-1)^n (n-2)! \avgrhoN{\prod_{k=1}^{n}\supbracket{Z}{k} }
      \label{SMeq:useful_full_contraction},
    \end{align}
    implicitly summing over all repeated indices on the left-hand side.
    
    \textit{Proof:}
    \begin{align*}
        \supbracket{A}{n}_{i_1 i_2 {\ldots} i_{n}} \prod_{k=1}^{n} \left(\rho_{i_k} \supbracket{Z}{k}_{i_k}\right) &= (n-2)! (-1)^n \left[ \prod_{k=1}^{n-1}\supbracket{Z}{k}_{i_n}  - \prod_{k=1}^{n-1} \supbracket{Z}{k}_N \right] \rho_{i_n} \supbracket{Z}{n}_{i_n} =\\
        &=(n-2)! (-1)^n \left[ \underbrace{\sum_{l=1}^{N-1} \rho_l \prod_{k=1}^n \supbracket{Z}{k}_l}_{\avgrhoN{\prod_{k=1}^n \supbracket{Z}{k}} - (1-\Rho) \prod_{k=1}^n \supbracket{Z}{k}_N} - \left( \underbrace{\sum_{l=1}^{N-1} \rho_l \supbracket{Z}{n}_l}_{\avgrhoN{\supbracket{Z}{n}} - (1-\Rho) \supbracket{Z}{n}_N}\right) \prod_{k=1}^{n-1} \supbracket{Z}{k}_N \right] =\\
        &= (n-2)! (-1)^n \avgrhoN{\prod_{k=1}^n \supbracket{Z}{k}},
    \end{align*}
    where we used the partial contraction, Eq.~\ref{SMeq:useful_partial_contraction}, in the first equality, and that $\avgrhoN{\supbracket{Z}{n}}=0$.
    
\end{description}

\subsection{Pseudoinverse of the Hessian matrix}

\begin{description}
    \item[Pseudo-inverse $(\supbracket{A}{2}_{\text{pseudo}})^{-1}$ acting on a vector] 
    For the pseudo-inverse of the Hessian $(\supbracket{A}{2}_{\text{pseudo}})^{-1}$, Eq.~\ref{SMeq:pseudo_inverse}, and a vector $Z$ with $Z_N=0$ we find
    \begin{align}
        \left( (\supbracket{A}{2}_{\text{pseudo}})^{-1} \right)_{ij} Z_j= \rho_i \left( Z_i - \avgrhoN{Z} + \sum_{\gamma=2}^R \frac{1}{1-\supbracket{\lambda}{\gamma}} \supbracket{E}{\gamma}_i \avgrhoN{\supbracket{E}{\gamma} Z}\right).
        \label{SMeq:pseudo_inv_vector}
    \end{align}
    
    \textit{Proof:}
    \begin{align}
    \left( (\supbracket{A}{2}_{\text{pseudo}})^{-1} \right)_{ij} Z_j &= \left(\delta_{ij} \rho_j - \rho_i \rho_j + \sum_{\delta=2}^R  \frac{1}{1-\supbracket{\lambda}{\delta}} \supbracket{e}{\delta}_i \supbracket{e}{\delta}_j\right) Z_j = \\
    &= \rho_i Z_i - \rho_i \avgrhoN{Z} + \rho_i \sum_{\delta=2}^R \frac{1}{1-\supbracket{\lambda}{\delta}} \supbracket{E}{\delta}_i \avgrhoN{\supbracket{E}{\delta}Z}.
    \end{align}
    
\end{description}

\begin{description}
    \item[Product of the Hessian $\supbracket{A}{2}$ and its pseudo-inverse $(\supbracket{A}{2}_{\text{pseudo}})^{-1}$] 
    For the Hessian $\supbracket{A}{2} = H$, Eq.~\ref{SMeq:Hessian}, and its pseudo-inverse $(\supbracket{A}{2}_{\text{pseudo}})^{-1}$, Eq.~\ref{SMeq:pseudo_inverse}, we find
    \begin{align}
        \left(\supbracket{A}{2} (\supbracket{A}{2}_{\text{pseudo}})^{-1} \right)_{ij} = \delta_{ij} - \rho_j \supbracket{E}{1}_j \sum_{\gamma=1}^R \supbracket{V}{1}_{\gamma} \supbracket{r}{\gamma}_i = \left( (\supbracket{A}{2}_{\text{pseudo}})^{-1} \supbracket{A}{2}\right)_{ji}.
        \label{SMeq:prod_pseudoinv}
    \end{align}
    
    \textit{Proof:}
    \begin{align}
        &\left(\supbracket{A}{2} (\supbracket{A}{2}_{\text{pseudo}})^{-1} \right)_{ij} = \left( \delta_{ik} \frac{1}{\rho_i} + 1_{ik} \frac{1}{1-\Rho} - \sum_{\gamma=1}^R \supbracket{r}{\gamma}_i \supbracket{r}{\gamma}_k \right) \left(\delta_{kj} \rho_j - \rho_j \rho_k + \sum_{\delta=2}^R  \frac{1}{1-\supbracket{\lambda}{\delta}} \supbracket{e}{\delta}_j \supbracket{e}{\delta}_k\right) = \\
        &= \delta_{ij} + \sum_{\delta=2}^R \frac{1}{1-\supbracket{\lambda}{\delta}} \left[\supbracket{E}{\delta}_i - \supbracket{E}{\delta}_N \right] \supbracket{e}{\delta}_j - \rho_j \sum_{\gamma=1}^R \supbracket{r}{\gamma}_i \left[ \supbracket{r}{\gamma}_j - \avgrho{\supbracket{r}{\gamma}} \right] - \sum_{\delta=2}^R \frac{\supbracket{\lambda}{\delta}}{1-\supbracket{\lambda}{\delta}} \supbracket{e}{\delta}_j \sum_{\gamma=1}^R \supbracket{V}{\delta}_{\gamma} \supbracket{r}{\gamma}_i,
        \label{SMeq:prod_pseudoinv_1}
    \end{align}
    where we used that $\avgrho{\supbracket{E}{
    \delta}} = 0 = \supbracket{r}{\gamma}_N$ and
    \begin{align}
        \sum_{k=1}^{N-1} \supbracket{r}{\gamma}_k \supbracket{e}{\delta}_k = \avgrho{\supbracket{r}{\gamma} \supbracket{E}{\delta}} = \sum_{\alpha=1}^R \supbracket{V}{\delta}_{\alpha} \avgrho{\supbracket{r}{\gamma} \left(\supbracket{r}{\alpha} - \avgrho{\supbracket{r}{\alpha}} \right)} = \sum_{\alpha=1}^R \supbracket{V}{\delta}_{\alpha} \text{Cov}_{\gamma \alpha} = \supbracket{\lambda}{\delta} \supbracket{V}{\delta}_{\gamma}.
        \label{SMeq:avg_r_e}
    \end{align}
    Using that $\supbracket{E}{\delta}_i - \supbracket{E}{\delta}_N = \sum_{\alpha=1}^R \supbracket{V}{\delta}_{\alpha} \supbracket{r}{\alpha}_i$, we simplify Eq.~\ref{SMeq:prod_pseudoinv_1} to
    \begin{align}
        &\left(\supbracket{A}{2} (\supbracket{A}{2}_{\text{pseudo}})^{-1} \right)_{ij} = \delta_{ij} + \sum_{\delta=2}^R \supbracket{e}{\delta}_j  \sum_{\gamma=1}^R \supbracket{V}{\delta}_{\gamma} \supbracket{r}{\gamma}_i  -
        \rho_j \sum_{\gamma=1}^R \supbracket{r}{\gamma}_i \left[ \supbracket{r}{\gamma}_j - \avgrho{\supbracket{r}{\gamma}} \right].
        \label{SMeq:prod_pseudoinv_2}
    \end{align}
    As a next step, we rewrite 
    \begin{align}
        \sum_{\delta=2}^R \supbracket{V}{\delta}_{\gamma} \supbracket{E}{\delta}_j &= \sum_{\delta=2}^R \sum_{\alpha=1}^R \supbracket{V}{\delta}_{\gamma} \supbracket{V}{\delta}_{\alpha} \left(\supbracket{r}{\alpha}_j - \avgrho{\supbracket{r}{\alpha}} \right) = \sum_{\alpha=1}^R \left(\supbracket{r}{\alpha}_j - \avgrho{\supbracket{r}{\alpha}} \right) \sum_{\delta=2}^R \supbracket{V}{\delta}_{\gamma} \supbracket{V}{\delta}_{\alpha} = \\
        &=\sum_{\alpha=1}^R \left(\supbracket{r}{\alpha}_j - \avgrho{\supbracket{r}{\alpha}} \right) \left( \delta_{\alpha \gamma} - \supbracket{V}{1}_{\gamma} \supbracket{V}{1}_{\alpha} \right) = \\
        &=\supbracket{r}{\gamma}_j - \avgrho{\supbracket{r}{\gamma}} - \sum_{\alpha=1}^R \left(\supbracket{r}{\alpha}_j - \avgrho{\supbracket{r}{\alpha}} \right) \supbracket{V}{1}_{\gamma} \supbracket{V}{1}_{\alpha}.
    \end{align}
    Substituting this into Eq.~\ref{SMeq:prod_pseudoinv_2} yields
    \begin{align}
        \left(\supbracket{A}{2} (\supbracket{A}{2}_{\text{pseudo}})^{-1} \right)_{ij} = \delta_{ij} - \rho_j \supbracket{E}{1}_j \sum_{\gamma=1}^R \supbracket{V}{1}_{\gamma} \supbracket{r}{\gamma}_i.
        \label{SMeq:prod_pseudoinv_3}
    \end{align}
    Since both $\supbracket{A}{2}$ and $(\supbracket{A}{2}_{\text{pseudo}})^{-1}$ are symmetric, we have $\left(\supbracket{A}{2} (\supbracket{A}{2}_{\text{pseudo}})^{-1} \right)_{ij} = \left((\supbracket{A}{2}_{\text{pseudo}})^{-1} \supbracket{A}{2}  \right)_{ji}$.

    \item[Projection properties of $ (\supbracket{A}{2}_{\text{pseudo}})^{-1} \supbracket{A}{2}$]
    The product satisfies a few projection-operator like properties:
    \begin{enumerate}
        \item Applying it to the (non-orthogonal) basis $\{ \supbracket{e}{\gamma} \}_{\gamma=1,\ldots,R}$ maps to zero or leaves the vector invariant: $ \left((\supbracket{A}{2}_{\text{pseudo}})^{-1} \supbracket{A}{2}\right)_{ij} \supbracket{e}{\gamma}_j = \begin{cases} 0 \hspace{10pt} &\gamma=1 \\
        \supbracket{e}{\gamma}_i \hspace{10pt} &\gamma\neq 1\end{cases}$.
        \item Applying it several times, yields the same result: $\left((\supbracket{A}{2}_{\text{pseudo}})^{-1} \supbracket{A}{2}\right)^2 = (\supbracket{A}{2}_{\text{pseudo}})^{-1} \supbracket{A}{2}$.
    \end{enumerate}
    Thus, $(\supbracket{A}{2}_{\text{pseudo}})^{-1} \supbracket{A}{2}$ can be interpreted as a projection operator along $\supbracket{e}{1}$ with respect to the (non-orthogonal) basis $\{ \supbracket{e}{\gamma} \}_{\gamma=1,\ldots,R}$
    
    \textit{Proof:} 
    \begin{enumerate}
        \item Using Eqs.~\ref{SMeq:prod_pseudoinv_3} and~\ref{SMeq:avg_r_e} we have
        \begin{align}
            \left((\supbracket{A}{2}_{\text{pseudo}})^{-1} \supbracket{A}{2}\right)_{ij} \supbracket{e}{\delta}_j  &= \left( \delta_{ij} - \rho_i \supbracket{E}{1}_i \sum_{\gamma=1}^R \supbracket{V}{1}_{\gamma} \supbracket{r}{\gamma}_j \right) \supbracket{e}{\delta}_j = \supbracket{e}{\delta}_i -  \supbracket{e}{1}_i \sum_{\gamma=1}^R \supbracket{V}{1}_{\gamma} \sum_{j=1}^{N-1} \supbracket{r}{\gamma}_j \supbracket{e}{\delta}_j = \\
            &=\supbracket{e}{\delta}_i -  \supbracket{e}{1}_i \sum_{\gamma=1}^R \supbracket{V}{1}_{\gamma} \supbracket{\lambda}{\delta} \supbracket{V}{\delta}_{\gamma} =\supbracket{e}{\delta}_i -  \supbracket{e}{1}_i \delta_{1\delta}.
        \end{align}
        \item According to Eq.~\ref{SMeq:prod_pseudoinv}, $(\supbracket{A}{2}_{\text{pseudo}})^{-1} \supbracket{A}{2} = \mathbf{1} - C$ is equal to the identity matrix minus a matrix $C$, so the square is $(\mathbf{1}-C)^2 = \mathbf{1} - 2C + C^2$.
        Furthermore, using Eq.~\ref{SMeq:avg_r_e},
        \begin{align}
            C_{ij} C_{jk} &= \left(\rho_i \supbracket{E}{1}_i \sum_{\gamma=1}^R \supbracket{V}{1}_{\gamma} \supbracket{r}{\gamma}_j \right) \left( \rho_j \supbracket{E}{1}_j \sum_{\delta=1}^R \supbracket{V}{1}_{\delta} \supbracket{r}{\delta}_k \right) = \\
            &=
            \rho_i \supbracket{E}{1}_i \sum_{\delta=1}^R \supbracket{V}{1}_{\delta} \supbracket{r}{\delta}_k  \sum_{\gamma=1}^R \supbracket{V}{1}_{\gamma} \avgrho{\supbracket{r}{\gamma} \supbracket{E}{1}} = 
            C_{ik} \sum_{\gamma=1}^R \supbracket{V}{1}_{\gamma} \supbracket{V}{1}_{\gamma} = C_{ik}.
        \end{align}
        
        Thus, 
        \begin{align}
            \left((\supbracket{A}{2}_{\text{pseudo}})^{-1} \supbracket{A}{2} \right)^2 = (\supbracket{A}{2}_{\text{pseudo}})^{-1} \supbracket{A}{2}.
        \end{align}
    \end{enumerate}
    
\end{description}

\clearpage

\footnotesize
\section{Glossary of symbols}

\subsection{Original system}

\begin{center}
	\begin{tabular}{c|c}
	\mbox{ } \hspace{20pt}\textbf{symbol} \hspace{20pt} \mbox{ } & \mbox{ } \hspace{20pt} \textbf{meaning} \hspace{20pt} \mbox{ } \\
	\hline
	greek letters &  feature indices (from 1 to $R$) \\
	\hline
	roman letters & component indices (from 1 to $N$ or $N-1$) \\
	\hline
	$\rho_i$ & density of component $i$ \\
	\hline
	$\Rho= \sum_{i=1}^{N-1} \rho_i = 1-\rho_N$ & density of all components except $N$ \\
	\hline
	$z$ & lattice coordination number \\
	\hline
	$k_B, T$ & Boltzmann constant, temperature \\
	\hline
	$\supbracket{J}{\gamma}$ & interaction strength of $\gamma$-th feature \\
	\hline
	$\supbracket{C}{\gamma}:= \frac{z \supbracket{J}{\gamma}}{2 k_B T}$ & rescaled interaction strength of $\gamma$-th feature\\
	\hline
	$\supbracket{s}{\gamma}_i$ & $\gamma$-th feature (spin) of component $i$ \\
	\hline
	$\supbracket{r}{\gamma}_i:= \sqrt{2 \supbracket{C}{\gamma}} (\supbracket{s}{\gamma}_i-\supbracket{s}{\gamma}_N)$  & rescaled  and shifted $\gamma$-th feature (spin) of component $i$\\
	\hline
	$\chi_{ij}:= \sum_{\gamma=1}^R \supbracket{C}{\gamma} \supbracket{s}{\gamma}_i \supbracket{s}{\gamma}_j$ & interaction matrix\\
	\hline
	$H_{ij}:= \frac{\partial^2 f_{N-1}}{\partial \rho_i \partial \rho_j}, \ i,j = 1, \ldots, N-1$ & \shortstack[c]{Hessian matrix of free energy as function of components $i=1,\ldots,N-1$\\ using $\rho_N = \rho_N (\rho_1, \ldots, \rho_{N-1}) = 1-\Rho$}\\
	\hline
	$K_{ij}:= H_{ij} + \sum_{\gamma=1}^R \supbracket{r}{\gamma}_i \supbracket{r}{\gamma}_j, \ i,j = 1, \ldots, N-1$ & non-interacting part of Hessian matrix\\
	\hline
	$\supbracket{A}{n}_{i_1, \ldots, i_n}:= \frac{\partial^n f_{N-1}}{\partial \rho_{i_1} \ldots \partial \rho_{i_n}}, \ i_k = 1, \ldots, N-1$ & higher-order derivatives of free energy\\
	\hline
	$U_{i\alpha} := \supbracket{r}{\alpha}_i =: W_{\alpha i}, \ \alpha=1, \ldots, R, \ i=1,\ldots,N-1$ & matrix of rescaled features\\
	\hline
	$\avgrhoN{X}:=\avgrho{X}:=\sum_{i=1}^N \rho_i X_i$ & averages with respect to the overall mixture composition\\
	\hline
	$\mathrm{Cov}_{\alpha\beta} := \avgrho{\supbracket{r}{\alpha} \supbracket{r}{\beta}} - \avgrho{\supbracket{r}{\alpha}} \avgrho{ \supbracket{r}{\beta}}, \ \alpha,\beta=1, \ldots, R$ & covariance matrix between features\\
	\hline
	$\mathrm{Cov}_{\alpha\beta} := \sum_{\gamma=1}^R \supbracket{\lambda}{\gamma} \supbracket{V}{\gamma}_{\alpha} \supbracket{V}{\gamma}_{\beta}$ & \shortstack[c]{eigendecomposition of covariance matrix into (descending) \\ eigenvalues $\supbracket{\lambda}{\gamma}$, $\gamma=1, \ldots, R$ and corresponding eigenvectors $\supbracket{\vec{V}}{\gamma}$}\\
	\hline
	$\supbracket{e}{\gamma}_i := \rho_i \supbracket{E}{\gamma}_i := \rho_i \sum_{\alpha=1}^R \supbracket{V}{\gamma}_{\alpha} (\supbracket{r}{\alpha}_i - \avgrho{\supbracket{r}{\alpha}}), \ \gamma=1, \ldots, R$ & features projected onto $\gamma$-th eigenvector of covariance matrix\\
	\hline
	$\rho_i (\epsilon) = \rhocp_i + \delta \rho_i (\epsilon):= \rhocp_i (1+\sum_{m=1}^{\infty} \frac{\epsilon^m}{m!} \supbracket{\omega}{m}_i ) =: \rhocp_i G(\epsilon) $ & curve in density space around the critical density $\rhocp$, parameterized by $\epsilon$\\
	\hline
	$\rho_i (\epsilon) = \rhocp_i \frac{e^{h(\epsilon)}}{\avgcp{e^{h(\epsilon)}}} $ & curve expressed in terms of a ``Hamiltonian-like" function $h$\\
	\hline
	$\avgrhoNcp{X}:=\avgcp{X}:=\sum_{i=1}^N \rhocp_i X_i$ & averages with respect to critical density\\
	\hline
	$\supbracket{\omega}{m}$ & $m$-th derivative of curve (wrt $\epsilon$), relative to critical density $\rhocp$\\
	\hline
	$\supbracket{\omega}{m} = \supbracket{\Omega}{m} = \supbracket{G}{m} (\epsilon)|_{\epsilon=0}$ & derivatives defining optimal path\\
	\hline
	$\supbracket{\upsilon}{m}_i = \rhocp_i \supbracket{\omega}{m}_i $ & $m$-th derivative of curve
	\\
	\hline
	$\supbracket{\Upsilon}{m}_i = \rhocp_i \supbracket{\Omega}{m}_i $ & $m$-th derivative of optimal curve\\
	\hline
	$(\supbracket{A}{2}_{\mathrm{pseudo}})^{-1}_{ij} :=\delta_{ij} \rho_i - \rho_i \rho_j + \sum_{\gamma=2}^R \frac{1}{1-\supbracket{\lambda}{\gamma}} \supbracket{e}{\gamma}_i \supbracket{e}{\gamma}_j $ & pseudo-inverse of Hessian matrix\\
	\hline
	$B_{n,l}$ & exponential Bell polynomial\\
	\hline
	$\supbracket{A}{n} \prod_{j=1}^n \supbracket{Z}{j} := \sum_{i_1, \ldots, i_n} \supbracket{A}{n}_{i_1 \ldots i_n} \prod_{j=1}^n \supbracket{Z}{j}_{i_j}$ & contraction\\
	\hline
	$(\supbracket{A}{n} \prod_{j=1}^{n-1} \supbracket{Z}{j})_k := \sum_{i_1, \ldots, i_{n-1}} \supbracket{A}{n}_{k i_1 \ldots i_{n-1}} \prod_{j=1}^{n-1} \supbracket{Z}{j}_{i_j}$ & partial contraction into a vector\\
	\hline
	$[XY]_k := X_k Y_k$ & element-wise multiplication of two vectors\\
	\hline
	$\supbracket{b}{m}:=\sum_{l=3}^{m+1} \supbracket{A}{l} B_{m,l-1}(\supbracket{\Upsilon}{1}, \ldots, \supbracket{\Upsilon}{m-1})$ & used for recursion relation of the vectors $\supbracket{\Upsilon}{i}$\\
	\hline
	$K_r (\epsilon) := \log \avgcp{e^{\epsilon r}}$ & cumulant generating function for the feature $r=\supbracket{r}{1}$ if $R=1$\\
	\hline
	$\kappa_i := \left.\frac{\mathrm{d}^i}{\mathrm{d} \epsilon^i} K_r (\epsilon)\right|_{\epsilon=0}$ & $i$-th cumulant
	\end{tabular}
\end{center}
	
\subsection{System with components of different sizes}

\begin{center}
	\begin{tabular}{c|c}
	\mbox{ } \hspace{20pt}\textbf{symbol} \hspace{20pt} \mbox{ } & \mbox{ } \hspace{20pt} \textbf{meaning} \hspace{20pt} \mbox{ } \\
	\hline
	$\tilde{X}$ & denotes analogous quantities $X$ of original system\\
	\hline
	$l_i$ &  length of component $i$\\
	\hline
	$\bar{l}:=\sum_{i=1}^N \rho_i l_i$ & average length\\
	\hline
	$\phi_m:= \rho_m l_m/\bar{l}$ &  effective probability measure\\
	\hline
	$\avgphi{X}:=\sum_{i=1}^N \phi_i X_i$ &  averages with respect to $\phi$\\
	\hline
	$\supbracket{\mathrm{Cov}}{\phi}_{\alpha \beta}:=\avgphi{\supbracket{r}{\alpha} \supbracket{r}{\beta}}- \avgphi{\supbracket{r}{\alpha}}\avgphi{ \supbracket{r}{\beta}} =: \sum_{\gamma=1}^R \supbracket{\tilde{\lambda}}{\gamma} \supbracket{\tilde{V}}{\gamma}_{\alpha} \supbracket{\tilde{V}}{\gamma}_{\beta}$ &  covariance matrix with respect to $\phi$ and its eigendecomposition
	\end{tabular}
\end{center}

\clearpage

\subsection{System with negative interaction strengths}

\begin{center}
	\begin{tabular}{c|c}
	\mbox{ } \hspace{20pt}\textbf{symbol} \hspace{20pt} \mbox{ } & \mbox{ } \hspace{20pt} \textbf{meaning} \hspace{20pt} \mbox{ } \\
	\hline
	$i=1,\ldots,R^+$ & \shortstack[c]{indices of features with positive interaction strengths\\ (``positive features")}\\
	\hline
	$i=R^+ +1,\ldots, R$ &  \shortstack[c]{indices of features with negative interaction strengths\\ (``negative features")}\\
	\hline
	$\supbracket{\tilde{r}}{\gamma} = \begin{cases} \supbracket{r}{\gamma} \ \ \gamma=1,\ldots,R^+\\ -i \supbracket{r}{\gamma} \ \ \gamma=R^+ +1,\ldots,R = R^+ + R^- \end{cases}$ & real version of the features \\
	\hline
	$\supbracket{\mathrm{Cov}}{\tilde{r}}$ &  covariance matrix of the real features\\
	\hline
	$\supbracket{C}{\pm\pm/\pm\mp}$ &  \shortstack[c]{submatrix of $\supbracket{\mathrm{Cov}}{\tilde{r}}$ with respect to\\ positive $+$ and/or negative $-$ features} \\
	\hline
	$\bar{C}:=\cpp - \cpm (\mathbf{1}+\cmm)^{-1} \cmp$ &  effective covariance matrix (dimension $R^+\times R^+$)\\
	\hline
	$\bar{C}_{\alpha\beta}:=\sum_{\gamma=1}^{R^+} \supbracket{\bar{\lambda}}{\gamma} \supbracket{\phi}{\gamma}_{\alpha} \supbracket{\phi}{\gamma}_{\beta}$ &  eigendecomposition of $\bar{C}$ in terms of descending eigenvalues\\
	\hline
	$\supbracket{\pi}{\gamma}_i := \supbracket{\tilde{r}}{\gamma}_i - \avgrho{\supbracket{\tilde{r}}{\gamma}}, \ \gamma=1,\ldots, R^+$ &  rescaled and shifted positive features\\
	\hline
	$\supbracket{\nu}{\gamma}_i := \supbracket{\tilde{r}}{R^+ +\gamma}_i - \avgrho{\supbracket{\tilde{r}}{R^+ +\gamma}}, \ \gamma=1,\ldots,R^-$ &  rescaled and shifted negative features (counted from 1 to $R^- {=} R{-}R^+$)\\
	\hline
	$\supbracket{\bar{e}}{1}_i := \rho_i \supbracket{\bar{E}}{1}_i := \rho_i \sum_{\alpha=1}^{R^+} \supbracket{\phi}{1}_{\alpha} [\supbracket{\pi}{\alpha}_i - \sum_{\beta,\delta=1}^{R^-} \cpm_{\alpha \beta} (\mathbf{1} + \cmm)^{-1}_{\beta \delta} \supbracket{\nu}{\delta}_i ]$ &  direction of instability
	\end{tabular}
\end{center}

\end{document}